\documentclass[aps,prb,epsfigure,twocolumn,nofootinbib]{revtex4} 
\usepackage{dcolumn}    
\usepackage{bm} 
\usepackage{graphicx}
\usepackage{amsmath}    
\usepackage{latexsym}
\usepackage{amsfonts}   
\usepackage{amssymb}
\usepackage{array}      
\usepackage{epsfig}
\usepackage{txfonts}
\usepackage[colorlinks=true,linkcolor=blue,urlcolor=blue,citecolor=blue]{hyperref}
\usepackage{color}
\usepackage{hyperref}
\usepackage{mathrsfs}
\usepackage{calligra}

\usepackage{caption}
\usepackage{subcaption}

\makeatletter
\newlength{\captionindent}
\long\def\@makecaption#1#2{%
  \vskip\abovecaptionskip
  \sbox\@tempboxa{#1: #2}%
  \ifdim \wd\@tempboxa >\hsize
    \hspace*{\captionindent}#1: #2\par
  \else
    \global \@minipagefalse
    \hb@xt@\hsize{\hfil\box\@tempboxa\hfil}%
  \fi
  \vskip\belowcaptionskip}
\makeatother

\usepackage{soul}
\definecolor{applegreen}{rgb}{0.55, 0.71, 0.0}

\usepackage[export]{adjustbox}
\usepackage{amsmath}

\graphicspath{{Figures/}}

\newcommand{\ee}{\mathrm{e}}
\newcommand{\ii}{\mathrm{i}}

\definecolor{amber}{rgb}{1.0, 0.49, 0.0}

\newcommand{\blue}[1]{{\color{blue} #1 }}

\makeatletter
\newcommand{\manuallabel}[2]{\def\@currentlabel{#2}\label{#1}}
\makeatother

\begin{document}
\title{The Rhombi-Chain Bose-Hubbard Model:\\  
geometric frustration and interactions}

\author{Christine Cartwright} 
\affiliation{Centre for Theoretical Atomic, Molecular and Optical Physics, 
Queen's University Belfast, Belfast BT7 1NN, United Kingdom}
\author{Gabriele De Chiara} 
\affiliation{Centre for Theoretical Atomic, Molecular and Optical Physics, Queen's University Belfast, Belfast BT7 1NN, United Kingdom}
\author{Matteo Rizzi}
\affiliation{Institut f\"ur Physik, Johannes Gutenberg Universit\"at, Staudingerweg 7, 55099 Mainz, Germany}

\begin{abstract}
We explore the effects of geometric frustration within a one-dimensional Bose-Hubbard model using a chain of rhombi subject to a magnetic flux. 
The competition of tunnelling, self-interaction and magnetic flux gives rise to the emergence of a pair-superfluid (pair-Luttinger liquid) phase
besides the more conventional Mott-insulator and superfluid (Luttinger liquid) phases. 
We compute the complete phase diagram of the model by identifying characteristic properties of the pair-Luttinger liquid phase 
such as pair correlation functions and structure factors 
and find that the pair-Luttinger liquid phase is very sensitive to changes away from perfect frustration (half-flux). We provide some proposals to make the model more resilient to variants away from perfect frustration.
We also study the bipartite entanglement properties of the chain. 
We discover that, while the scaling of the block entropy pair-superfluid and of the single-particle superfluid leads to the same central charge,
the properties of the low-lying entanglement spectrum levels reveal their fundamental difference.
\end{abstract}

\date{\today}

\maketitle

\section{Introduction}\label{sectintro} 

Largely degenerate low-energy manifolds appear in diverse physical contexts,
ranging from frustrated spin systems~\cite{Moessner2006,Balents2010} to disordered media with random impurities~\cite{Lagendijk2009},
passing by the celebrated Landau levels of a 2D electron gas in a transverse magnetic field~\cite{Ezawa2008}.
In all these cases, the flatness of the energy landscape gives rise to intriguing phenomena, 
like spin liquids (i.e. stable phases with no broken symmetry at all), localization phenomena of various kinds, and the quantum Hall effect.
A central role in the creation of such peculiar degeneracies is played by constraints which
descend from geometrical reasons (like the dimensionality of the system and the form of the underlying
lattice structure, when present) or from gauge potentials (e.g. the vector potential of the above mentioned
magnetic field), and often by their interplay.
Interestingly, localization can be achieved even in the absence of disorder by simply mixing these latter two ingredients,
as in the case of Aharanov-Bohm cages~\cite{Sutherland1986, ABcage1998}, 
or by properly tuned long-range hopping terms without any net magnetic field~\cite{Tang2011,Sun2011,Neupert2011}.

The insertion of interactions among the system components on top of flat dispersion bands leads to even richer scenarios for many-body physics,
often strongly correlated and profoundly non-perturbative.
The two archetypical ones are arguably 
i) the occurrence of ferromagnetism in repulsive flat-band Hubbard models as guaranteed
by rigorous results by Lieb~\cite{Lieb1989}, Mielke~\cite{Mielke1991a, Mielke1991b}, and Tasaki~\cite{Tasaki1992, Tasaki1998};
and ii) the emergence of a many-body spectrum which is itself nearly-degenerate, 
as is the case in fractional quantum Hall effect(s) and with anyonic quasi-excitations~\cite{Tsui1982, Laughlin1983, Moore1991, Wilczek1990}.
More recently, a flurry of interest has been blowing about possible realisations of flat-band topological insulators with nontrivial Chern numbers~\cite{Tang2011,Sun2011,Neupert2011,LiZhaoLiu2013,MRizzi2017}.

The interest in frustrated lattices for mobile quantum particles has been further enhanced by the availability of platforms for tailoring so-called \emph{synthetic quantum matter}~\cite{Leykam2018}: 
e.g., quantum-dot lattices for electrons~\cite{ReimannManninen2002}, Josephson junction arrays for Cooper pairs~\cite{FazioVanderZant2001}, 
photonic lattices~\cite{Haldane2008,Hafezi2013,Ningyuan2015,Deng2016,Ozawa2018}
and optical lattices for cold atoms~\cite{BlochDalibardZwerger2008,LewensteinSanpera2012,DalibardNascimbene2012}. 
In the latter, despite the charge neutrality of the constituents, it is nowadays routine to produce synthetic gauge fields,
via laser-assisted tunnelling~\cite{Goldman2014, Dalibard2011}
and/or via shaking of the lattice structure~\cite{Holthaus1997,Sias2008,Struck2011,Eckardt2017}
Moreover, it is possible to load the lattices with fermionic or bosonic particles, and even with mixtures,
while also tuning the interactions among them via Fano-Feshbach resonances.

For bosons, the dominance of interaction over kinetic terms leads to incompressible Wigner-crystal-like ground states at certain fractional fillings. These are determined by the possibility of occupying non-overlapping localized eigenstates and may also lead to the appearence of supersolid phases~\cite{Altman2010,MoellerCooper2012}. Here it is possible to obtain a pair of particles by adding a particle to the critical density which forms the Wigner crystals \cite{Mielke2018}. In the opposite limit of large occupation number, known as the quantum rotor limit and particularly relevant for Josephson junction arrays,
Dou\c{c}ot and Vidal highlighted the possibility of obtaining a coherent transport of particle pairs with the corresponding absence of single particle transport~\cite{VidalDoucot2000,Doucot2002,Rizzi2006JJ}. 
In this case there is no need to invoke three-body hardcore constraints to obtain pair superfluids in cold atomic systems as used in other studies~\cite{Daley2009,Daley2010,DiehlDaley2010,MazzaRizzi2010,Daley2014,Lewenstein2015}.

Flat bands and unconventional pairing coincide well with fermions.
For attractive interactions, an intriguing connection has recently been highlighted between the quantum metric of the bands (which is distinct from, but related to, the Chern number)
and the BCS superfluid density of the system. This arises even in the absence of a Fermi surface~\cite{PeottaTorma2015,Huber2016}. For the repulsive case, a superfluid appears at fillings lower than that which stabilises a crystalline insulating phase (similar to the one mentioned for bosons)~\cite{AokiKobayashi2016}.

We point the reader to a very recent contribution by Tovmasyan et al., 
which presents a unifying picture for flat-bands loaded with particles of quantum statistics~\cite{Tovmasyan2018}.
We also notice that comparative studies of the dynamics of few-particle fermionic and bosonic quasi one-dimensional systems with flat bands have been carried out in Ref.~\cite{Hyrkas2013}. Scattering processes throughout a flat band system have been examined in Ref.~\cite{Dutta2017}.

\begin{figure}[t] 
\includegraphics[width=1\columnwidth]{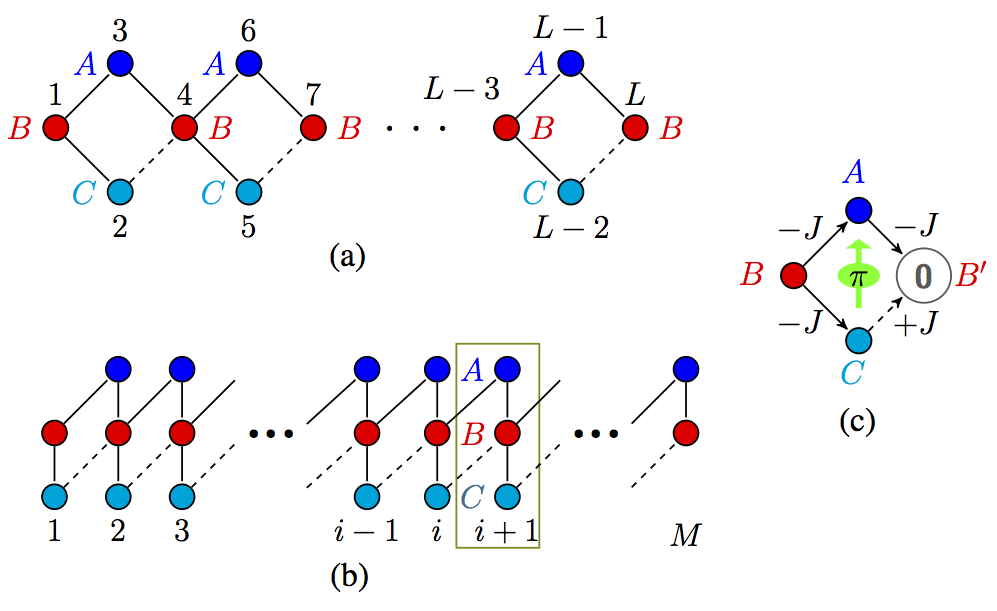}
\caption{One-dimensional lattice of rhombi with $M$ cells and $L$ sites. Each cell contains three sites labelled $A,B,C$. Solid (dashed) line connections indicate a tunnelling amplitude $-J$ $(-J \ee^{i\phi})$. \textbf{(a)} is the a representation of the model in real space showing the number of sites whereas \textbf{(b)} is a schematic diagram to illustrate the cells more clearly. \textbf{(c)} is an illustration of a restricted single particle tunnelling for the fully-frustrated system.}
\label{model}
 \end{figure}

In this work, we focus on a Bose-Hubbard model for a one-dimensional (1D) lattice of rhombi 
(also often referred to as a diamond lattice or $AB_2$ lattice), 
where each rhombus is pierced by a tunable magnetic flux $\phi$ (see Fig.~\ref{model}).
When $\phi$ is an odd multiple of $\pi$, all three bands of the single particle dispersion relation become flat (Fig.~\ref{momentumpic}) and a complete basis of fully localized Ahranov-Bohm cages (see Fig.~\ref{model}c) is present (Sec.~\ref{sectmod}).
This specific lattice structure (namely the same as in the paper by Dou\c{c}ot and Vidal~\cite{Doucot2002})
has very recently received a revival in attention, 
due to both its experimental realization in two distinct photonic waveguide platforms~\cite{Kremer2018,Mukherjee2018}
and to a couple of novel theoretical insights which describe the formation of pairs~\cite{Tovmasyan2018} 
and a hidden topological character of the bands~\cite{Kremer2018}.
We notice that the latter could possibly have a relation to the hidden $\mathbb{Z}_2$ symmetry,
that led people in the community of Josephson arrays to propose
this kind of geometry for building a topologically protected quantum memory~\cite{DoucotJJ2009,Ioffe2002, Doucot2012}.
We focus on a commensurate integer filling of the lattice, i.e., one (or two) particle(s) per site. We explore the phase diagram as a function of the magnetic flux and the on-site repulsive interaction (Sec.~\ref{sectphasediag}), using density matrix renormalisation group (DMRG) with Matrix Product States (MPS)~\cite{DMRG1,DMRG2,DMRG3} simulations, to address some of the questions (re-)opened by these recent contributions.\\
\indent \emph{First} we confirm the expectations in the two extremal regimes of frustration, at zero and $\pi$ flux.
At zero flux, the well-known Mott insulator (MI) to superfluid 
-- or Luttinger liquid (LL) as we are in quasi-1D -- phase takes place,
at a critical coupling  slightly renormalized due to the microscopic geometry of the lattice.
At magnetic flux $\pi$, single particle transport is absent and the LL cannot occur but the Mott lobe still closes (Sec.~\ref{sectphase}) and transport occurs by the flow of boson-pairs,
which constitute a more exotic pair-Luttinger liquid (PLL)~\cite{VidalDoucot2000, Doucot2002}.
Incidentally, we address the reader to other recent references for fractional fillings~\cite{Huber2013,Takayoshi2013,PhillipsPRB2015,Anisimovas2016}
where pair Luttinger liquids have also been reported.\\
\indent \emph{Then} we show that the PLL unfortunately gets destroyed very quickly at imperfect frustration (i.e., $\phi = \pi (1 - \epsilon)$). This is consistent with the qualitative exponential prediction by Dou\c{c}ot and Vidal~\cite{Doucot2002}, 
i.e., $\epsilon \simeq \exp(-J/U)$.
Noticeably, the size of the region can be (marginally) extended by using higher bosonic filling,
as this pushes the Mott fluctuations further back to a smaller $J/U$
(as we exemplify with two particles per site).
It would thus be interesting to find a way to predict an optimal filling to get reasonable experimental errors in the flux
(larger than the present $0.5\%$), allowing for a realistic detection of PLL,
but this goes beyond the scope of our present work. We also propose an alternative method using amplitude modulation which is much more resilient to imperfect flux. This is achievable using a digital micromirror device (DMD) or a single atom microscope.\\
\indent \emph{Additionally}, we employ entanglement analysis to distinguish the two gapless Luttinger phases (for recent reviews see Refs.\cite{CalabreseCardyReview2009,LaflorencieReview,DeChiaraSanpera2018}). Although the conformal field theory (CFT) central charge extracted from the scaling of the entanglement entropy
is not able to distinguish them, a noticeable difference emerges when looking at the degeneracy pattern of the low-lying entanglement spectrum levels (Sec.~\ref{sectentang}).
Finally, we state the conclusions of our main findings and offer our perspective on further works to be carried out. (Sec~\ref{sectconc}).

\section{The Model} \label{sectmod}

The Bose-Hubbard model for a quasi-one-dimensional chain of rhombi is our focus, as mentioned in the Introduction (Sec~\ref{sectintro}). The geometric formulation is depicted in Fig.~\ref{model}.
The unit cell of such a lattice is made of three sites that we label $A, B$ and $C$. The coordination number of $A,C$ is two, while for $B$ it is four \footnote{The reader is asked to note that this notation differs from that of Vidal and Dou\c{c}ot \cite{Doucot2002} and of Mukherjee et al. \cite{Mukherjee2018} }.
We consider only nearest-neighbour hopping and on-site interactions, which is common in cold-atomic setups. 
The Hamiltonian is:
\begin{eqnarray} 
\label{BHeqn}
\hat{H}_{BH}&=& \hat{H}_0 + \hat{H}_U  \equiv 
\hat{H}_0 + \frac{U}{2}\sum_j \sum_{\alpha} \hat{n}_{j,\alpha}(\hat{n}_{j,\alpha}-1) \\
\hat{H}_0 & = & - J \sum_j \sum_\ell \sum_{\alpha,\beta} 
T^{(\ell)}_{\alpha ,\beta} \ \hat{b}^{\dagger}_{j+\ell,\alpha} \hat{b}^{\phantom{\dagger}}_{j,\beta} \ ,
\end{eqnarray}
where the index $j$ denotes the lattice cell, the Greek letters label the basis inside a cell, i.e., $\alpha, \beta \in \{A,B,C\}$, and $\ell \in \{0, \pm 1\}$ represents the (relative coordinate of the) cell where the particle is hopping to. 
We have introduced the annihilation (creation) operators $\hat{b}^{(\dagger)}_{j,\alpha}$, 
and the number operator 
$\hat{n}^{\phantom{\dagger}}_{j,\alpha}=\hat{b}_{j,\alpha}^{\dagger} \hat{b}^{\phantom{\dagger}}_{j,\alpha}$. 
In this manuscript we consider a chain with open boundary conditions, as shown in Fig.~\ref{model}. In order to avoid spurious effects at the edges, we deal with $M$ cells, of which only $M-2$ are complete, i.e., we take $M-1$ full rhombi that correspond to $L=3 M - 2$ sites in total. 

In order to accommodate a piercing (synthetic) magnetic flux $\phi$ through each rhombus,
we choose the hopping matrices to be:
\begin{equation}\label{eq:tunnell}
{T}^{(0)} = \left(\begin{array}{ccc}
0&1&0\\
1&0&1\\
0&1&0\\
\end{array} \right),
\quad
{T}^{(+1)}=\left(\begin{array}{ccc}
0&1&0\\
0&0& \ee^{\ii\phi}\\
0&0&0\\
\end{array} \right),
\quad
{T}^{(-1)}= \left({T}^{(+1)}\right)^\dagger
\end{equation}
The dashed connections in Fig.~\ref{model} denote a tunnelling coefficient of $- J \ee^{\ii\phi}$,
whereas the solid lines have tunnelling coefficient $- J$. 
We stress here that this is one of the many possible gauge choices:
e.g., distributing homogeneously the flux as $\ee^{\pm \ii \phi/4}$ on each link 
might be even more convenient for experimental purposes~\cite{Pannetier1984, Sooyeul1998, Polini2005}. We notice that the already mentioned recent photonic implementations make use of the single-link~\cite{Kremer2018} and the four-link gauge~\cite{Mukherjee2018}, respectively.
 
\begin{figure}[tb] \center
\includegraphics[width=0.84\columnwidth]{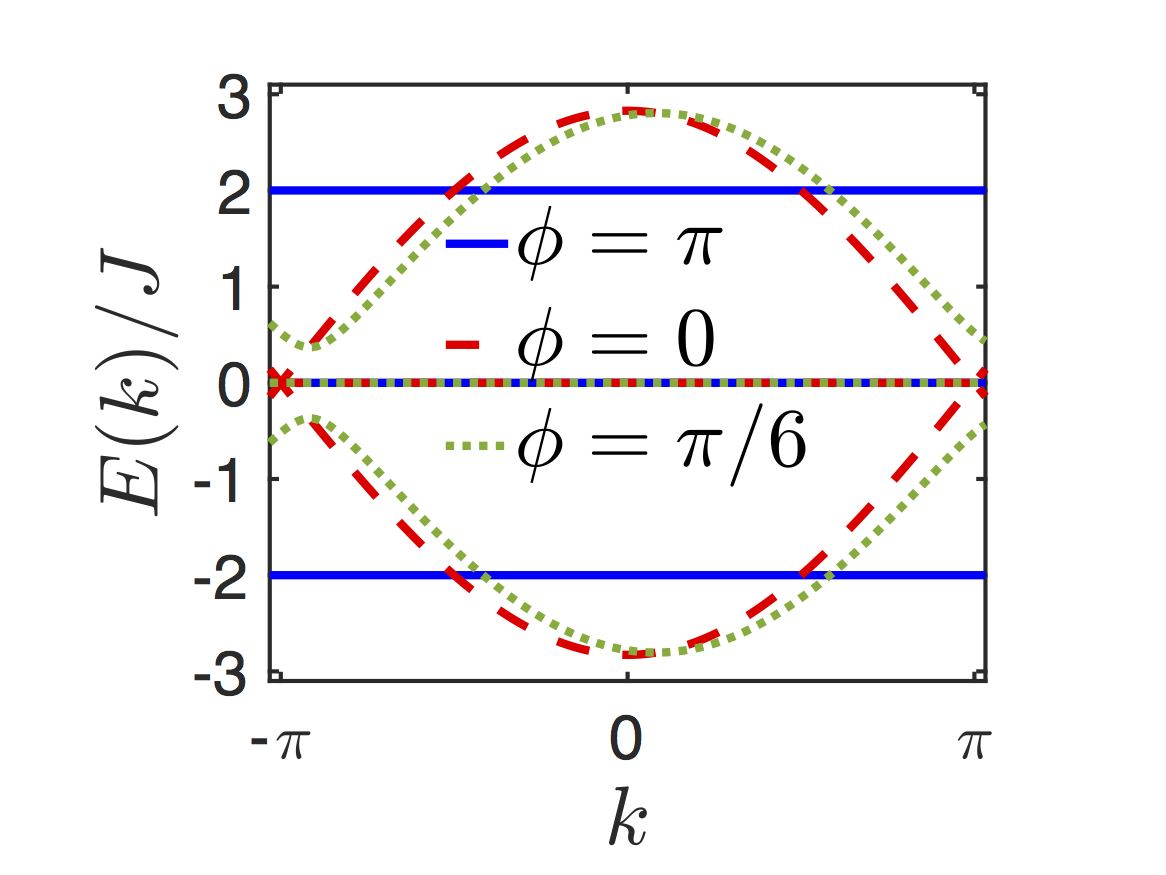}
 \caption{The single particle energy band $E(k)$ as a function of the lattice momentum $k$. The fully-frustrated $\phi=\pi$ bands are the solid lines, whereas the non-frustrated $\phi=0$ are shown by dashed lines. An intermediate frustration (at $\phi=\frac{\pi}{6}$ ) is shown by the dotted lines.}
\label{momentumpic}
 \end{figure}

Let us focus first on the non-interacting Hamiltonian $H_0$ and its band structure
in the infinite, perfectly translational invariant regime (i.e., with no edges):
\begin{equation}
	E_\tau (k) = 2J \tau \sqrt{1+\cos\left(k-\frac{\phi}{2}\right)\cos\left(\frac{\phi}{2}\right)}\ ,
\end{equation} 
where $\tau = 0, \pm1$ denotes the three bands.
A chiral (sub-lattice) symmetry operator $\Gamma = \mathrm{diag}\{-1,+1,-1\}$,
such that $\Gamma^2 = \mathbb{I}$ and $\Gamma H_0(k) \Gamma = - H_0(k)$,
is robust with respect to the gauge choice
(while other choices could also display lattice-inversion symmetry, for example).
This does not, however, constrain the three bands with well-defined topological invariants. These could instead emerge by dealing with the squared Hamiltonian, as very recently commented in Ref.~\cite{Kremer2018}. As the band structure is invariant under the insertion of integer flux-quanta (i.e. under $\phi \to \phi+2\pi$), we restrict ourselves to the range $\phi \in [-\pi, \pi$].
Interestingly, the flat middle band $\tau=0$ is insensitive to $\phi$ 
and occurs purely due to geometrical reasons~\cite{Travkin2000}.
In particular, the corresponding eigenmodes $\hat{w}^{(\dagger)}_{j,0}$ 
have zero amplitude on the $B$~site of cell~$j$ around which they are centred (see Fig.~\ref{model}).
As visible in Fig.~\ref{momentumpic}, the curvature of the other two bands decreases with growing flux until they become perfectly flat at \emph{full frustration}, i.e., $\phi=\pi$: 
$E^{(\phi=\pi)}_\tau (k) = 2J\tau $~\cite{Doucot2002}.

The simultaneous flatness of \emph{all} bands can be understood in terms of Aharanov-Bohm cages~\cite{ABcage1998,Doucot2002,Altman2010}, i.e., of perfectly localized eigenmodes $\hat{w}^{(\dagger)}_{j,\pm}$, which occur due to destructive interference preventing the movement of \emph{single} particles from one $B$~site (in cell~$j$)
to another (as illustrated in Fig.~\ref{model}).
In our gauge, these localised modes are:
\begin{equation}\label{eq:bulkmodes}
\hat{w}_{j,\tau} = 
\frac{(-1)^\tau \, \hat{b}_{j-1,C} - (-1)^\tau \, \hat{b}_{j,A} - 2\tau \, \hat{b}_{j,B} + \hat{b}_{j,C} + \hat{b}_{j+1,A}}{\left(2^{1+|\tau|/2}\right)} \ .
\end{equation}
The presence of the edges in our open boundary setup gives rise to two extra mid-gap modes 
$\hat{e}_{s,\sigma}$ per side ($s = \mathrm{L},\mathrm{R}$ and $\sigma=\pm1$) at energies $\sigma \sqrt{2} J$: 
\begin{equation}\label{eq:edgemodes}
\begin{split}
\hat{e}_{\mathrm{L},\sigma} = & \frac{- \sigma \, \sqrt{2} \, \hat{b}_{1,B} + \hat{b}_{1,C} + \hat{b}_{2,A}}{2}
\\
\hat{e}_{\mathrm{R},\sigma} =  & \frac{- \sigma \, \sqrt{2} \, \hat{b}_{M,B} - \hat{b}_{M-1,C} + \hat{b}_{M,A}}{2} \ .
\end{split}
\end{equation}
Moreover, it restricts the running of the cell index $j$ in Eq.~\eqref{eq:bulkmodes}
to the full cells, i.e., $j=2, \cdots, M - 1$.
Summarising, the non-interacting Hamiltonian at $\pi$ flux can be written as:
\begin{equation}\label{eq:cagebasis}
\hat{H}_0^{(\phi=\pi)} = 
\sum_{j=2}^{M-1} \sum_{\tau \in \{0,\pm\}} 2 J \tau \, \hat{w}^{\dagger}_{j,\tau} \hat{w}^{\phantom{\dagger}}_{j,\tau}
+ \sum_{s \in \{L,R\}} \sum_{\sigma \in \{\pm\}} \sigma \, \sqrt{2} J \, \hat{e}^{\dagger}_{s,\sigma} \hat{e}^{\phantom{\dagger}}_{s,\sigma} \ .
\end{equation}

Due to the overlapping nature of the Wannier eigenmodes in Eqs.~\eqref{eq:bulkmodes}-\eqref{eq:edgemodes}~\cite{Aoki1996}, the Hubbard interaction term $\hat{H}_U$ in Eq.~\eqref{BHeqn} can be rewritten as~\cite{Doucot2002,Kazymyrenko2005,Lopes2011,MRizzi2017}:
\begin{eqnarray}
\hat{H}^{(\mathrm{bulk})}_U & = & \sum_{j=2}^{M-1} \left( \phantom{\sum_{\ell \in\{\pm1\}}}
\widetilde{U}_{\tau_1,\tau_2,\tau_3,\tau_4} 
\hat{w}^{\dagger}_{j,\tau_1} \hat{w}^{\dagger}_{j,\tau_2} 
\hat{w}^{\phantom{\dagger}}_{j,\tau_3} \hat{w}^{\phantom{\dagger}}_{j,\tau_4} \right. \label{eq:Uterm}\\
& & \quad + \sum_{\ell \in\{\pm1\}}
\widetilde{V}^{(\ell)}_{\tau_1,\tau_2,\tau_3,\tau_4} 
\hat{w}^{\dagger}_{j+\ell,\tau_1} \hat{w}^{\dagger}_{j,\tau_2} 
\hat{w}^{\phantom{\dagger}}_{j,\tau_3} \hat{w}^{\phantom{\dagger}}_{j+\ell,\tau_4} \label{eq:Vterm}\\
& & \quad \left. + \sum_{\ell \in\{\pm1\}}
\widetilde{J}^{(\ell)}_{\tau_1,\tau_2,\tau_3,\tau_4} 
\hat{w}^{\dagger}_{j+\ell,\tau_1} \hat{w}^{\dagger}_{j+\ell,\tau_2} 
\hat{w}^{\phantom{\dagger}}_{j,\tau_3} \hat{w}^{\phantom{\dagger}}_{j,\tau_4} \right), \label{eq:Jterm} \phantom{aa} 
\end{eqnarray}
plus similar terms for the edges, with all amplitudes linear in $U$.
The full form of the different terms in $\hat{H}^{(\mathrm{bulk})}$ can be found in the supplementary Mathematica script~\footnote{The Mathematica script and output accompanies the arXiv version of this paper}. As an example of the nature of these terms, we show the Hamiltonian with terms restricted to the lowest band:
\begin{equation}\label{eq:lowband}
\begin{aligned}
\hat{H}^{(bulk)}_{U,-1}=&\frac{5}{32}\sum_{j=2}^{M-2} n_{j,-1}\big(n_{j,-1}-1\big)\\
 &+\frac{1}{16}\sum_{j=3}^{M-2}(n_{j-1,-1} n_{j,-1} )\\
&+\frac{1}{64}\sum_{j=3}^{M-2} \big(w_{j-1,-1}^{\dagger^2}w_{j,-1}^2 + w_{j,-1}^{\dagger^2}w_{j-1,-1}^2 \big)
\end{aligned}
\end{equation}
where $n_{j,-1} = (w_{j,-1}^{\dagger} w_{j,-1})$. The appearance of a pair-tunnelling term, with the minimum of the dispersion relation at momentum $k=\pi$ can be seen in Eq.~\eqref{eq:lowband}.

\begin{figure}
\includegraphics[width=0.9\columnwidth]{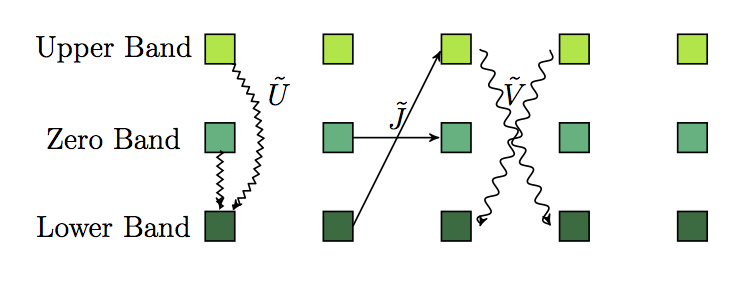}
\caption{Illustration of the type of cage terms that control the movement and the interaction of the pairs of particles. $\tilde{U}$ is the on-cell interaction of pairs, $\tilde{J}$ is the hopping of pairs between cells and $\tilde{V}$ can be interpreted as nearest-neighbour interaction or correlated swapping of 2 particles across neighbouring sites.}

\label{cageflavours}
\end{figure}
We stress that a number of other studies~\cite{Altman2010, Huber2013, MoellerCooper2012}
employed projection on lowest-band states, 
applicable as long as interactions are smaller than the gap between the bands,
in analogy to the lowest Landau level projection in quantum Hall systems.
Here instead, we retain the full description of the model. 
This formulation of the interactions, makes it evident that a local $\mathbb{Z}_2$ symmetry 
is preserved by the Hamiltonian, 
namely the parity of the population of all three kinds of cages localized around each hub~B, i.e., 
\begin{equation} \label{eq:parity}
\left[\hat{H}_\mathrm{BH}, \hat{P}_j \right] =0 \quad \forall j \quad \mbox{with} \quad 
\hat{P}_j \equiv \exp\left[\ii \pi \sum_{\tau} \hat{w}^\dagger_{j,\tau} \hat{w}^{\phantom{\dagger}}_{j,\tau}\right] .
\end{equation}
In particular, the interaction effects can be sorted out in three different kinds:
i) $\widetilde{U}$ interactions and cage flavour-flips around a given hub;
ii) $\widetilde{V}$ interactions and correlated flips between nearest-neighbouring hubs;
iii) $\widetilde{J}$ pair-tunnelling (possibly with flips) between nearest-neighbouring hubs. An example of each of these terms is pictorially shown in Fig~\ref{cageflavours}.
The $\widetilde{J}$ terms explicitly show that a delocalisation of particle (bound) pairs is possible,
in spite of the single particle perfect localization.
When this takes place and how robust this pair coherent phase actually is, 
forms the core subject of our work, which aims to extend the seminal results by Dou\c{c}ot and Vidal~\cite{Doucot2002} and the most recent generalization by Tovmasyan et al.~\cite{Tovmasyan2018}.

Before delving into the phase diagram analysis at commensurate filling  in the rest of the paper,
we comment here briefly about possible experimental schemes for cold atomic setups.
The two main procedures available are using either real space geometries or synthetic dimensions. 
a) Real space geometries can be made of lasers intersecting at $\pm 45$ degrees with the lattice dimension
	plus additional superlattices transverse to it to isolate single rhombi chains~\cite{Tarruell2011,Hyrkas2013};
	or by either digital micro-mirror devices (DMD)~\cite{Gauthier2016DMD} 
	or magic / anti-magic trapping of alkaline-earth atoms~\footnote{F. Gerbier, private communication}.
b) Synthetic dimensions would involve exploiting three internal hyperfine states of some atom to map them
	into the three basis sites of the unit cell~\cite{Boada2012,Celi2014} 
	-- the main difference with the present analysis being the range of interactions, 
	extending over the whole unit cell. 

In both schemes, the phase imprinting on the tunnelling matrix elements can then be achieved
by laser-assisted hopping~\cite{Goldman2014, Dalibard2011, Sias2008, Ma2011} and/or shaking of the lattice barrier amplitude~\cite{Holthaus1997,Sias2008,Struck2011,Eckardt2010}.
We envision that imprinting the phase on a single link (as the gauge choice in this work) 
could be achieved by shaking only the corresponding lattice barriers.
Other choices, like a Landau gauge with a $j$-cell dependent $T^{(0)}_{\alpha,\alpha \pm 1} = \exp(\mp \ii j \phi)$
and $T^{(+1)}_{\alpha,\alpha+1} =1$ or a symmetric one with $T^{(0)}_{\alpha,\alpha \pm 1} = \exp(\mp \ii j \phi/2)$ 
and $T^{(+1)}_{\alpha,\alpha + 1} = \exp(- \ii j \phi/2)$ might be more suitable for laser-assisted schemes via a running wave, as often realised in experiments~\cite{BlochDalibardZwerger2008,Goldman2014}.
In the two recent experiments conducted on photonic waveguides, the tunnelling coefficients were engineered
in a similar spirit, either by insertion of extra elements with different refractive index~\cite{Kremer2018}  or by Floquet schemes~\cite{Mukherjee2018}.
In these, however, interactions between the photons are a bit more difficult to obtain and tune with current technologies. There is effort being put into this and we might expect some progress in the near future.
%


\section{Complete Phase Diagram} \label{sectphasediag}

We focus here on a commensurate filling of one particle per physical site, i.e., three particles per lattice cell ($N=L=3M-2$).
At infinite interactions we expect a Mott Insulator (MI) independently of the piercing flux, 
since the kinetic energy does not play any role. 
At zero interactions and full frustration, we also expect an insulating state, though of a different kind,
since all single-particle wavefunctions are localized. 
This is, however, not the case as soon as a tiny interaction is present.
As shown above (Sec.~\ref{sectmod}), the on-site Hubbard interaction induces pair-hopping terms between neighbouring 
sites (Eq.~\eqref{eq:Jterm}), which in turn can be shown to lead to a pair quasi-condensation~\cite{Doucot2002,Tovmasyan2018} (a pair-Luttinger liquid, PLL).
These occur without violating the extensive collection of local $\mathbb{Z}_2$ invariants of Eq.~\eqref{eq:parity}.

At small enough fluxes, instead, it is legitimate to assume that the presence of rim sites (A and C) appears as a decoration to a pure 1D lattice. These simply change the single-particle band curvature and therefore renormalise the critical coupling between the MI and the ``standard" Luttinger liquid (LL)~\cite{MonienWhiteBHmodel}.
To the best of our knowledge, however, the questions about the nature of the LL-PLL transition and the position of the triple point at finite or perfect frustration, i.e., about the robustness of the exotic PLL, still remain open.

\begin{figure}[t!]
\begin{subfigure}{\columnwidth}
\includegraphics[width=0.63\columnwidth,trim={3cm 0cm 3cm 0cm}]{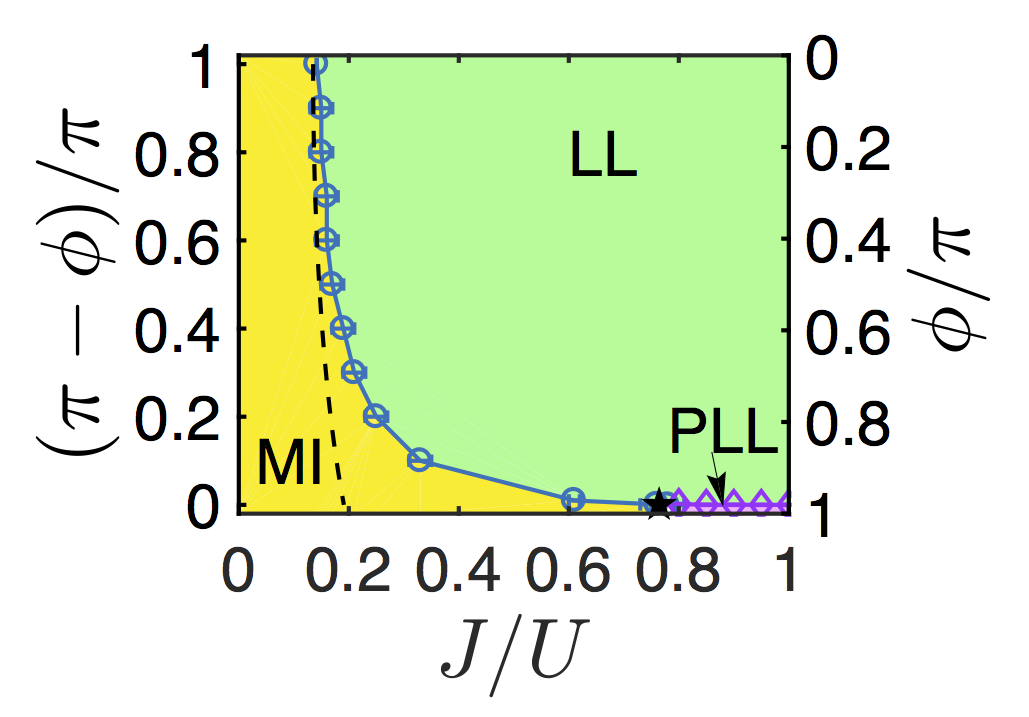} 
\caption{}
\label{phasefill1full}
\end{subfigure}\\
\begin{subfigure}{0.49\columnwidth}
\includegraphics[width=4.1cm,trim={2cm 0cm 0cm 0cm}]{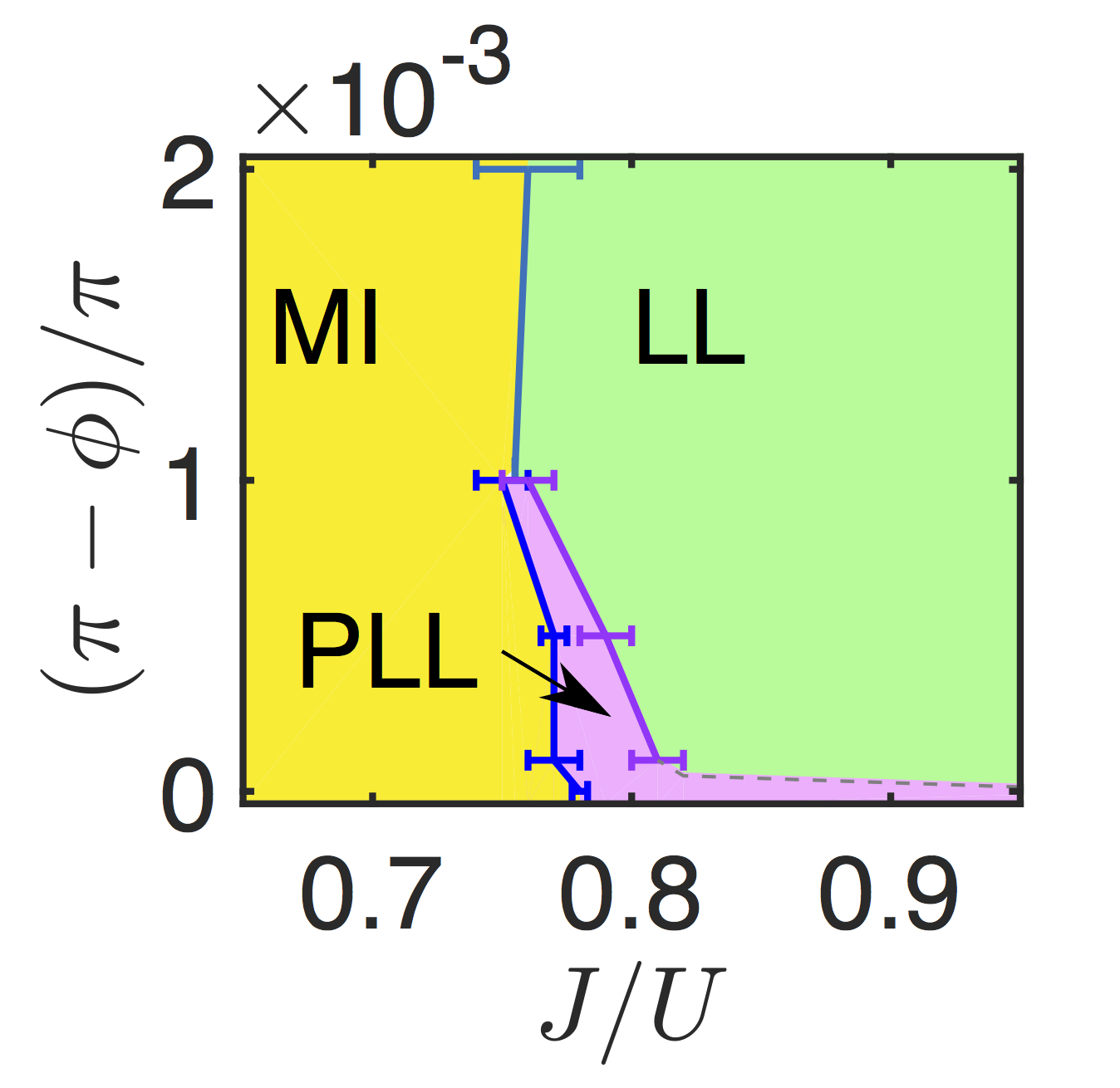} 
\caption{}
\label{phasefill1}
\end{subfigure}
\begin{subfigure}{0.49\columnwidth}
\includegraphics[width=4.42cm,trim={2cm 0.21cm 0cm 0cm}]{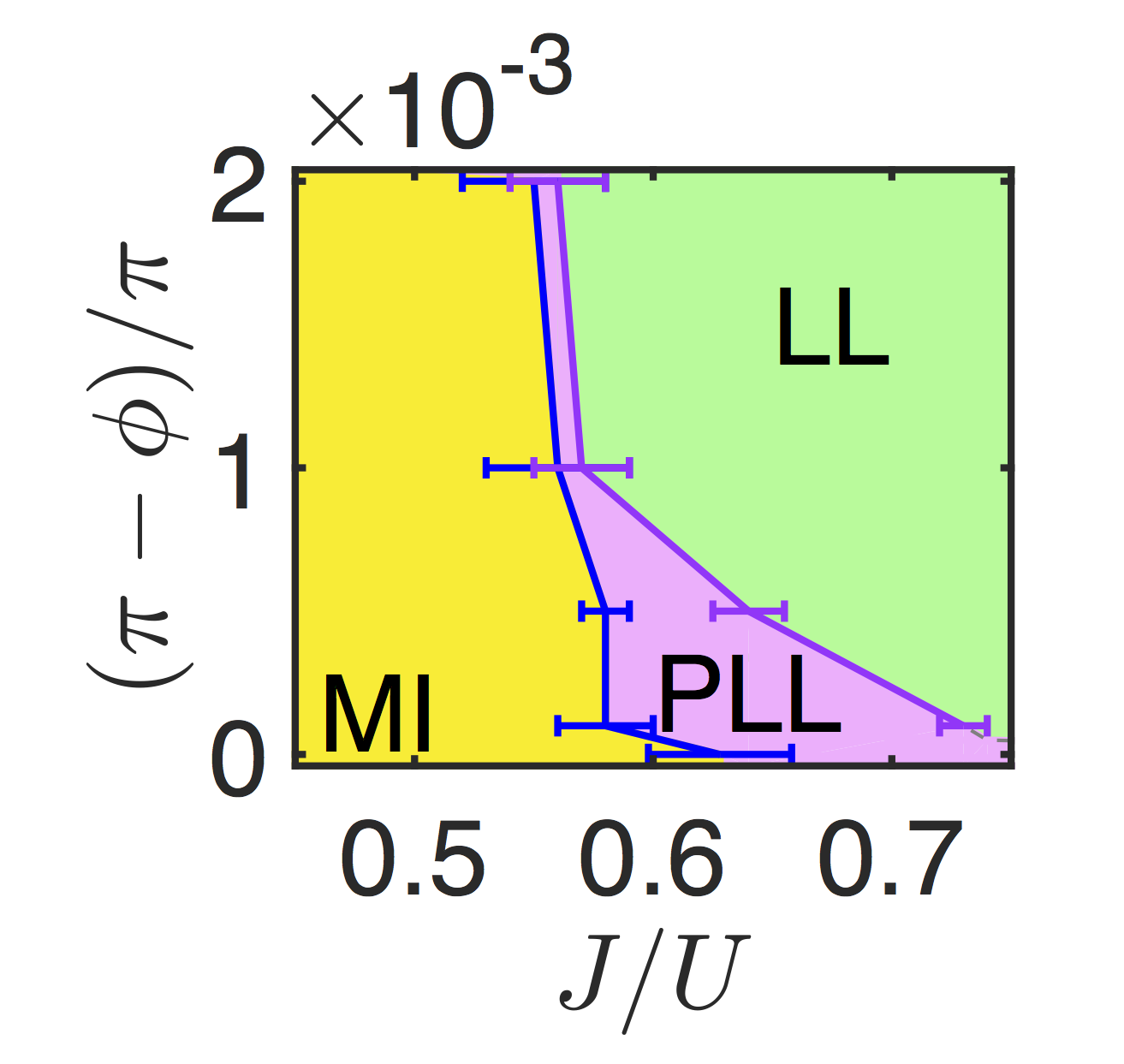} 
\caption{}
\label{phasefill2}
\end{subfigure}
\caption{The phase-diagram with different phase shifts $\phi$ against the tunnelling coefficient $J/U$. The Luttinger liquid (LL), pair-Luttinger liquid (PLL) and the Mott insulator (MI) regions are labelled. The critical points delimiting the MI region are obtained from the energy gap as in Sec.~\ref{sectphase}. The critical points separating the LL and PLL phases are obtained by looking at the decay of the single and pair correlation functions. The error bars have been omitted when they are smaller than the marker size. \textbf{(a)} is the full variation of $\phi$ using a filling=1, whilst \textbf{(b)} and \textbf{(c)} are regions close to full frustration for filling=1 and filling=2 respectively. The $\star$ in \textbf{(a)} denotes $\tilde{G}(\pi)$, where Vidal and Dou\c{c}ot predicted the LL-PLL transition to be at $\phi=\pi$. The dashed line represents their MI-LL transition prediction, $G^{*}(\phi)$ in Eq.~\eqref{eq:critpred}, which is only valid for small $\phi$ values.}
\label{phasephi}
\end{figure}

In Fig.~\ref{phasephi} we show the complete phase diagram of the model as $J/U$ and $\phi$ vary, which constitutes our main result. 
The transitions from MI to the corresponding gapless phase were obtained by evaluating the compressibility gap as shown in Sec.~\ref{sectphase},
while the LL-PLL transition was determined by examining the correlation decay as done in Sec.~\ref{sectPLL}. Single-particle Green's functions decay (at least) exponentially fast in the PLL, while pair-correlations display quasi-long range ordering via an algebraic behaviour
(just as the single-particle ones do in the usual LL).

In our simulations we observe a small intermediate region between the  LL and PLL phases in which it is indistinguishable whether the single correlations better fit an exponential or power law scenario. To display this behaviour we have added error bars in the numerical data for the LL-PLL transition. This could be related to finite size effects and the nature of the transition between these two gapless phases remains an open problem.

We have performed a comparison of our estimates with those given by Vidal and Dou\c{c}ot \cite{Doucot2002} for small values of $\phi$, in terms of   $g=\sqrt{E_{\rm c}/E_{\rm J}} \leftrightarrow \sqrt{U \langle n\rangle /J}$. Defining $G=J/U$ we obtain:
\begin{equation}\label{eq:critpred}
G^{*}(\phi) = \frac{G^{*}(0)} { \cos (\frac{\phi}{4})} 
\end{equation}
where $G^*(0)= 4 / (3\pi^2) \simeq 0.135$, in good agreement with our numerically found $J_c^{[\phi=0]}=0.14 \pm 0.01$ (see Sec.~\ref{sectphase}). This curve \eqref{eq:critpred} is shown by the dashed line in Fig. \ref{phasefill1full}, which is within our error bars up to $\phi=0.3\pi$ and displays only slight discrepancies up to $\phi=0.5\pi$.
For large fluxes further corrections are expected and the prediction at perfect frustration reads $\tilde{G}(\pi)= 4 G^*(\pi) \simeq 0.764$
(shown by the $\star$ symbol in Fig.~\ref{phasefill1full}),
again in nice agreement with our numerical estimates $(J/U)_c^{[\pi]}=0.78 \pm 0.03$.\\
It turns out that the PLL only exists in a very narrow region at imperfect frustration, 
consistent with the qualitative prediction $| \pi - \phi_c | \simeq \exp(-J/U)$ by Dou\c{c}ot and Vidal~\cite{Doucot2002}. 
The presence of a large MI region at unit filling prevents such an exponential from growing large enough. A possible strategy to increase the stability of PLL, therefore, is to reduce the MI by resorting to higher filling factors 
(which, incidentally, should also allow for better signal-to-noise ratio in the experimental detection).
We tested it by using filling $N/L=2$, as shown in Fig~\ref{phasefill2}.
Despite the sensible shrinking of the MI region (by almost 25\%), 
it seems that the prefactor of the exponential also changes, resulting in quite a marginal overall increase of the PLL region.
Determining an optimal filling for PLL detection under common experimental constraints
could constitute an interesting extension for future works.
It is possible that the best scenario is indeed the original large $N/L$, quantum rotor, limit 
of Josephson junction arrays or perhaps coupled extended condensates or even photonic waveguides.

Incidentally, we recall that the local $\mathbb{Z}_2$ symmetry could also be interpreted in terms of the two possible directions of the persistent current induced by the flux $\phi$ around each rhombus~\cite{Doucot2002,Ioffe2002}.
The configurations with zero or one fluxoid per rhombus are indeed perfectly degenerate at $\pi$ flux.
In our gauge, however, all matrix elements are real at $\pi$-flux and
time-invariance is apparently restored (despite the magnetic field). The numerical algorithm tends to pick up real-valued solutions
with no spontaneous local current.
The problem could in principle be overcome by looking at current-current correlators at a distance, 
in order to detect a possible (anti-) ferromagnetic ordering of the rhombi chirality:
a related Ising transition should then discriminate LL from PLL~\cite{Doucot2002,Ioffe2002}.\\

We will show in Sec.~\ref{sectPLL}, however, that the PLL region turns out to be so narrow that we cannot
accumulate a reasonable region of points to perform a precise enough finite-size scaling to identify the universality class of the transition. 
We will show that neither the entanglement entropy scaling of Sec.~\ref{sectentang} will be able 
to discriminate the predicted $c=3/2$ conformal central charge of the critical line. 
Thus, the final answer about the critical behaviour of this $U(1) \times \mathbb{Z}_2$ system 
still remains as elusive as for the square ladder incarnations~\cite{Granato1986,Granato1991,Teitel1983,Halsey1985, Korshunov1986, Li1994}.

Before describing the data analysis, we notice here a certain similarity between this phase diagram and 
the one found for a fermionic (imbalanced) Creutz-Hubbard ladder~\cite{MRizzi2017},
although there all phases are of reasonable size and insulating (see also Ref.~\cite{Tovmasyan2018}).
It would be interesting to see whether the robustness of the PLL towards band curvature
might be different against different deformations of the model, 
and whether this has any relation to the (emergent) topological character~\cite{Kremer2018}.
For example, by substituting the $e^{i\phi}$ phase factor with an amplitude modulation $\cos(\phi)$,
our preliminary numerical data (see \ref{sec:app}) indicate a considerably more robust PLL. In practice this setup requires the ability to imprint a different local tunnelling on one connection within each rhombus. This formulation can be achieved using a single atom microscope, digital micromirror devices (DMDs), or an adaptation of other methods. We will now analyse the different phases in detail, examining terms of physical observables and entanglement.

\section{The Mott-Insulator Lobe} \label{sectphase}

\begin{figure}[t]\center
\includegraphics[width=0.79\columnwidth , trim={1cm, 0cm, 1cm, 0cm}]{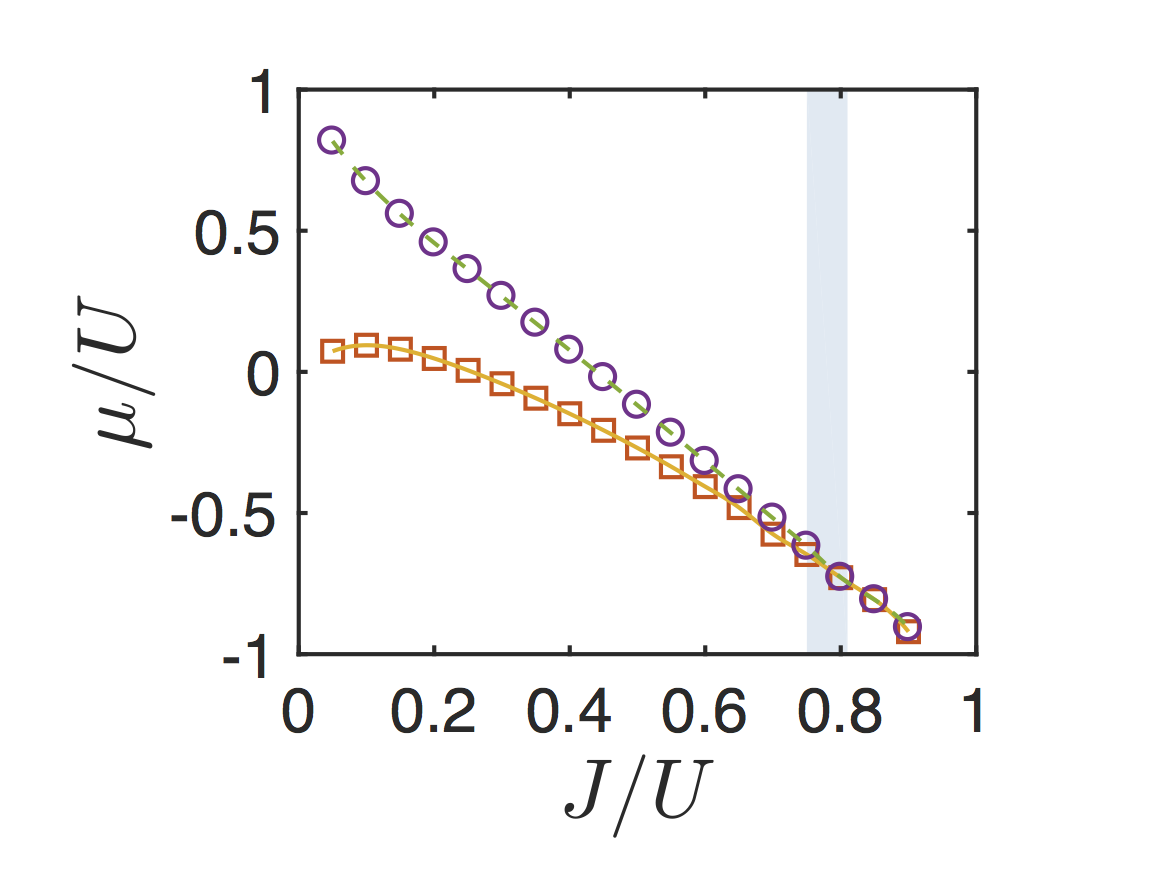} 
\caption{Ground state phase diagram at unit filling $N=L$ and full frustration $\phi=\pi$ in the $\mu/U-J/U$ plane. 
Circles and squares represent the numerical data for $\mu_+$ and $\mu_-$, respectively. 
The region $\mu_-<\mu<\mu_+$ is the Mott-insulator lobe. 
The two lines are cubic splines approximations. 
Their crossing occurs at the BKT point $J_c \approx (0.78 \pm 0.03) U$ indicated by the shaded region.}
\label{phasediag}
\end{figure}
\begin{figure}[t!]
\includegraphics[width=0.725\columnwidth, trim={1.5cm 0cm 2cm 0cm}]{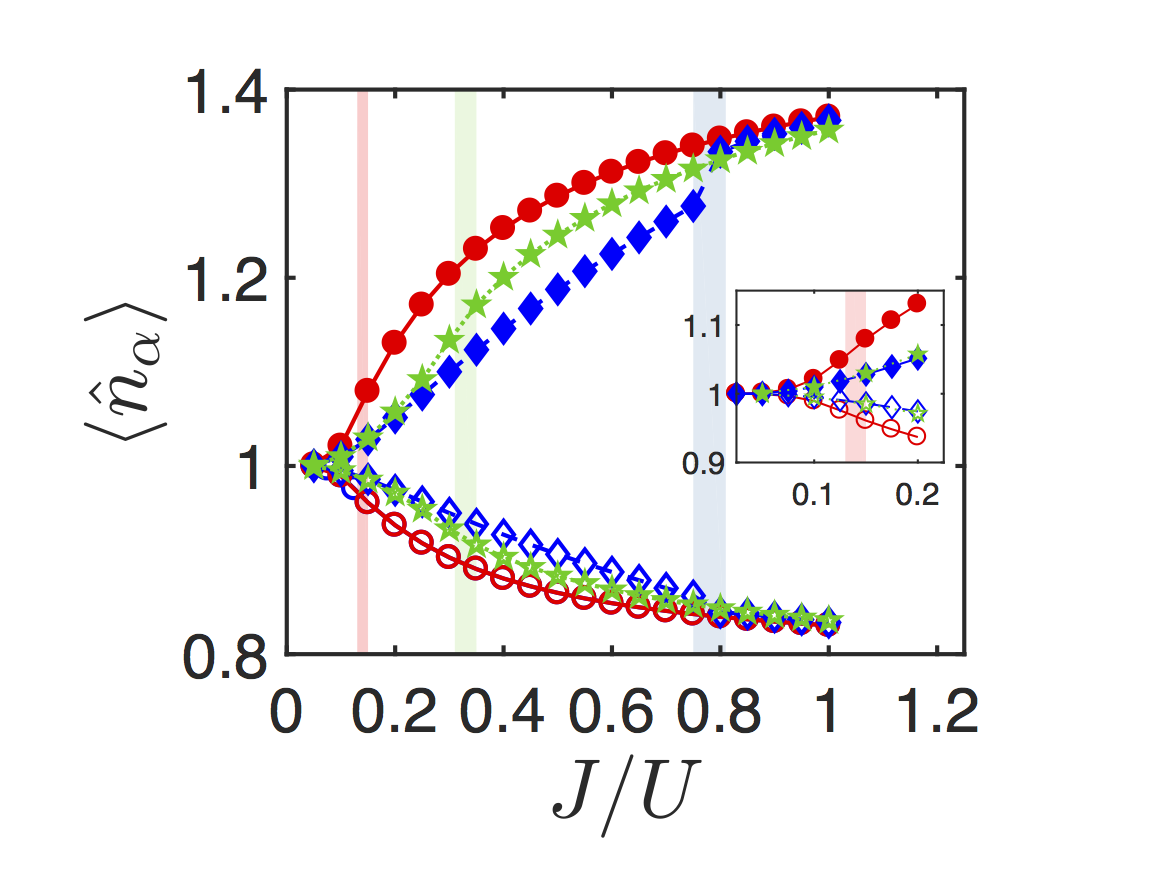} 
\caption{Illustration of the spread of the on-site density ${\langle \hat n_{\alpha}\rangle}$ averaged over the lattice of 226 sites for $\chi=300$ as the tunnelling and frustration is varied. Here the non-frustrated ($\phi=0$) is shown by the red circles ($\circ$), the intermediate frustration ($\phi=0.9\pi$) is shown by the green dotted pentagrams ($\star$) and  the fully frustrated ($\phi=\pi$) are the blue dashed diamonds ($\diamondsuit$). The density of the $B$ sites is shown by filled markers and  sites $A$ and $C$ (which have equal on-site density) are shown by the empty markers. Again the approximate critical points are highlighted by the shaded regions at $J_c^{[\phi=0]}=0.14$(red), $J_c^{[\phi=0.9\pi]}=0.33$ (green) and $J_c^{[\phi=\pi]}=0.78$ (blue). The inset is a zoomed in version close to the $\phi=0$ MI-LL transition.}
\label{osdens}
\end{figure}

The dominance of the onsite repulsion $U$ over the tunnelling coefficient $J$ leads,
for unit filling, to the gapped Mott-insulator (MI) phase,
with the noticeable difference of the particle distribution being uniform across different cells, 
but not within them. The hubs $B$ host some extra density with respect to the rims $A$ and $C$ (see Fig.~\ref{osdens}). 

The position of the Berezinski-Kosterlitz-Thouless (BKT) transition from MI to the compressible gapless phase, 
be it the LL or the PLL one, can be reasonably estimated by the vanishing of the compressibility gap.
If we denote the energy cost for adding or removing $n$ particles by:
\begin{equation}\label{mueqn}
\mu_{+n}=\lim_{L\to\infty} \frac{E_{L+n}-E_{L}}{n};\quad
\mu_{-n}=\lim_{L\to\infty} \frac{E_{L}-E_{L-n}}{n} ,
\end{equation}
where $E_N$ represents the ground-state energy of $N$ particles.
The MI-LL transition happens as soon as $\mu_{+1}=\mu_{-1}$~\cite{MonienWhiteBHmodel}.
The Mott lobe, i.e., the stability region of the MI in terms of the chemical potential, $\mu_{-1}<\mu<\mu_{+1}$,
is illustrated in Fig.~\ref{phasediag}.
We notice that since we do not explicitly impose the local $\mathbb{Z}_2$ constraints, 
this same criterion works also for the MI-PLL transition,
coinciding exactly with the apparently more appropriate definition of $\mu_{+2}=\mu_{-2}$.

The ground state energies at fixed number of particles have been obtained 
via numerical MPS/DMRG simulations~\cite{DMRG1,DMRG2,DMRG3} 
on finite-size systems with open-boundary conditions,
explicitly preserving the Abelian $U(1)$ symmetry.
The local Hilbert space has been truncated to $n_{\max} = 4$ bosons per site and the bond dimension was increased until convergence:
typically $\chi=200$ was sufficient to obtain a maximal discarded weight of $\mathcal{O} (10^{-7})$ or better.

A finite-size scaling of the $\mu$'s has been performed, according to the prediction
$\pm (E_{L\pm n}-E_{L})/n - \mu_{\pm n} \simeq \mathcal{O}(1/L)$ for chains up to 75 rhombi (i.e., $226$ sites).
Finally, the BKT tip of the lobe is estimated by cubic splines interpolation of the functions $\mu_\pm (J/U)$,
as depicted in Fig.~\ref{phasediag} for the fully frustrated case, $\phi=\pi$:
the result is $(J/U)_c^{[\phi=\pi]}=0.78\pm 0.03$ as indicated by shaded region.
As predicted, this frustrated value is considerably larger than the completely unfrustrated one, $(J/U)_c^{[\phi=0]}=0.14 \pm 0.01$,
which in turn is roughly one half of the purely 1D-chain value $(J/U)_c^{[1D]}=0.30 \pm 0.01$~\cite{MonienWhiteBHmodel},
due to the presence of the rhombi (which enlarge the bandwidth by a factor $\sqrt{2}$).

\begin{figure*}[t]\center
\begin{subfigure}{0.245\textwidth}
\includegraphics[width=4.7cm, trim={2cm 0.74cm 1cm 2cm}]{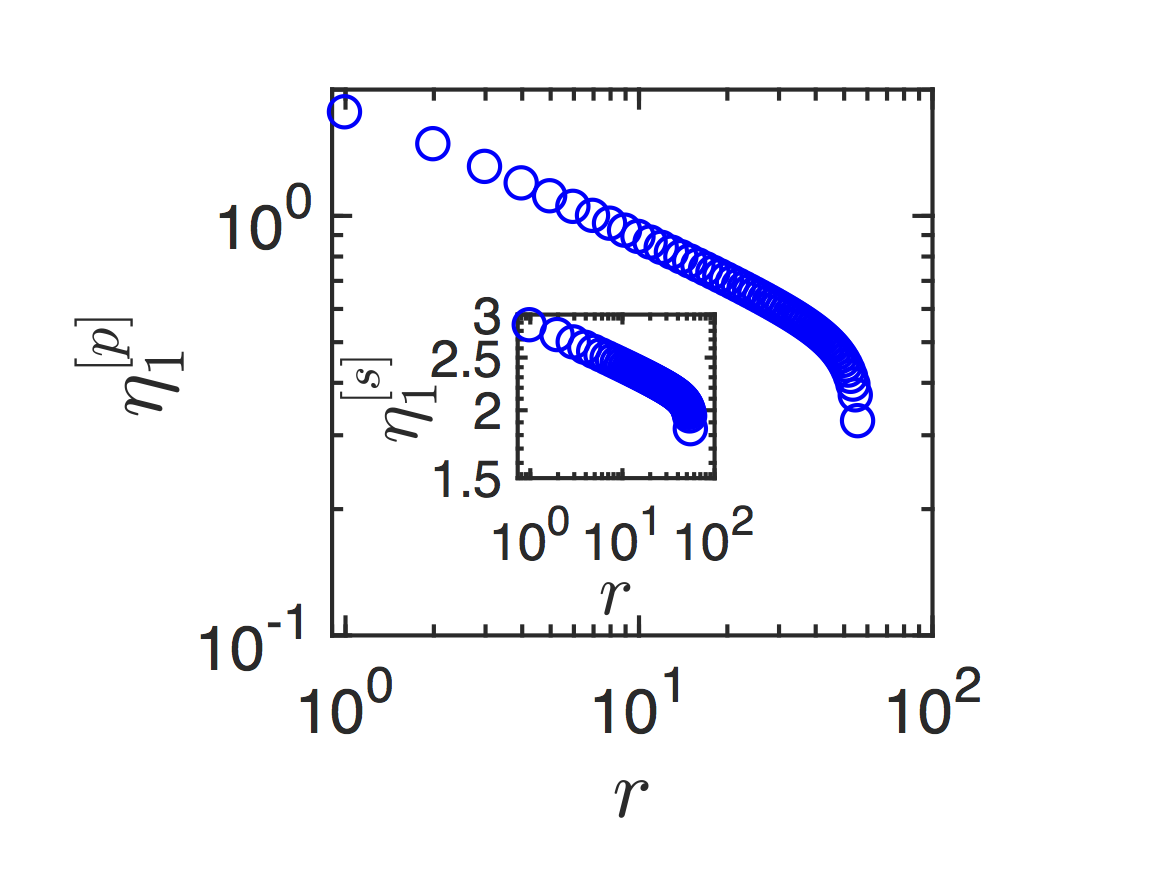} 
\caption{}
\label{halffrust}
\end{subfigure}
\begin{subfigure}{0.247\textwidth}
\includegraphics[width=4.55cm, trim={2cm 0.4cm 1cm 2.2cm}]{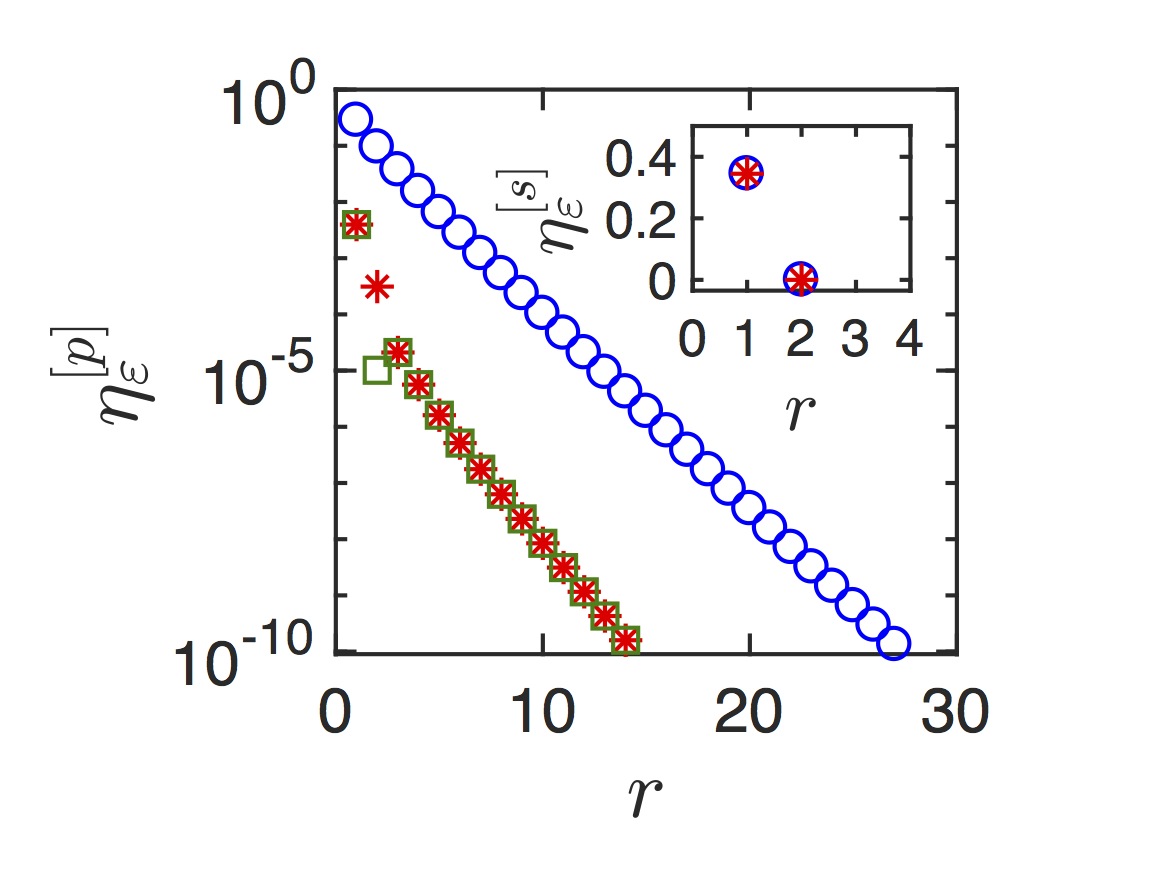} 
\caption{}
\label{sqdecayMI}
\end{subfigure}
\begin{subfigure}{0.245\textwidth}
\includegraphics[width=4.7cm,trim={2cm 0.66cm 1cm 2.2cm}]{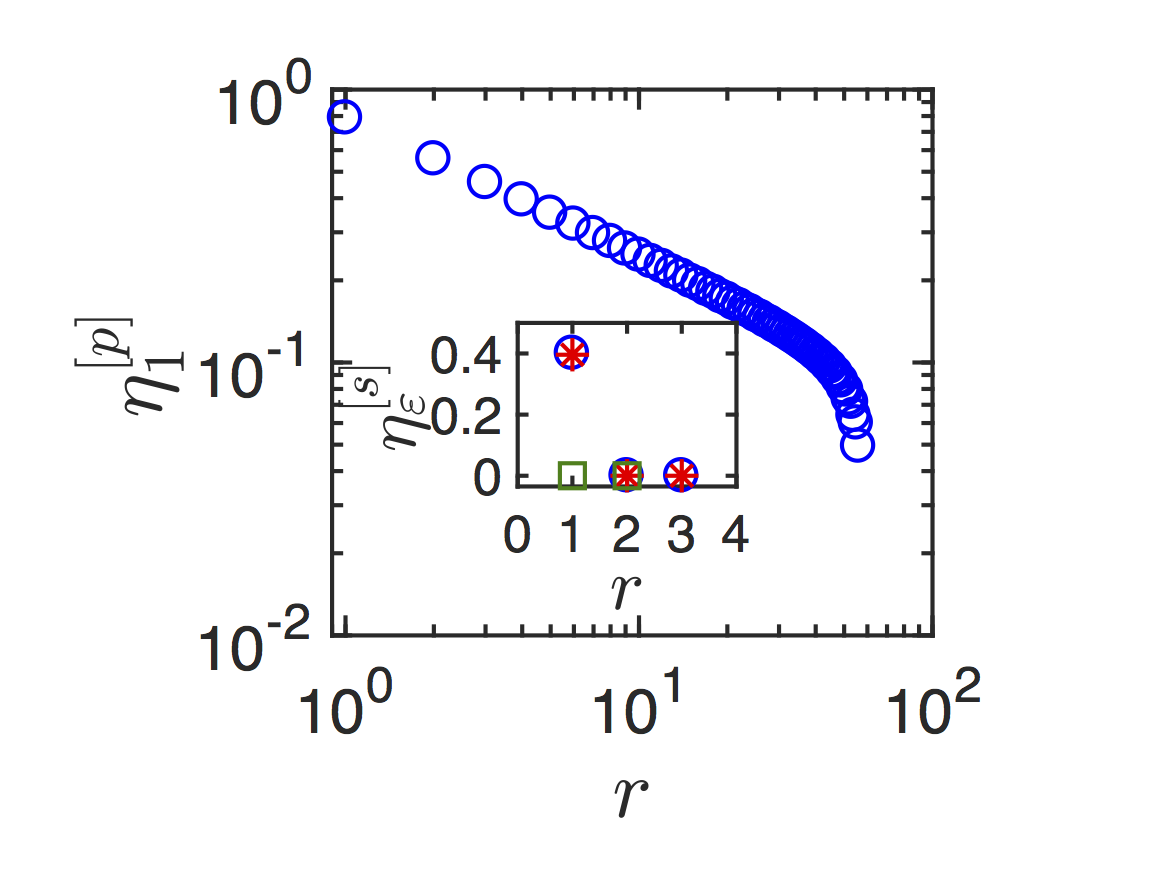} 
\caption{}
\label{sqdecayPSF}
\end{subfigure}
\begin{subfigure}{0.245\textwidth}
\includegraphics[width=4.55cm, trim={2cm 0.5cm 1cm 1.8cm}]{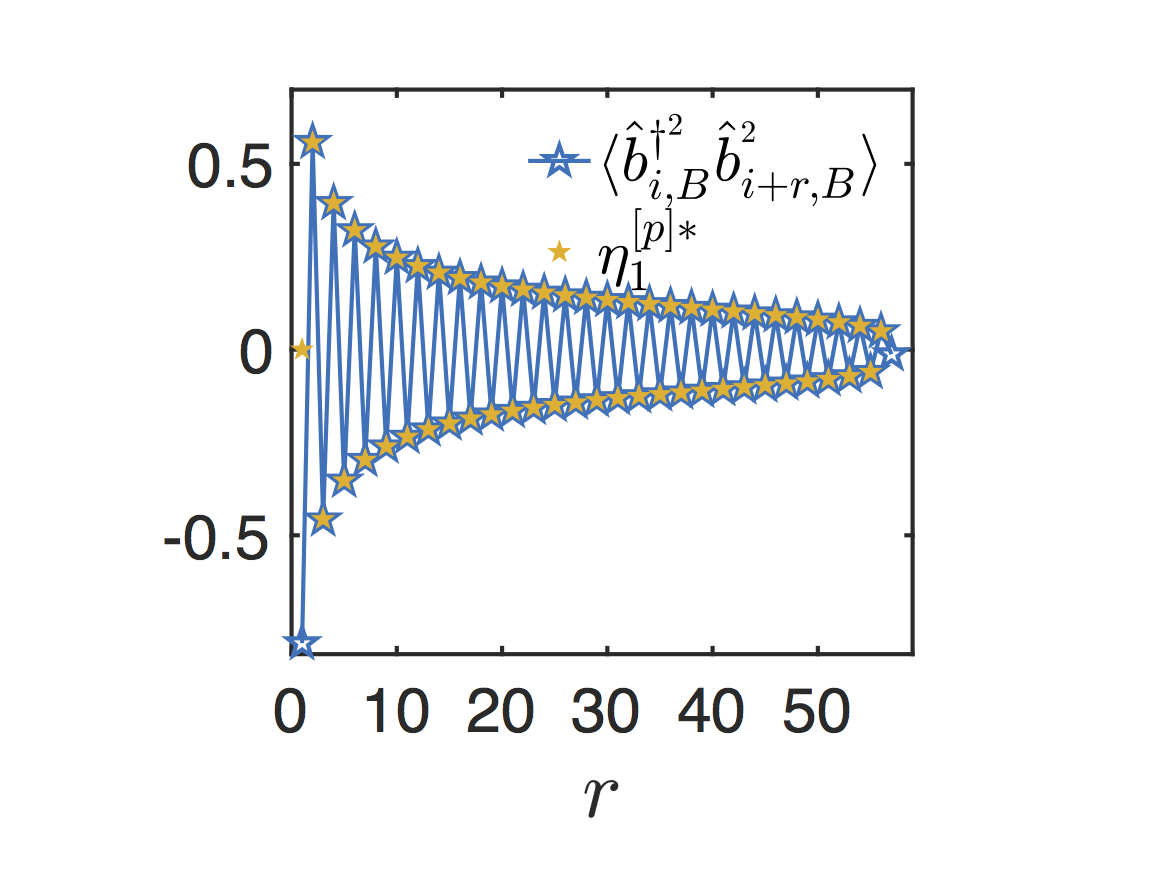} 
\caption{}
\label{paircorrB}
\end{subfigure}
\caption{The spatial correlation functions against the intersite distance for a intermediate frustrated case ($\phi=0.5\pi$ and $J=0.9U$) in (\textbf{a}) and for the fully frustrated case ($\phi=\pi$) in (\textbf{b-d}) for a chain of length $L=226$ with DMRG bond dimension $\chi=300$. 
In \textbf{(a-c)} values below a threshold of $10^{-10}$, which constitutes numerical error, have been excluded.
In the diagrams, $\eta_1$ (dark blue $\bigcirc$) is the largest eigenvalue, $\eta_2$ (red $*$) is the second largest and $\eta_3$ (green $\square$) is the smallest eigenvalue. 
\textbf{(b-c):} The decay of the eigenvalues of the pair correlations $\langle \hat b_i^{\dagger^2}\hat b_{i+r}^{^2}\rangle$ for \textbf{(b)} $J=0.4U$ in semi-logarithmic scale and for \textbf{(c)}  $J=0.9U$ in a double logarithmic scale. The inset is the decay of the eigenvalues of the correlation matrices for the corresponding single correlations $\langle \hat b_i^{\dagger}\hat b_{i+r} \rangle$ at $J=0.4$ (MI phase) and $J=0.9U$ (PLL phase). 
\textbf{(d):} The pair correlations for the $B$ sites $\langle \hat b_i^{\dagger^2}\hat b_{i+r}^{^2}\rangle$ from $i \approx L/4$ for $J/U=0.9$ and the corresponding $\eta_1^{[p]*}$ values, which are the maximum eigenvalues retaining their sign.}
\label{corrdecay}
\end{figure*}

As mentioned above, we have analysed the on-site density distribution, 
which turns out to be reasonably uniform across the cells within the chain, 
with boundary effects only slightly affecting the two more external cells on either side. In all cases (except for when $J/U$ is very small) the density on the $B$ sites is always larger than uniform filling (one per site) and has the same approximate magnitude away from the borders.
In Fig.~\ref{osdens} we therefore plot the average ${\langle \hat n_{\alpha}\rangle}$ of a chain of 226 sites (75 rhombi),
and we notice a certain number of features:
i) the different connectivity of the sub-lattices leads to an enhanced density ${\langle \hat n_{\alpha}\rangle}$
on the hubs $\alpha=B$ with respect to the rims $\alpha=A,C$;
ii) the effect of the magnetic flux $\phi$ is most evident at intermediate $J/U$ where one system is in a different phase to the other (gapped MI or gapless LL/PLL). Deep in the MI phase there is very little discrepancy between the unfrustrated ($\phi=0$), intermediate frustrated ($\phi=0.9\pi$) and the fully frustrated ($\phi=\pi$). Before the first transition out of MI at $J_c^{[\phi=0]}=0.14$ the discrepancy is as small as $0.2\%$ between the fluxes examined, though here $U$ dominates so strongly that any effects due to flux are negligible;
iii) finally, the growth of the hub/rim imbalance appears to depend mostly on the competition between $J$ over $U$, with the hub density experiencing an increase as $J/U$ increases. The function has a less pronounced curve (closer to a linear dependence) in the MI phases, due to the restrictiveness of the phases. In the fully frustrated case (filled markers) there is a more pronounced jump around the transition point, which does not occur in either of the other cases at $\phi=0$ and $\phi=0.9\pi$ (empty markers).
Points i) and iii) seem to make manifest the absence of the so-called ``uniform pairing condition'' of Ref.~\cite{Huber2016}.

\section{Gapless Phases} \label{sectPLL}
In order to characterise the gapless phase(s) outside the Mott insulator lobe, we resort here to spatial correlations 
of single $[s]$ and pair $[p]$ operators, and their Fourier transform.
We formed $3 \times 3$ matrices of the different combinations across the $i$ and $i+r$ cells:
\begin{equation}
\label{singcorreq}
D^{(i,i+r)[s]}_{\alpha,\beta}=\langle \hat{b}_{i,\alpha}^{\dagger} \hat{b}^{\phantom{\dagger}}_{i+r,\beta} \rangle,
\quad  
D^{(i,i+r)[p]}_{\alpha,\beta}=\langle (\hat{b}_{i,\alpha}^{{\dagger}})^2 (\hat{b}^{\phantom{\dagger}}_{i+r,\beta})^2 \rangle,
\end{equation}
and the corresponding structure factors~\cite{Ejima2012}:
\begin{equation}\label{structeqn}
S^{[\gamma]}_{\alpha,\beta}(k)=\sum_{i \neq j} \frac{ \ee^{ \mathrm{i}k(i-j)}}{M-2}{D}_{\alpha,\beta}^{[\gamma](i,j)} ,
\end{equation}
where $k \in [-\pi,\pi]$ and $M-2$ is the number of full cells. 
Then we evaluated their eigenvalues (and eigenvectors):
\begin{eqnarray}
\label{eigeqn}
D^{(i,i+r)[\gamma]}\textbf{v}^{(i,i+r)[\gamma]}_{\varepsilon} & = & \eta^{(i,i+r)[\gamma]}_{\varepsilon} \textbf{v}^{(i,i+r)[\gamma]}_{\varepsilon} , \\
\label{structeigeqn}
S^{[\gamma]}(k) \, \textbf{w}^{[\gamma]}_{\varepsilon}(k) & = & \zeta_{\varepsilon}^{[\gamma]}(k) \, \textbf{w}_{\varepsilon}^{[\gamma]}(k)
\end{eqnarray}
where $\varepsilon=1,2,3$ in decreasing order and $\gamma=s,p$.
We chose this strategy to better illustrate the behaviour of the correlations as a whole as opposed 
to focussing individually on all different matrix elements: such deeper analysis could be the subject of future extensions of this work.

As already discussed qualitatively, we identify the pair Luttinger liquid as the phase exhibiting 
quasi-long range order (QLRO) in the $\eta^{[p]}$ eigenvalues, while the $\eta^{[s]}$ are disordered~\cite{Takayoshi2013}. We have fixed $i \approx L/4$ in Eq.~\eqref{eigeqn} to suppress boundary effects and looked for power-law versus exponential 
decay of the different correlation eigenvalues, as illustrated in Fig.~\ref{corrdecay}. The structure factor of Eq.~\eqref{structeqn} was instead computed by including all complete cells, excluding only the two incomplete ones at the edges (see Fig.~\ref{model}).\\

Firstly, we confirm that cases with the LL phase display QLRO in both the single and pair particle correlations. This was done to classify the phases occurring reasonably far from full frustration. In Fig.~\ref{halffrust} this is shown for $\phi=0.5\pi$, $J=0.9$ and $L=226$, where the QLRO is evident in the double logarithmic scale.
Secondly we confirm that, in the perfectly frustrated case $\phi=\pi$, the single particle correlations are always short-ranged
and the system cannot possibly enter the LL phase.
Noticeably, we do not explicitly impose the emergent extensive collection of  local $\mathbb{Z}_2$ invariants in our numerics. 
All eigenvalues $\eta_\varepsilon^{[s]}$ vanish completely at a distance $r=2$, 
thus displaying perfect Aharanov-Bohm caging (see Eq.~\eqref{eq:bulkmodes}). 
This is visible in the insets of panels~\ref{sqdecayMI}-\ref{sqdecayPSF} .

\begin{figure}[t!]
\begin{subfigure}{1\columnwidth}
\includegraphics[width=0.75\columnwidth,trim={2cm 0.8cm 1cm 1.5cm}]{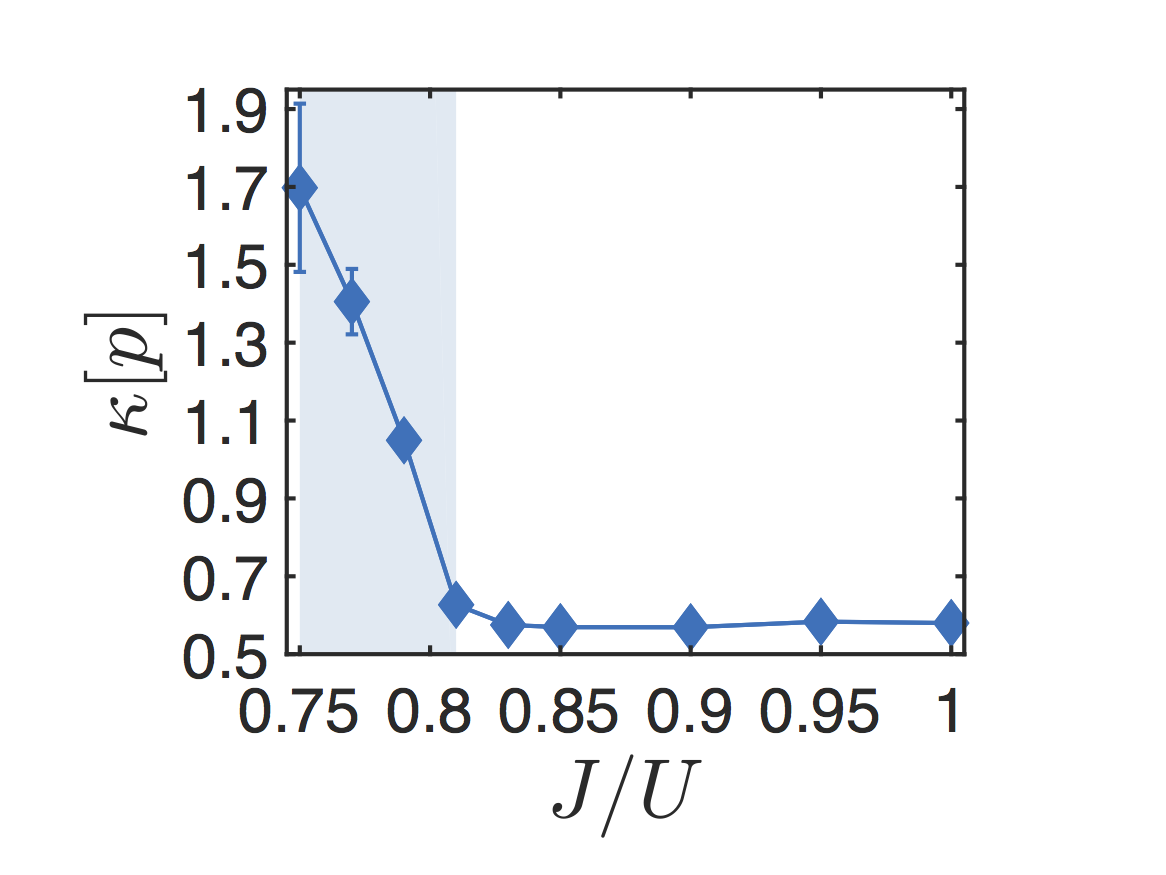} 
\caption{}
\label{fig:omegapair}
\end{subfigure}
\begin{subfigure}{1\columnwidth}
\includegraphics[width=0.75\columnwidth,trim={2cm 0.8cm 1cm 0.2cm}]{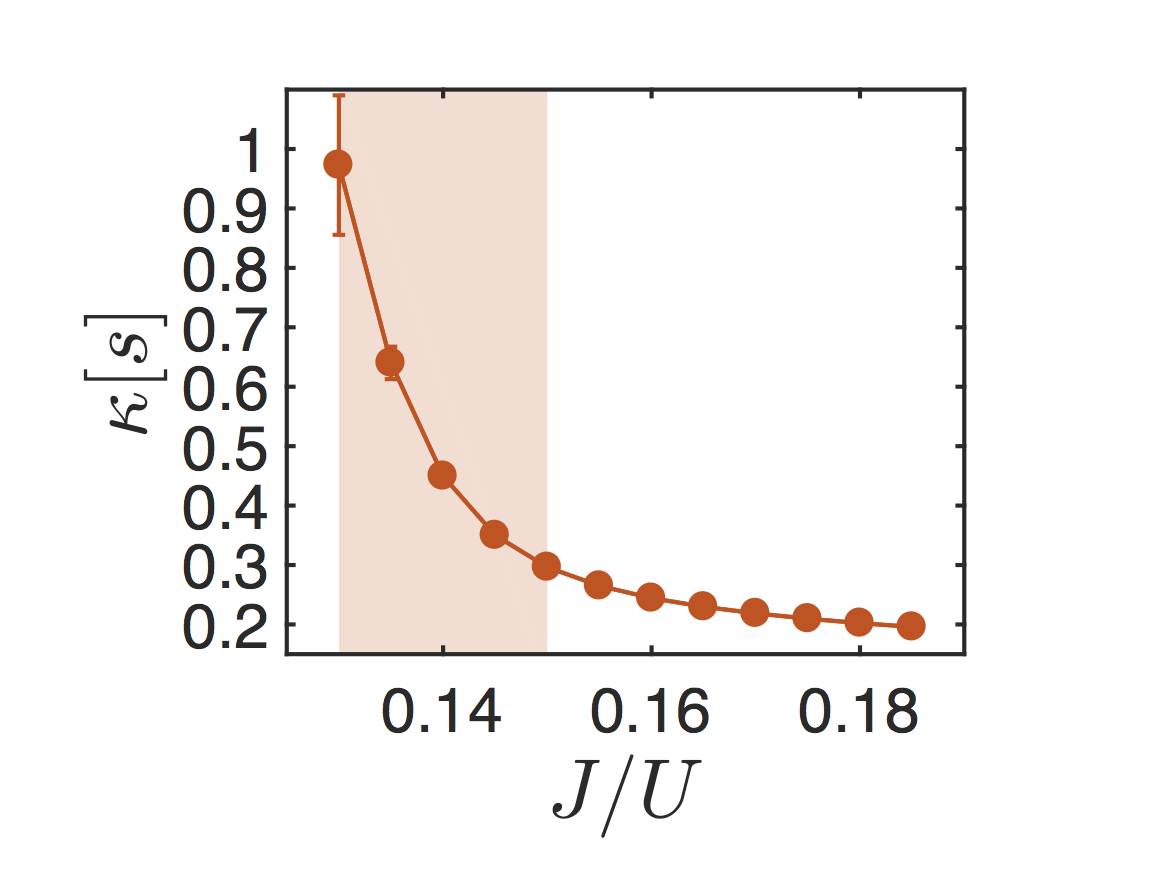} 
\caption{}
\label{fig:omegasing}
\end{subfigure}
\caption{\textbf{(a-b):} Parameter $\kappa[p]$ ($\kappa[s]$) obtained from fitting the pair(single)-correlation function for $\phi=\pi$ ($\phi=0$) with a power law (see Eq.\eqref{corrdecayeqn}) as a function of $J/U$ for L=226. The shaded region indicates the Mott-PLL(Mott-LL) transition region of uncertainty obtained in Sec.~\ref{sectphasediag}.}
\label{fig:omegagraph}
\end{figure}

Concerning the pair correlations, we find that the second and third eigenvalues are always substantially smaller than the dominant one. The second and third eigenvalue are less than 1.5\% and 1\% of the first respectively at their maximal point, which occurs at the start and then they decay exponentially fast.
Therefore we focus on $\eta_1^{[p]}$ (blue circles):
the semi-logarithmic plot of Fig.~\ref{sqdecayMI} shows the exponential decay well within the MI ($J=0.4U<J_c$),
while the log-log plot of Fig.~\ref{sqdecayPSF} highlights the algebraic decay a bit beyond the transition to PLL ($J=0.9U>J_c$).
The decay can be fitted using the following power-law:
\begin{equation} \label{corrdecayeqn}
\eta^{(i,i+r)[\gamma]}_1 \simeq A r^{-\kappa[\gamma]},
\end{equation}
where $\gamma=[s]$  and $\gamma=[p]$ for the single and pair power law fit respectively. The exponent $\kappa[\gamma]$ is then plotted in Fig.~\ref{fig:omegagraph} with $\gamma=p$ for $\phi=\pi$ and $\gamma=s$ for $\phi=0$.
In Fig. \ref{fig:omegapair} a drastic change in the fitted exponent for $\phi=\pi$ is evident around the critical value $J/U=0.78\pm0.03$,
obtained above in Sec.~\ref{sectphase} via the closure of the compressibility gap.
This reminds us of the usual $K=2\kappa[s]<1/2$ criterion for the MI-LL transition~\cite{Giamarchi2004}. The system's ground state can be seen to satisfy this criterion for the single correlation functions with $\kappa[s]$ passing through 0.25 around the transition in Fig. \ref{fig:omegasing} at $\phi=0$. Moreover,  the value $\kappa=0.577 \pm 0.007$ seems to describe very well the PLL, at least in the examined interval $J\in[0.82,1]U$.

Additionally, we can look at the eigenvector $\textbf{v}^{(i,i+r) [p]}_{1}$,
which we find to weakly depend on $r$ and on $J$ inside the PLL.
For $J=0.9U$ it reads approximately $v_1^{[p]}=(0.01, 0.98, 0.01)^T$,
which highlights a largely predominant role of the hubs $B$ for the QLRO.
This is evident by comparing the $\eta_1^{[p]*}$ values (the previously defined $\eta_1^{[p]}$ with their associated sign) with the $B-B$ correlations in Fig.~\ref{paircorrB}.
For $\eta_1^{[p]*}$ we notice the alternating sign for even-odd distances and that the magnitude of this oscillation is practically equal to that of the $B-B$ correlations, except for a case at each end.

\begin{figure}[t]\center
\begin{subfigure}{0.99\columnwidth}
\includegraphics[width=6.8cm,trim={1cm 0.8cm 1cm 1.2cm}, clip=true]{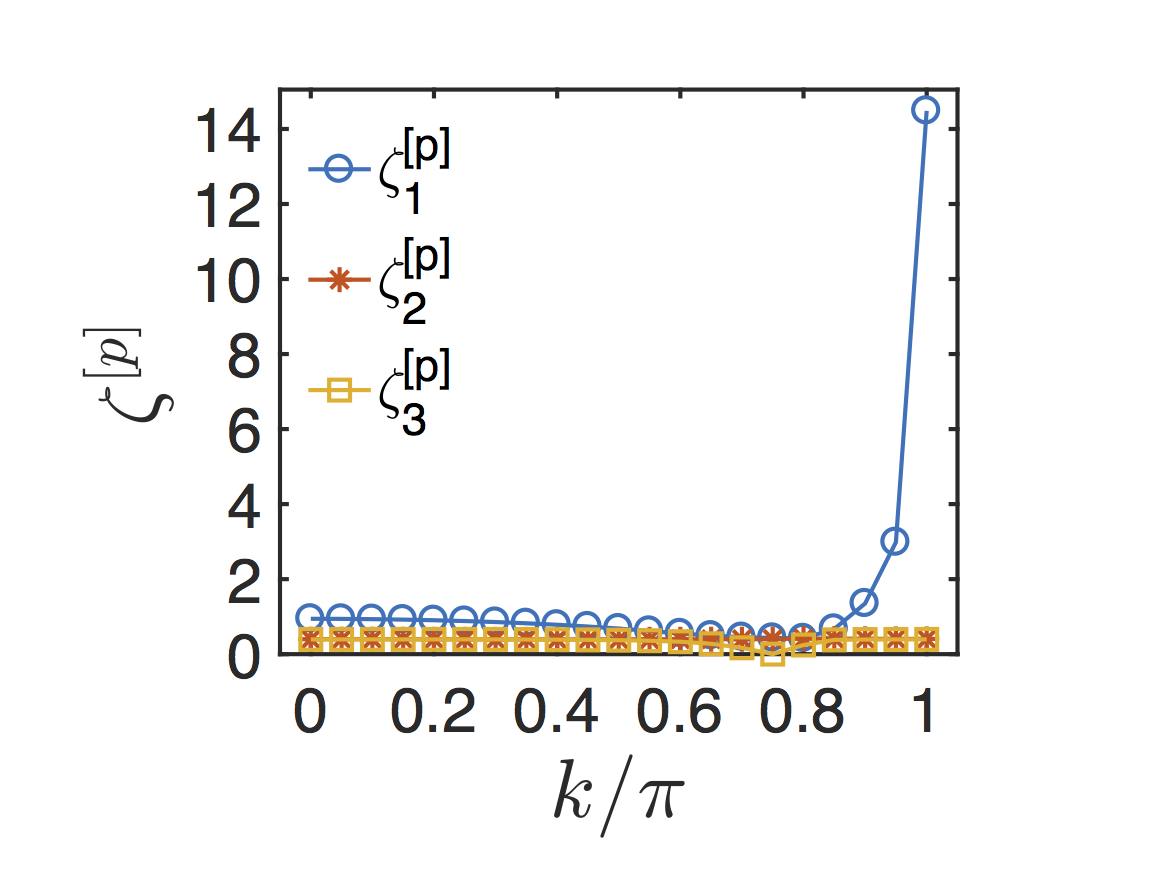}  
\caption{}
\label{structsq}
\end{subfigure}
\begin{subfigure}{0.992\columnwidth}
\includegraphics[width=6.8cm,trim={1cm 0.8cm 1cm 0.2cm}, clip=true]{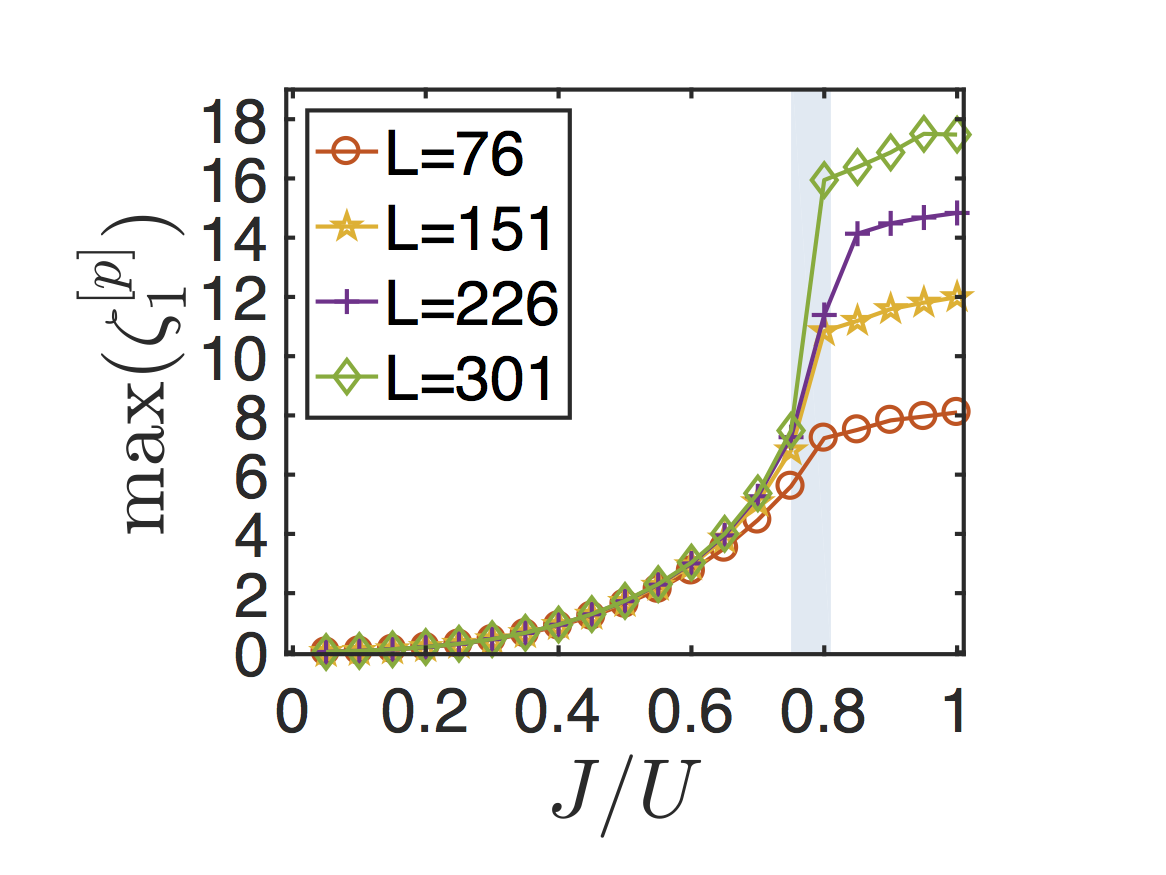}  
\caption{}
\label{struckpeak}
\end{subfigure}
  \caption{Eigenvalues $\zeta_\varepsilon^{[p]}$ of the structure factor matrix for the pair correlations for $J=0.9U$, $\chi=300$ for \textbf{(a)} $L=226$ and as a function of the crystal momentum $k$ and \textbf{(b)} different lengths of the peak at $k=\pi$.}
 \label{structsq}
\end{figure}

Such alternating character of the pair correlations gets reflected in a macroscopic peak at $k=\pi$ 
of the largest structure factor eigenvalue $\zeta_1$, as shown in Fig.~\ref{structsq} for $J=0.9U$ and $L=226$ sites.
The corresponding eigenvector for the largest eigenvalue reads $\textbf{w}_{\varepsilon}^{[\gamma]}(\pi) \simeq (0.1, 0.8, 0.1)^T$,
displaying again the dominance of $B$ sites in the pairing mechanism, while the $A$ and $C$ sites have an equal but very small effect. It should be noted that this only differs from the pair correlations eigenvector due to the fact that the structure factor is calculated for all full cells; if it is considered only from the quarter cell then we once again obtain the average $\textbf{w}_{\varepsilon}^{[\gamma]}(\pi) \simeq (0.01, 0.98, 0.01)^T$. 
The scaling of this peak at $k=\pi$ with the system size $L$ can also be taken as an indicator of the phase transition:
as shown in Fig.~\ref{struckpeak}. Indeed, it starts to become macroscopic (i.e., to diverge with the increasing length $L$)
in the PLL phase ($J\geq0.75U$), while it stays finite in the Mott region (as indicated by data collapse).

\section{Entanglement entropy and spectrum} \label{sectentang}

Here we employ bipartite entanglement as a supplementary detection tool for the different gapless phases.
To this end, we consider the reduced density matrix $\rho_\ell$ of a bipartition of the rhombi chain into two segments of lengths $\ell$ and $L-\ell$,
and we examine its entanglement entropy $S_L(\ell)$ and spectrum $\lambda_i$ (sorted in decreasing order):
\begin{eqnarray}
\label{VNenteqn}
S_L(\ell) & = & -\mathrm{Tr} \left(\rho_\ell \ln\rho_\ell\right)  = - \sum_i \lambda_i \ln \lambda_i ,
\end{eqnarray}
where we drop the dependence of the eigenvalues on $\ell$ and $L$ for the sake of simplicity.

\subsection{Entropy}
\begin{figure}[t]
\begin{subfigure}{0.8\columnwidth}
\includegraphics[width=5.2cm,height=4.9cm, trim={2cm 0.5cm 2cm 1.2cm}]{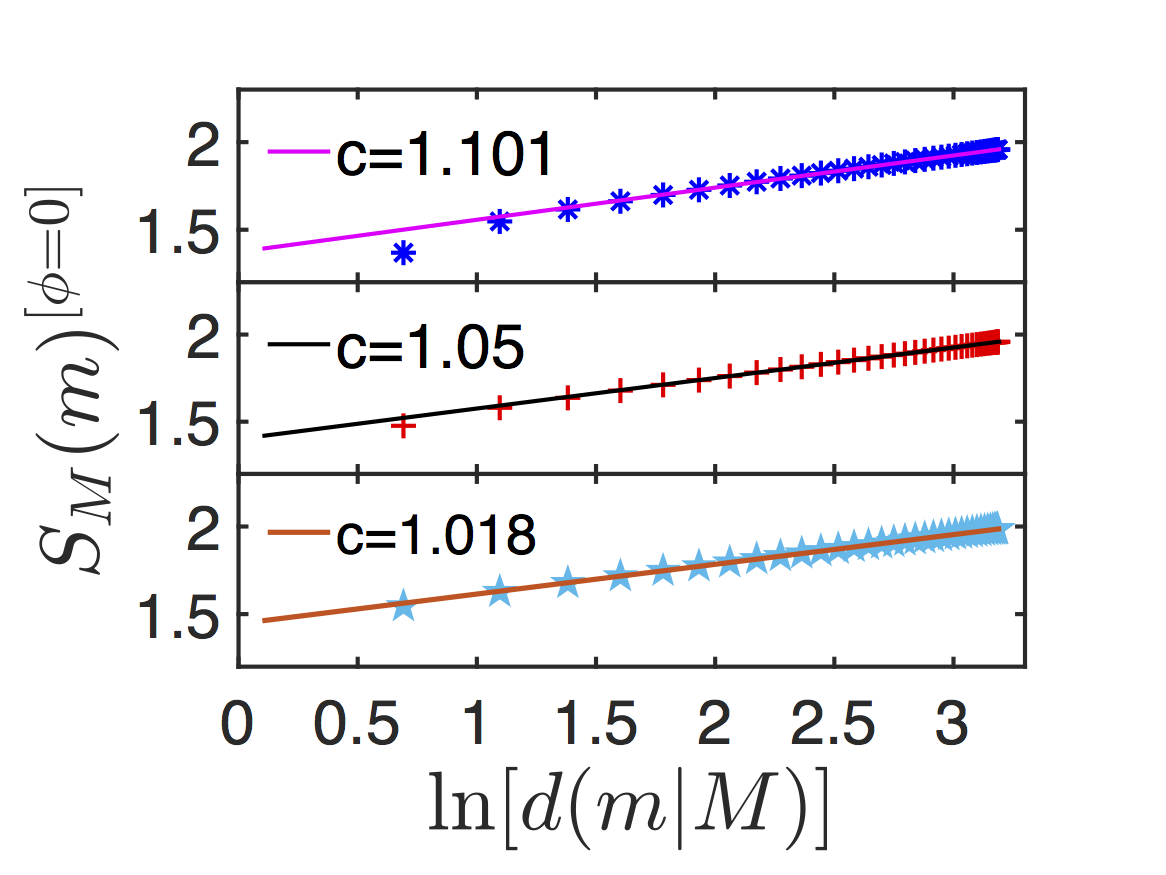}
\caption{}
\label{entnonfrust}
\end{subfigure}
\vspace{0.5cm}
\begin{subfigure}{0.5\columnwidth}
\includegraphics[width=3.5cm,trim={2cm 0.7cm 2cm 0cm}]{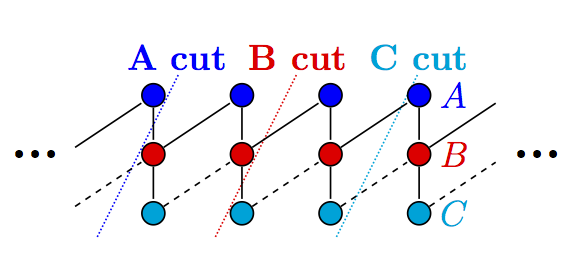}
\caption{}
\label{fig:modelcut}
\end{subfigure}
\begin{subfigure}{0.8\columnwidth}
\includegraphics[width=5.2cm,height=4.9cm, trim={2cm 0.5cm 2cm 1.2cm}]{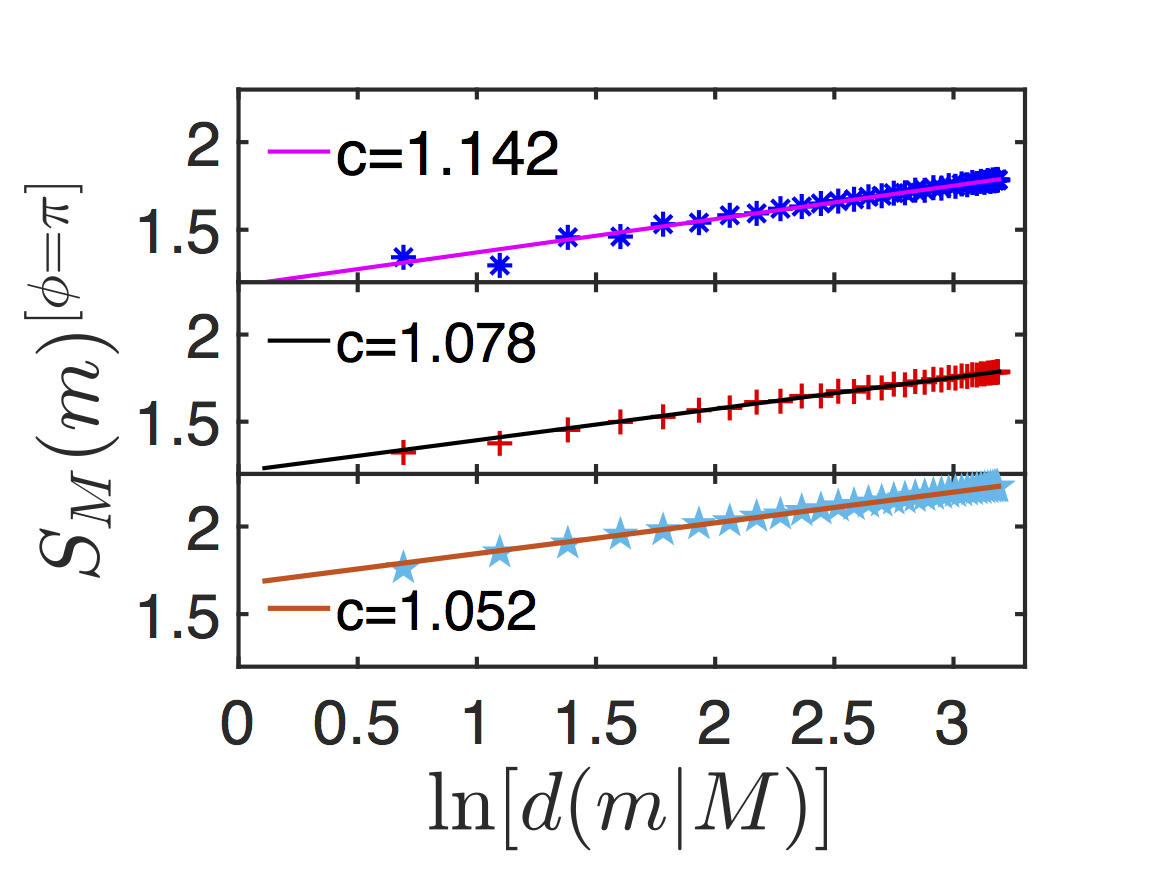}
\caption{}
\label{entfrust}
\end{subfigure}
\caption{Block entanglement entropy for $J=0.9U$ and $L=226$ as a function of the chord length $\ln(d(m|M))=\ln\left(\left[\frac{M}{\pi}\sin\left(\frac{m\pi}{M}\right)\right]\right)$ and compared to the CFT prediction Eq.~\eqref{blockentropy} for \textbf{(c)} the non-frustrated case ($\phi=0$) and \textbf{(a)} the fully frustrated case ($\phi=\pi$). \textbf{(b)} is an illustration of the different cuts that can be made on the model. The cuts after A,B and C are distinguished using the symbols $\ast,+, \star$ respectively. }
\label{entropy}
\end{figure}

For a critical system with open boundary conditions, conformal field theory (CFT) predicts that the von Neumann entanglement entropy scales as:
\begin{equation}\label{blockentropy}
S_L(\ell)=\frac{c}{6}\ln\left[\frac{L}{\pi}\sin\left(\frac{\pi \ell}{L}\right)\right] + A + \mathcal{O} \left(\frac{1}{\ell}\right) ,
\end{equation}
where $c$ is the central charge, which can be used as an indicator of the universality class of the corresponding field theory, 
and $A$ is a model dependent (i.e., non-universal) constant~\cite{Vidal2003,Calabrese2004}. Here we calculate this based on which cell each site occupies, so $L$ is replaced by $M$ in Eq.~\eqref{blockentropy} and $m$ is used to show which cell we are cutting.

In Fig.~\ref{entropy} we distinguish three different cuts of the chain, according to the sub-lattice after which they take place (see \ref{fig:modelcut}),
and perform the fit of Eq.~\eqref{blockentropy} on each separately.
Data are shown for $J/U=0.9$ and we introduced the chord distance $d(m |M)=\left[\frac{M}{\pi}\sin\left(\frac{m\pi}{M}\right)\right]$ for convenience.
The $C$-cut splits a rhombus in half and therefore gives rise to a higher entropy with respect to the $A,B$-cuts.
Alternatively, we can understand this by considering that the $C$-cut separates two cells and that correlations have a strong oscillatory character between neighbouring cells, causing a supplementary amount of entanglement.

For the LL of the unfrustrated regime $\phi=0$, in Fig.~\ref{entnonfrust}, we find that the cut after the maximal cut C has the central charge $c_C^{[\phi=0]}= 1.0180 \pm 0.0003$. The cuts after $A$ and $B$ have central charges $c_A^{[\phi=0]}=1.101 \pm 0.001$ and $c_B^{[\phi=0]}= 1.050 \pm 0.001$ respectively. For the PLL of the frustrated regime $\phi=\pi$, in Fig.~\ref{entfrust}, we find that the maximal cut fits such that the central charge $c_C^{[\phi=\pi]}= 1.052 \pm 0.002 $, which is comparable to the unfrustrated case. For the cuts after $A$ and $B$ the central charges are  $c_A^{[\phi=\pi]}=1.142 \pm 0.006$ and $c_B^{[\phi=\pi]}=1.078  \pm 0.005$ respectively.

\begin{figure*}[t!]
\includegraphics[width=0.78\textwidth,trim={0cm 0cm 0cm 0cm}]{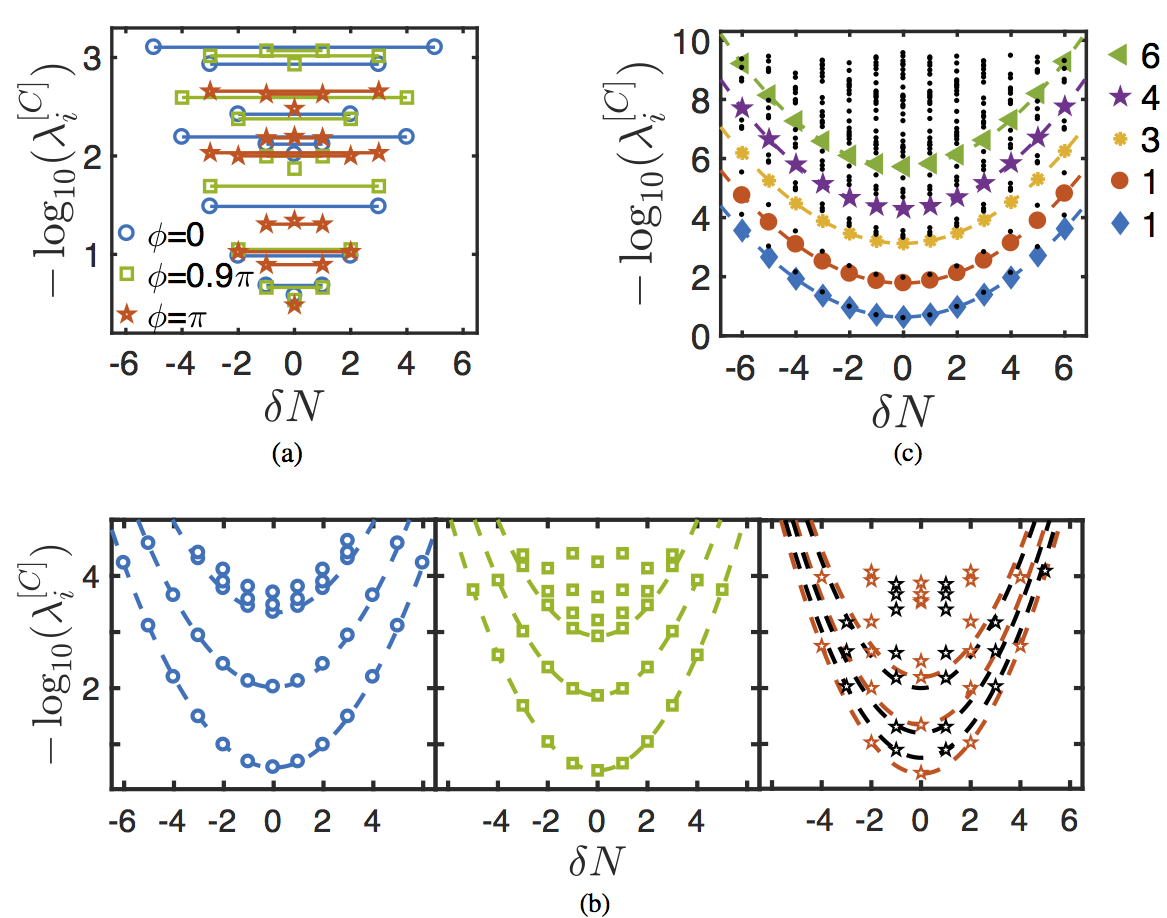}
\caption{\textbf{(a):} The entanglement spectrum as a function of the dispersion from uniform filling $\delta N$ of the number of bosons for the cut after $C$.
$\phi=0$ denotes the unfrustrated case  ($\circ$), $\phi=0.9\pi$ the intermediate frustration ($\square$) and $\phi=\pi$ denotes the fully frustrated case ($\star$), simulated at $J=0.9U$. In \textbf{(a)} a solid line is used to join the degenerate eigenvalues in all cases. \textbf{(b):} Approximate parabolas for $\phi=0$, $\phi=0.9\pi$ and $\phi=\pi$ (left to right) based on the length $L=226$. In the third panel different colours denote the possible curve fitting to even and odd. \textbf{(c):} The unfrustrated ES for a cut after $C$ at $L=226$ with the thermodynamic limit approximation shown by the parabolas. The legend shows the degeneracy of each parabola.}
\label{ESplot}
\end{figure*}
Values for both the frustrated and the unfrustrated are thus fully compatible with the well-known result for the LL phase of the Bose-Hubbard model on a purely 1D chain~\cite{Vidal2003, Calabrese2004, Ejima2012}, i.e., $c=1$.
This confirms that only one bosonic component (out of three possible ones) becomes gapless, in either case, as we have already seen via the correlations in the previous section. This holds regardless of which cut in the system is fitted, i.e. even if we fit every cut after A or B which is a cut across cells we get $c\simeq 1$, once we have considered that finite size effects are taking place.
We, therefore conclude that $c$ is not a good indicator to distinguish PLL from LL. The low-lying levels of the entanglement spectrum, however, may allow this as shown in the next section.\\

Before turning to the entanglement spectrum analysis, let us mention that the entropy scaling across the PLL-LL transition 
at finite deviations from $\phi=\pi$ is not displaying any clear signature of a $c=3/2$ CFT line,
as one would expect from its predicted Ising character~\cite{Doucot2002,Ioffe2002}. The difficulties in analysing transitions between gapless phases has already been noticed in spin models\cite{DeChiaraPRB2011}.

\subsection{Entanglement Spectrum}

Despite having the same central charge, we expect qualitative differences between the wavefunction structure inside the LL and PLL phase. 
We, therefore, resort to the entanglement spectrum, which is capable of revealing key properties about the system, 
such as symmetries and excitations, which the von Neumann entropy, being a single number, is unable to provide
~\cite{LiHaldane, Oshikawa, ESDeng, DeChiaraPRL2012,ESDeChiara2013}. 
Here we choose to focus on the $C$-cut which leaves $(M-1)/2$ rhombi on each side,
so that the bipartition is perfectly symmetric, at least concerning the number of sites.
Thanks to the conservation of the total number of particles in the system, each Schmidt eigenvalue $\lambda_i$ can be associated to
an eigenvector of the reduced density matrix with a fixed number of particles.
In Fig.~\ref{ESplot}\blue{a} we plot the $\lambda_i$'s for a chain of $75$ rhombi, 
according to the excess number of particles $\delta N$
with respect to a homogeneously distributed unit filling (i.e., $L/2$ particles on both sides of the bipartition), 
similarly to Ref.~\cite{Lauchli2013Operator}.
The tunnelling value we consider is $J/U=0.9$, inside both the LL and PLL phases (see Fig.~\ref{phasefill1full}).

The eigenvalues are clearly symmetric with respect to $\delta N=0$ regardless of the amount of frustration. 
For the unfrustrated LL at $\phi=0$ (blue $\circ$) and the intermediate frustrated LL at $\phi=0.9\pi$ (green $\square$), it is easy to recognise both the $-\log\lambda_i \propto \delta N^2$ dependence and also the starting of the equally spaced CFT tower within each distinct $\delta N$,
as predicted for the standard 1D Bose-Hubbard chain~\cite{Lauchli2013Operator}.
Both features apparently disappear for the PLL at full frustration $\phi=\pi$ (red $\star$), 
thus signalling a dramatic change in the underlying wavefunction, undetected by the entropy scaling analysis.
In order to examine this more clearly we plot fitted curves at length $L=226$ of the same curvature for given $\phi$ in Fig. \ref{ESplot}\blue{b} for $\phi=0$, $\phi=0.9\pi$ and $\phi=\pi$ from left to right. It is evident from this that the unfrustrated cases can be extrapolated to the typical curves. For $\phi=\pi$, however, it is impossible to fit the eigenvalues with functions of the same curvature. For example, if the first 5 points are examined closely it can be seen that a parabola would not be able to fit adequately both 1 to 2 and 3 and 1 to 4 and 5. Instead it seems that two distinct parabola sets are appearing at the even and odd $\delta_N$'s as shown by the red and black curves.
In Fig.~\ref{ESplot}\blue{c} we present the results of a finite-size scaling towards the thermodynamic limit for the unfrustrated case ($\phi=0$) shown by the parabolas. A modified degeneracy counting and the appearance of a secondary tower, 
both possibly related to the internal structure of the lattice, are evident. Examining these at $\delta_N=0$ the spacing of the parabolas between every second one is approximately equal i.e., $1-3 \approx 2-4 \approx 3-5$, whereas the spacing between neighbouring parabolas differs. Higher parabolas are excluded as they fall below the accuracy of our results.
For the fully frustrated case, instead, where it seems that two distinct parabola sets are appearing, a reasonable finite-size scaling procedure is not possible without conserving explicitly the local $\mathbb{Z}_2$ quantities. Both these aspects go beyond the scopes of the present work and deserve future investigation.

We do, however, notice that the resulting pattern of (quasi-) degenerate multiplets in the entanglement spectrum changes quite radically in the fully frustrated case from the unfrustrated and intermediate frustrated cases. So we are confident that the entanglement spectrum can be used to distinguish between the two cases, where the entropy cannot.


\section{Conclusions} \label{sectconc}

In this paper we have analysed the ground state phase diagram of a system of interacting bosons on a geometrically frustrated lattice;
namely, we have considered a quasi-1D chain of rhombi pierced by magnetic flux.
For unit filling and a sufficiently low tunnelling amplitude the system is in the Mott insulator phase as expected.
For larger tunnelling values, we have numerically confirmed that when full geometric frustration prevents the movement of single particles (i.e., the bands become flat),
the system still enters into a gapless phase where the elementary moving objects are pairs of particles.
We have explored the regime where the frustration is not perfect, highlighting that the pair fluid can only be obtained for a very small region, 
making this quite challenging for an experimental realisation, especially at low particle filling.
It is, however, possible to extend this region by a small amount using higher filling and even more by using amplitude modulation within the system instead of the phase shifts we applied, which is also possible experimentally.
From a different perspective, we have highlighted that,
whilst the central charge obtained from the entropy cannot be used to distinguish between the PLL and LL phase,
the features and quantum numbers of the entanglement spectrum do have noticeable differences between the two.

There are a number of directions this work could be expanded upon, for example:
i) compare the robustness of the PLL phase with respect to other deformations of the flat bands (such as amplitude modulation), 
and compare it to other flat band models (e.g., the Creutz ladder~\cite{MRizzi2017,Tovmasyan2018}),
to see whether the (here hidden~\cite{Kremer2018}) topological character plays any role;
ii) work out an explicit mapping to the effective Ising model predicted by Dou\c{c}ot and Vidal~\cite{Doucot2002},
in terms of measurable quantities (as done for Creutz ladder fermions~\cite{MRizzi2017}), 
in order to shed new light on the nature of the PLL-LL transition
(possibly once the PLL region is also extended to simplify things);
iii) deepen the understanding of the striking change in the entanglement spectrum, 
possibly by also explicitly enforcing the extensive number of $\mathbb{Z}_2$ symmetries~\cite{Doucot2002,Tovmasyan2018} in the numerics.\\
Moreover, it would be very interesting to examine the dynamics of our interacting chain, 
in order to formulate experimental detection strategies, 
now that platforms for artificial flat-band systems are flourishing again~\cite{Leykam2018,Kremer2018,Mukherjee2018}.

\acknowledgments 
The authors would like to acknowledge useful discussions with K. McAlpine, A. Trombettoni, F. Gerbier, B. Dou\c{c}ot, J. Vidal, S. Al-Assam and A. Haller.
The authors are grateful for the computational time from the Mogon cluster of the JGU (made available by the CSM and AHRP), and S. Montangero for a long-standing collaboration on the flexible Abelian Symmetric Tensor Networks Library employed here.
CC wishes to thank the EPSRC and the Professor Caldwell Travel Studentship for support.\\

\bibliographystyle{apsrev4-1}
\bibliography{ref}

\begin{thebibliography}{101}%
\makeatletter
\providecommand \@ifxundefined [1]{%
 \@ifx{#1\undefined}
}%
\providecommand \@ifnum [1]{%
 \ifnum #1\expandafter \@firstoftwo
 \else \expandafter \@secondoftwo
 \fi
}%
\providecommand \@ifx [1]{%
 \ifx #1\expandafter \@firstoftwo
 \else \expandafter \@secondoftwo
 \fi
}%
\providecommand \natexlab [1]{#1}%
\providecommand \enquote  [1]{``#1''}%
\providecommand \bibnamefont  [1]{#1}%
\providecommand \bibfnamefont [1]{#1}%
\providecommand \citenamefont [1]{#1}%
\providecommand \href@noop [0]{\@secondoftwo}%
\providecommand \href [0]{\begingroup \@sanitize@url \@href}%
\providecommand \@href[1]{\@@startlink{#1}\@@href}%
\providecommand \@@href[1]{\endgroup#1\@@endlink}%
\providecommand \@sanitize@url [0]{\catcode `\\12\catcode `\$12\catcode
  `\&12\catcode `\#12\catcode `\^12\catcode `\_12\catcode `\%12\relax}%
\providecommand \@@startlink[1]{}%
\providecommand \@@endlink[0]{}%
\providecommand \url  [0]{\begingroup\@sanitize@url \@url }%
\providecommand \@url [1]{\endgroup\@href {#1}{\urlprefix }}%
\providecommand \urlprefix  [0]{URL }%
\providecommand \Eprint [0]{\href }%
\providecommand \doibase [0]{http://dx.doi.org/}%
\providecommand \selectlanguage [0]{\@gobble}%
\providecommand \bibinfo  [0]{\@secondoftwo}%
\providecommand \bibfield  [0]{\@secondoftwo}%
\providecommand \translation [1]{[#1]}%
\providecommand \BibitemOpen [0]{}%
\providecommand \bibitemStop [0]{}%
\providecommand \bibitemNoStop [0]{.\EOS\space}%
\providecommand \EOS [0]{\spacefactor3000\relax}%
\providecommand \BibitemShut  [1]{\csname bibitem#1\endcsname}%
\let\auto@bib@innerbib\@empty
\bibitem [{\citenamefont {Moessner}\ and\ \citenamefont
  {Ramirez}(2006)}]{Moessner2006}%
  \BibitemOpen
  \bibfield  {author} {\bibinfo {author} {\bibfnamefont {R.}~\bibnamefont
  {Moessner}}\ and\ \bibinfo {author} {\bibfnamefont {A.~P.}\ \bibnamefont
  {Ramirez}},\ }\href {\doibase 10.1063/1.2186278} {\bibfield  {journal}
  {\bibinfo  {journal} {Phys. Today}\ }\textbf {\bibinfo {volume} {59}},\
  \bibinfo {pages} {24} (\bibinfo {year} {2006})}\BibitemShut {NoStop}%
\bibitem [{\citenamefont {Balents}(2010)}]{Balents2010}%
  \BibitemOpen
  \bibfield  {author} {\bibinfo {author} {\bibfnamefont {L.}~\bibnamefont
  {Balents}},\ }\href {\doibase 10.1038/nature08917} {\bibfield  {journal}
  {\bibinfo  {journal} {Nature}\ }\textbf {\bibinfo {volume} {464}},\ \bibinfo
  {pages} {199} (\bibinfo {year} {2010})}\BibitemShut {NoStop}%
\bibitem [{\citenamefont {Lagendijk}\ \emph {et~al.}(2009)\citenamefont
  {Lagendijk}, \citenamefont {Van~Tiggelen},\ and\ \citenamefont
  {Wiersma}}]{Lagendijk2009}%
  \BibitemOpen
  \bibfield  {author} {\bibinfo {author} {\bibfnamefont {A.}~\bibnamefont
  {Lagendijk}}, \bibinfo {author} {\bibfnamefont {B.}~\bibnamefont
  {Van~Tiggelen}}, \ and\ \bibinfo {author} {\bibfnamefont {D.~S.}\
  \bibnamefont {Wiersma}},\ }\href {\doibase 10.1063/1.3206091} {\bibfield
  {journal} {\bibinfo  {journal} {Phys. Today}\ }\textbf {\bibinfo {volume}
  {62}},\ \bibinfo {pages} {24} (\bibinfo {year} {2009})}\BibitemShut {NoStop}%
\bibitem [{\citenamefont {Ezawa}(2008)}]{Ezawa2008}%
  \BibitemOpen
  \bibfield  {author} {\bibinfo {author} {\bibfnamefont {Z.~F.}\ \bibnamefont
  {Ezawa}},\ }\href@noop {} {\emph {\bibinfo {title} {Quantum Hall effects:
  Field theoretical approach and related topics}}}\ (\bibinfo  {publisher}
  {World Scientific Publishing Co Inc},\ \bibinfo {address} {Singapore},\
  \bibinfo {year} {2008})\BibitemShut {NoStop}%
\bibitem [{\citenamefont {Sutherland}(1986)}]{Sutherland1986}%
  \BibitemOpen
  \bibfield  {author} {\bibinfo {author} {\bibfnamefont {B.}~\bibnamefont
  {Sutherland}},\ }\href {\doibase 10.1103/PhysRevB.34.5208} {\bibfield
  {journal} {\bibinfo  {journal} {Phys. Rev. B}\ }\textbf {\bibinfo {volume}
  {34}},\ \bibinfo {pages} {5208} (\bibinfo {year} {1986})}\BibitemShut
  {NoStop}%
\bibitem [{\citenamefont {Vidal}\ \emph {et~al.}(1998)\citenamefont {Vidal},
  \citenamefont {Mosseri},\ and\ \citenamefont {Dou\ifmmode~\mbox{\c{c}}\else
  \c{c}\fi{}ot}}]{ABcage1998}%
  \BibitemOpen
  \bibfield  {author} {\bibinfo {author} {\bibfnamefont {J.}~\bibnamefont
  {Vidal}}, \bibinfo {author} {\bibfnamefont {R.}~\bibnamefont {Mosseri}}, \
  and\ \bibinfo {author} {\bibfnamefont {B.}~\bibnamefont
  {Dou\ifmmode~\mbox{\c{c}}\else \c{c}\fi{}ot}},\ }\href {\doibase
  10.1103/PhysRevLett.81.5888} {\bibfield  {journal} {\bibinfo  {journal}
  {Phys. Rev. Lett.}\ }\textbf {\bibinfo {volume} {81}},\ \bibinfo {pages}
  {5888} (\bibinfo {year} {1998})}\BibitemShut {NoStop}%
\bibitem [{\citenamefont {Tang}\ \emph {et~al.}(2011)\citenamefont {Tang},
  \citenamefont {Mei},\ and\ \citenamefont {Wen}}]{Tang2011}%
  \BibitemOpen
  \bibfield  {author} {\bibinfo {author} {\bibfnamefont {E.}~\bibnamefont
  {Tang}}, \bibinfo {author} {\bibfnamefont {J.-W.}\ \bibnamefont {Mei}}, \
  and\ \bibinfo {author} {\bibfnamefont {X.-G.}\ \bibnamefont {Wen}},\ }\href
  {\doibase 10.1103/PhysRevLett.106.236802} {\bibfield  {journal} {\bibinfo
  {journal} {Phys. Rev. Lett.}\ }\textbf {\bibinfo {volume} {106}},\ \bibinfo
  {pages} {236802} (\bibinfo {year} {2011})}\BibitemShut {NoStop}%
\bibitem [{\citenamefont {Sun}\ \emph {et~al.}(2011)\citenamefont {Sun},
  \citenamefont {Gu}, \citenamefont {Katsura},\ and\ \citenamefont
  {Das~Sarma}}]{Sun2011}%
  \BibitemOpen
  \bibfield  {author} {\bibinfo {author} {\bibfnamefont {K.}~\bibnamefont
  {Sun}}, \bibinfo {author} {\bibfnamefont {Z.}~\bibnamefont {Gu}}, \bibinfo
  {author} {\bibfnamefont {H.}~\bibnamefont {Katsura}}, \ and\ \bibinfo
  {author} {\bibfnamefont {S.}~\bibnamefont {Das~Sarma}},\ }\href {\doibase
  10.1103/PhysRevLett.106.236803} {\bibfield  {journal} {\bibinfo  {journal}
  {Phys. Rev. Lett.}\ }\textbf {\bibinfo {volume} {106}},\ \bibinfo {pages}
  {236803} (\bibinfo {year} {2011})}\BibitemShut {NoStop}%
\bibitem [{\citenamefont {Neupert}\ \emph {et~al.}(2011)\citenamefont
  {Neupert}, \citenamefont {Santos}, \citenamefont {Chamon},\ and\
  \citenamefont {Mudry}}]{Neupert2011}%
  \BibitemOpen
  \bibfield  {author} {\bibinfo {author} {\bibfnamefont {T.}~\bibnamefont
  {Neupert}}, \bibinfo {author} {\bibfnamefont {L.}~\bibnamefont {Santos}},
  \bibinfo {author} {\bibfnamefont {C.}~\bibnamefont {Chamon}}, \ and\ \bibinfo
  {author} {\bibfnamefont {C.}~\bibnamefont {Mudry}},\ }\href {\doibase
  10.1103/PhysRevLett.106.236804} {\bibfield  {journal} {\bibinfo  {journal}
  {Phys. Rev. Lett.}\ }\textbf {\bibinfo {volume} {106}},\ \bibinfo {pages}
  {236804} (\bibinfo {year} {2011})}\BibitemShut {NoStop}%
\bibitem [{\citenamefont {Lieb}(1989)}]{Lieb1989}%
  \BibitemOpen
  \bibfield  {author} {\bibinfo {author} {\bibfnamefont {E.~H.}\ \bibnamefont
  {Lieb}},\ }\href {\doibase 10.1103/PhysRevLett.62.1201} {\bibfield  {journal}
  {\bibinfo  {journal} {Phys. Rev. Lett.}\ }\textbf {\bibinfo {volume} {62}},\
  \bibinfo {pages} {1201} (\bibinfo {year} {1989})}\BibitemShut {NoStop}%
\bibitem [{\citenamefont {Mielke}(1991{\natexlab{a}})}]{Mielke1991a}%
  \BibitemOpen
  \bibfield  {author} {\bibinfo {author} {\bibfnamefont {A.}~\bibnamefont
  {Mielke}},\ }\href {\doibase 10.1088/0305-4470/24/2/005} {\bibfield
  {journal} {\bibinfo  {journal} {Journal of Physics A: Mathematical and
  General}\ }\textbf {\bibinfo {volume} {24}},\ \bibinfo {pages} {L73}
  (\bibinfo {year} {1991}{\natexlab{a}})}\BibitemShut {NoStop}%
\bibitem [{\citenamefont {Mielke}(1991{\natexlab{b}})}]{Mielke1991b}%
  \BibitemOpen
  \bibfield  {author} {\bibinfo {author} {\bibfnamefont {A.}~\bibnamefont
  {Mielke}},\ }\href {\doibase 10.1088/0305-4470/24/14/018} {\bibfield
  {journal} {\bibinfo  {journal} {Journal of Physics A: Mathematical and
  General}\ }\textbf {\bibinfo {volume} {24}},\ \bibinfo {pages} {3311}
  (\bibinfo {year} {1991}{\natexlab{b}})}\BibitemShut {NoStop}%
\bibitem [{\citenamefont {Tasaki}(1992)}]{Tasaki1992}%
  \BibitemOpen
  \bibfield  {author} {\bibinfo {author} {\bibfnamefont {H.}~\bibnamefont
  {Tasaki}},\ }\href {\doibase 10.1103/PhysRevLett.69.1608} {\bibfield
  {journal} {\bibinfo  {journal} {Phys. Rev. Lett.}\ }\textbf {\bibinfo
  {volume} {69}},\ \bibinfo {pages} {1608} (\bibinfo {year}
  {1992})}\BibitemShut {NoStop}%
\bibitem [{\citenamefont {Tasaki}(1998)}]{Tasaki1998}%
  \BibitemOpen
  \bibfield  {author} {\bibinfo {author} {\bibfnamefont {H.}~\bibnamefont
  {Tasaki}},\ }\href {\doibase 10.1143/PTP.99.489} {\bibfield  {journal}
  {\bibinfo  {journal} {Progress of Theoretical Physics}\ }\textbf {\bibinfo
  {volume} {99}},\ \bibinfo {pages} {489} (\bibinfo {year} {1998})}\BibitemShut
  {NoStop}%
\bibitem [{\citenamefont {Tsui}\ \emph {et~al.}(1982)\citenamefont {Tsui},
  \citenamefont {Stormer},\ and\ \citenamefont {Gossard}}]{Tsui1982}%
  \BibitemOpen
  \bibfield  {author} {\bibinfo {author} {\bibfnamefont {D.~C.}\ \bibnamefont
  {Tsui}}, \bibinfo {author} {\bibfnamefont {H.~L.}\ \bibnamefont {Stormer}}, \
  and\ \bibinfo {author} {\bibfnamefont {A.~C.}\ \bibnamefont {Gossard}},\
  }\href {\doibase 10.1103/PhysRevLett.48.1559} {\bibfield  {journal} {\bibinfo
   {journal} {Phys. Rev. Lett.}\ }\textbf {\bibinfo {volume} {48}},\ \bibinfo
  {pages} {1559} (\bibinfo {year} {1982})}\BibitemShut {NoStop}%
\bibitem [{\citenamefont {Laughlin}(1983)}]{Laughlin1983}%
  \BibitemOpen
  \bibfield  {author} {\bibinfo {author} {\bibfnamefont {R.~B.}\ \bibnamefont
  {Laughlin}},\ }\href {\doibase 10.1103/PhysRevLett.50.1395} {\bibfield
  {journal} {\bibinfo  {journal} {Phys. Rev. Lett.}\ }\textbf {\bibinfo
  {volume} {50}},\ \bibinfo {pages} {1395} (\bibinfo {year}
  {1983})}\BibitemShut {NoStop}%
\bibitem [{\citenamefont {Moore}\ and\ \citenamefont {Read}(1991)}]{Moore1991}%
  \BibitemOpen
  \bibfield  {author} {\bibinfo {author} {\bibfnamefont {G.}~\bibnamefont
  {Moore}}\ and\ \bibinfo {author} {\bibfnamefont {N.}~\bibnamefont {Read}},\
  }\href {\doibase 10.1016/0550-3213(91)90407-O} {\bibfield  {journal}
  {\bibinfo  {journal} {Nuclear Physics B}\ }\textbf {\bibinfo {volume}
  {360}},\ \bibinfo {pages} {362 } (\bibinfo {year} {1991})}\BibitemShut
  {NoStop}%
\bibitem [{\citenamefont {Wilczek}(1990)}]{Wilczek1990}%
  \BibitemOpen
  \bibfield  {author} {\bibinfo {author} {\bibfnamefont {F.}~\bibnamefont
  {Wilczek}},\ }\href@noop {} {\emph {\bibinfo {title} {Fractional statistics
  and anyon superconductivity}}},\ Vol.~\bibinfo {volume} {5}\ (\bibinfo
  {publisher} {World scientific},\ \bibinfo {address} {London},\ \bibinfo
  {year} {1990})\BibitemShut {NoStop}%
\bibitem [{\citenamefont {Li}\ \emph {et~al.}(2013)\citenamefont {Li},
  \citenamefont {Zhao},\ and\ \citenamefont {Liu}}]{LiZhaoLiu2013}%
  \BibitemOpen
  \bibfield  {author} {\bibinfo {author} {\bibfnamefont {X.}~\bibnamefont
  {Li}}, \bibinfo {author} {\bibfnamefont {E.}~\bibnamefont {Zhao}}, \ and\
  \bibinfo {author} {\bibfnamefont {W.~V.}\ \bibnamefont {Liu}},\ }\href
  {\doibase 10.1038/ncomms2523} {\bibfield  {journal} {\bibinfo  {journal}
  {Nature communications}\ }\textbf {\bibinfo {volume} {4}},\ \bibinfo {pages}
  {1523} (\bibinfo {year} {2013})}\BibitemShut {NoStop}%
\bibitem [{\citenamefont {J\"unemann}\ \emph {et~al.}(2017)\citenamefont
  {J\"unemann}, \citenamefont {Piga}, \citenamefont {Ran}, \citenamefont
  {Lewenstein}, \citenamefont {Rizzi},\ and\ \citenamefont
  {Bermudez}}]{MRizzi2017}%
  \BibitemOpen
  \bibfield  {author} {\bibinfo {author} {\bibfnamefont {J.}~\bibnamefont
  {J\"unemann}}, \bibinfo {author} {\bibfnamefont {A.}~\bibnamefont {Piga}},
  \bibinfo {author} {\bibfnamefont {S.-J.}\ \bibnamefont {Ran}}, \bibinfo
  {author} {\bibfnamefont {M.}~\bibnamefont {Lewenstein}}, \bibinfo {author}
  {\bibfnamefont {M.}~\bibnamefont {Rizzi}}, \ and\ \bibinfo {author}
  {\bibfnamefont {A.}~\bibnamefont {Bermudez}},\ }\href {\doibase
  10.1103/PhysRevX.7.031057} {\bibfield  {journal} {\bibinfo  {journal} {Phys.
  Rev. X}\ }\textbf {\bibinfo {volume} {7}},\ \bibinfo {pages} {031057}
  (\bibinfo {year} {2017})}\BibitemShut {NoStop}%
\bibitem [{\citenamefont {Leykam}\ \emph {et~al.}(2018)\citenamefont {Leykam},
  \citenamefont {Andreanov},\ and\ \citenamefont {Flach}}]{Leykam2018}%
  \BibitemOpen
  \bibfield  {author} {\bibinfo {author} {\bibfnamefont {D.}~\bibnamefont
  {Leykam}}, \bibinfo {author} {\bibfnamefont {A.}~\bibnamefont {Andreanov}}, \
  and\ \bibinfo {author} {\bibfnamefont {S.}~\bibnamefont {Flach}},\ }\href
  {\doibase 10.1080/23746149.2018.1473052} {\bibfield  {journal} {\bibinfo
  {journal} {Advances in Physics: X}\ }\textbf {\bibinfo {volume} {3}},\
  \bibinfo {pages} {1473052} (\bibinfo {year} {2018})}\BibitemShut {NoStop}%
\bibitem [{\citenamefont {Reimann}\ and\ \citenamefont
  {Manninen}(2002)}]{ReimannManninen2002}%
  \BibitemOpen
  \bibfield  {author} {\bibinfo {author} {\bibfnamefont {S.~M.}\ \bibnamefont
  {Reimann}}\ and\ \bibinfo {author} {\bibfnamefont {M.}~\bibnamefont
  {Manninen}},\ }\href {\doibase 10.1103/RevModPhys.74.1283} {\bibfield
  {journal} {\bibinfo  {journal} {Rev. Mod. Phys.}\ }\textbf {\bibinfo {volume}
  {74}},\ \bibinfo {pages} {1283} (\bibinfo {year} {2002})}\BibitemShut
  {NoStop}%
\bibitem [{\citenamefont {Fazio}\ and\ \citenamefont {Van
  Der~Zant}(2001)}]{FazioVanderZant2001}%
  \BibitemOpen
  \bibfield  {author} {\bibinfo {author} {\bibfnamefont {R.}~\bibnamefont
  {Fazio}}\ and\ \bibinfo {author} {\bibfnamefont {H.}~\bibnamefont {Van
  Der~Zant}},\ }\href {\doibase 10.1016/S0370-1573(01)00022-9} {\bibfield
  {journal} {\bibinfo  {journal} {Physics Reports}\ }\textbf {\bibinfo {volume}
  {355}},\ \bibinfo {pages} {235} (\bibinfo {year} {2001})}\BibitemShut
  {NoStop}%
\bibitem [{\citenamefont {Haldane}\ and\ \citenamefont
  {Raghu}(2008)}]{Haldane2008}%
  \BibitemOpen
  \bibfield  {author} {\bibinfo {author} {\bibfnamefont {F.~D.~M.}\
  \bibnamefont {Haldane}}\ and\ \bibinfo {author} {\bibfnamefont
  {S.}~\bibnamefont {Raghu}},\ }\href {\doibase 10.1103/PhysRevLett.100.013904}
  {\bibfield  {journal} {\bibinfo  {journal} {Phys. Rev. Lett.}\ }\textbf
  {\bibinfo {volume} {100}},\ \bibinfo {pages} {013904} (\bibinfo {year}
  {2008})}\BibitemShut {NoStop}%
\bibitem [{\citenamefont {Hafezi}\ \emph {et~al.}(2013)\citenamefont {Hafezi},
  \citenamefont {Mittal}, \citenamefont {Fan}, \citenamefont {Migdall},\ and\
  \citenamefont {Taylor}}]{Hafezi2013}%
  \BibitemOpen
  \bibfield  {author} {\bibinfo {author} {\bibfnamefont {M.}~\bibnamefont
  {Hafezi}}, \bibinfo {author} {\bibfnamefont {S.}~\bibnamefont {Mittal}},
  \bibinfo {author} {\bibfnamefont {J.}~\bibnamefont {Fan}}, \bibinfo {author}
  {\bibfnamefont {A.}~\bibnamefont {Migdall}}, \ and\ \bibinfo {author}
  {\bibfnamefont {J.}~\bibnamefont {Taylor}},\ }\href
  {http://dx.doi.org/10.1038/nphoton.2013.274} {\bibfield  {journal} {\bibinfo
  {journal} {Nature Photonics}\ }\textbf {\bibinfo {volume} {7}},\ \bibinfo
  {pages} {1001} (\bibinfo {year} {2013})}\BibitemShut {NoStop}%
\bibitem [{\citenamefont {Ningyuan}\ \emph {et~al.}(2015)\citenamefont
  {Ningyuan}, \citenamefont {Owens}, \citenamefont {Sommer}, \citenamefont
  {Schuster},\ and\ \citenamefont {Simon}}]{Ningyuan2015}%
  \BibitemOpen
  \bibfield  {author} {\bibinfo {author} {\bibfnamefont {J.}~\bibnamefont
  {Ningyuan}}, \bibinfo {author} {\bibfnamefont {C.}~\bibnamefont {Owens}},
  \bibinfo {author} {\bibfnamefont {A.}~\bibnamefont {Sommer}}, \bibinfo
  {author} {\bibfnamefont {D.}~\bibnamefont {Schuster}}, \ and\ \bibinfo
  {author} {\bibfnamefont {J.}~\bibnamefont {Simon}},\ }\href {\doibase
  10.1103/PhysRevX.5.021031} {\bibfield  {journal} {\bibinfo  {journal} {Phys.
  Rev. X}\ }\textbf {\bibinfo {volume} {5}},\ \bibinfo {pages} {021031}
  (\bibinfo {year} {2015})}\BibitemShut {NoStop}%
\bibitem [{\citenamefont {Deng}\ \emph {et~al.}(2016)\citenamefont {Deng},
  \citenamefont {Lai},\ and\ \citenamefont {Chien}}]{Deng2016}%
  \BibitemOpen
  \bibfield  {author} {\bibinfo {author} {\bibfnamefont {X.-H.}\ \bibnamefont
  {Deng}}, \bibinfo {author} {\bibfnamefont {C.-Y.}\ \bibnamefont {Lai}}, \
  and\ \bibinfo {author} {\bibfnamefont {C.-C.}\ \bibnamefont {Chien}},\ }\href
  {\doibase 10.1103/PhysRevB.93.054116} {\bibfield  {journal} {\bibinfo
  {journal} {Phys. Rev. B}\ }\textbf {\bibinfo {volume} {93}},\ \bibinfo
  {pages} {054116} (\bibinfo {year} {2016})}\BibitemShut {NoStop}%
\bibitem [{\citenamefont {Ozawa}\ \emph {et~al.}(2018)\citenamefont {Ozawa},
  \citenamefont {Price}, \citenamefont {Amo}, \citenamefont {Goldman},
  \citenamefont {Hafezi}, \citenamefont {Lu}, \citenamefont {Rechtsman},
  \citenamefont {Schuster}, \citenamefont {Simon}, \citenamefont {Zilberberg},\
  and\ \citenamefont {Carusotto}}]{Ozawa2018}%
  \BibitemOpen
  \bibfield  {author} {\bibinfo {author} {\bibfnamefont {T.}~\bibnamefont
  {Ozawa}}, \bibinfo {author} {\bibfnamefont {H.~M.}\ \bibnamefont {Price}},
  \bibinfo {author} {\bibfnamefont {A.}~\bibnamefont {Amo}}, \bibinfo {author}
  {\bibfnamefont {N.}~\bibnamefont {Goldman}}, \bibinfo {author} {\bibfnamefont
  {M.}~\bibnamefont {Hafezi}}, \bibinfo {author} {\bibfnamefont
  {L.}~\bibnamefont {Lu}}, \bibinfo {author} {\bibfnamefont {M.}~\bibnamefont
  {Rechtsman}}, \bibinfo {author} {\bibfnamefont {D.}~\bibnamefont {Schuster}},
  \bibinfo {author} {\bibfnamefont {J.}~\bibnamefont {Simon}}, \bibinfo
  {author} {\bibfnamefont {O.}~\bibnamefont {Zilberberg}}, \ and\ \bibinfo
  {author} {\bibfnamefont {L.}~\bibnamefont {Carusotto}},\ }\href@noop {}
  {\bibfield  {journal} {\bibinfo  {journal} {ArXiv e-prints}\ } (\bibinfo
  {year} {2018})},\ \Eprint {http://arxiv.org/abs/1802.04173}
  {arXiv:1802.04173} \BibitemShut {NoStop}%
\bibitem [{\citenamefont {Bloch}\ \emph {et~al.}(2008)\citenamefont {Bloch},
  \citenamefont {Dalibard},\ and\ \citenamefont
  {Zwerger}}]{BlochDalibardZwerger2008}%
  \BibitemOpen
  \bibfield  {author} {\bibinfo {author} {\bibfnamefont {I.}~\bibnamefont
  {Bloch}}, \bibinfo {author} {\bibfnamefont {J.}~\bibnamefont {Dalibard}}, \
  and\ \bibinfo {author} {\bibfnamefont {W.}~\bibnamefont {Zwerger}},\ }\href
  {\doibase 10.1103/RevModPhys.80.885} {\bibfield  {journal} {\bibinfo
  {journal} {Rev. Mod. Phys.}\ }\textbf {\bibinfo {volume} {80}},\ \bibinfo
  {pages} {885} (\bibinfo {year} {2008})}\BibitemShut {NoStop}%
\bibitem [{\citenamefont {Lewenstein}\ \emph {et~al.}(2012)\citenamefont
  {Lewenstein}, \citenamefont {Sanpera},\ and\ \citenamefont
  {Ahufinger}}]{LewensteinSanpera2012}%
  \BibitemOpen
  \bibfield  {author} {\bibinfo {author} {\bibfnamefont {M.}~\bibnamefont
  {Lewenstein}}, \bibinfo {author} {\bibfnamefont {A.}~\bibnamefont {Sanpera}},
  \ and\ \bibinfo {author} {\bibfnamefont {V.}~\bibnamefont {Ahufinger}},\
  }\href@noop {} {\emph {\bibinfo {title} {Ultracold atoms in optical
  lattices}}},\ Vol.\ \bibinfo {volume} {143}\ (\bibinfo  {publisher} {Oxford
  University Press USA},\ \bibinfo {address} {New York},\ \bibinfo {year}
  {2012})\BibitemShut {NoStop}%
\bibitem [{\citenamefont {Bloch}\ \emph {et~al.}(2012)\citenamefont {Bloch},
  \citenamefont {Dalibard},\ and\ \citenamefont
  {Nascimbene}}]{DalibardNascimbene2012}%
  \BibitemOpen
  \bibfield  {author} {\bibinfo {author} {\bibfnamefont {I.}~\bibnamefont
  {Bloch}}, \bibinfo {author} {\bibfnamefont {J.}~\bibnamefont {Dalibard}}, \
  and\ \bibinfo {author} {\bibfnamefont {S.}~\bibnamefont {Nascimbene}},\
  }\href {\doibase 10.1038/nphys2259} {\bibfield  {journal} {\bibinfo
  {journal} {Nature Physics}\ }\textbf {\bibinfo {volume} {8}},\ \bibinfo
  {pages} {267} (\bibinfo {year} {2012})}\BibitemShut {NoStop}%
\bibitem [{\citenamefont {Goldman}\ \emph {et~al.}(2014)\citenamefont
  {Goldman}, \citenamefont {Juzeli{\=u}nas}, \citenamefont {{\"O}hberg},\ and\
  \citenamefont {Spielman}}]{Goldman2014}%
  \BibitemOpen
  \bibfield  {author} {\bibinfo {author} {\bibfnamefont {N.}~\bibnamefont
  {Goldman}}, \bibinfo {author} {\bibfnamefont {G.}~\bibnamefont
  {Juzeli{\=u}nas}}, \bibinfo {author} {\bibfnamefont {P.}~\bibnamefont
  {{\"O}hberg}}, \ and\ \bibinfo {author} {\bibfnamefont {I.~B.}\ \bibnamefont
  {Spielman}},\ }\href {http://stacks.iop.org/0034-4885/77/i=12/a=126401}
  {\bibfield  {journal} {\bibinfo  {journal} {Reports on Progress in Physics}\
  }\textbf {\bibinfo {volume} {77}},\ \bibinfo {pages} {126401} (\bibinfo
  {year} {2014})}\BibitemShut {NoStop}%
\bibitem [{\citenamefont {Dalibard}\ \emph {et~al.}(2011)\citenamefont
  {Dalibard}, \citenamefont {Gerbier}, \citenamefont {Juzeli{\=u}nas},\ and\
  \citenamefont {\"Ohberg}}]{Dalibard2011}%
  \BibitemOpen
  \bibfield  {author} {\bibinfo {author} {\bibfnamefont {J.}~\bibnamefont
  {Dalibard}}, \bibinfo {author} {\bibfnamefont {F.}~\bibnamefont {Gerbier}},
  \bibinfo {author} {\bibfnamefont {G.}~\bibnamefont {Juzeli{\=u}nas}}, \ and\
  \bibinfo {author} {\bibfnamefont {P.}~\bibnamefont {\"Ohberg}},\ }\href
  {\doibase 10.1103/RevModPhys.83.1523} {\bibfield  {journal} {\bibinfo
  {journal} {Rev. Mod. Phys.}\ }\textbf {\bibinfo {volume} {83}},\ \bibinfo
  {pages} {1523} (\bibinfo {year} {2011})}\BibitemShut {NoStop}%
\bibitem [{\citenamefont {Drese}\ and\ \citenamefont
  {Holthaus}(1997)}]{Holthaus1997}%
  \BibitemOpen
  \bibfield  {author} {\bibinfo {author} {\bibfnamefont {K.}~\bibnamefont
  {Drese}}\ and\ \bibinfo {author} {\bibfnamefont {M.}~\bibnamefont
  {Holthaus}},\ }\href {\doibase
  http://dx.doi.org/10.1016/S0301-0104(97)00025-6} {\bibfield  {journal}
  {\bibinfo  {journal} {Chemical Physics}\ }\textbf {\bibinfo {volume} {217}},\
  \bibinfo {pages} {201 } (\bibinfo {year} {1997})},\ \bibinfo {note} {dynamics
  of Driven Quantum Systems}\BibitemShut {NoStop}%
\bibitem [{\citenamefont {Sias}\ \emph {et~al.}(2008)\citenamefont {Sias},
  \citenamefont {Lignier}, \citenamefont {Singh}, \citenamefont {Zenesini},
  \citenamefont {Ciampini}, \citenamefont {Morsch},\ and\ \citenamefont
  {Arimondo}}]{Sias2008}%
  \BibitemOpen
  \bibfield  {author} {\bibinfo {author} {\bibfnamefont {C.}~\bibnamefont
  {Sias}}, \bibinfo {author} {\bibfnamefont {H.}~\bibnamefont {Lignier}},
  \bibinfo {author} {\bibfnamefont {Y.~P.}\ \bibnamefont {Singh}}, \bibinfo
  {author} {\bibfnamefont {A.}~\bibnamefont {Zenesini}}, \bibinfo {author}
  {\bibfnamefont {D.}~\bibnamefont {Ciampini}}, \bibinfo {author}
  {\bibfnamefont {O.}~\bibnamefont {Morsch}}, \ and\ \bibinfo {author}
  {\bibfnamefont {E.}~\bibnamefont {Arimondo}},\ }\href {\doibase
  10.1103/PhysRevLett.100.040404} {\bibfield  {journal} {\bibinfo  {journal}
  {Phys. Rev. Lett.}\ }\textbf {\bibinfo {volume} {100}},\ \bibinfo {pages}
  {040404} (\bibinfo {year} {2008})}\BibitemShut {NoStop}%
\bibitem [{\citenamefont {Struck}\ \emph {et~al.}(2011)\citenamefont {Struck},
  \citenamefont {{\"O}lschl{\"a}ger}, \citenamefont {Le~Targat}, \citenamefont
  {Soltan-Panahi}, \citenamefont {Eckardt}, \citenamefont {Lewenstein},
  \citenamefont {Windpassinger},\ and\ \citenamefont {Sengstock}}]{Struck2011}%
  \BibitemOpen
  \bibfield  {author} {\bibinfo {author} {\bibfnamefont {J.}~\bibnamefont
  {Struck}}, \bibinfo {author} {\bibfnamefont {C.}~\bibnamefont
  {{\"O}lschl{\"a}ger}}, \bibinfo {author} {\bibfnamefont {R.}~\bibnamefont
  {Le~Targat}}, \bibinfo {author} {\bibfnamefont {P.}~\bibnamefont
  {Soltan-Panahi}}, \bibinfo {author} {\bibfnamefont {A.}~\bibnamefont
  {Eckardt}}, \bibinfo {author} {\bibfnamefont {M.}~\bibnamefont {Lewenstein}},
  \bibinfo {author} {\bibfnamefont {P.}~\bibnamefont {Windpassinger}}, \ and\
  \bibinfo {author} {\bibfnamefont {K.}~\bibnamefont {Sengstock}},\ }\href
  {\doibase 10.1126/science.1207239} {\bibfield  {journal} {\bibinfo  {journal}
  {Science}\ }\textbf {\bibinfo {volume} {333}},\ \bibinfo {pages} {996}
  (\bibinfo {year} {2011})}\BibitemShut {NoStop}%
\bibitem [{\citenamefont {Eckardt}(2017)}]{Eckardt2017}%
  \BibitemOpen
  \bibfield  {author} {\bibinfo {author} {\bibfnamefont {A.}~\bibnamefont
  {Eckardt}},\ }\href {\doibase 10.1103/RevModPhys.89.011004} {\bibfield
  {journal} {\bibinfo  {journal} {Rev. Mod. Phys.}\ }\textbf {\bibinfo {volume}
  {89}},\ \bibinfo {pages} {011004} (\bibinfo {year} {2017})}\BibitemShut
  {NoStop}%
\bibitem [{\citenamefont {Huber}\ and\ \citenamefont
  {Altman}(2010)}]{Altman2010}%
  \BibitemOpen
  \bibfield  {author} {\bibinfo {author} {\bibfnamefont {S.~D.}\ \bibnamefont
  {Huber}}\ and\ \bibinfo {author} {\bibfnamefont {E.}~\bibnamefont {Altman}},\
  }\href {\doibase 10.1103/PhysRevB.82.184502} {\bibfield  {journal} {\bibinfo
  {journal} {Phys. Rev. B}\ }\textbf {\bibinfo {volume} {82}},\ \bibinfo
  {pages} {184502} (\bibinfo {year} {2010})}\BibitemShut {NoStop}%
\bibitem [{\citenamefont {M\"oller}\ and\ \citenamefont
  {Cooper}(2012)}]{MoellerCooper2012}%
  \BibitemOpen
  \bibfield  {author} {\bibinfo {author} {\bibfnamefont {G.}~\bibnamefont
  {M\"oller}}\ and\ \bibinfo {author} {\bibfnamefont {N.~R.}\ \bibnamefont
  {Cooper}},\ }\href {\doibase 10.1103/PhysRevLett.108.045306} {\bibfield
  {journal} {\bibinfo  {journal} {Phys. Rev. Lett.}\ }\textbf {\bibinfo
  {volume} {108}},\ \bibinfo {pages} {045306} (\bibinfo {year}
  {2012})}\BibitemShut {NoStop}%
\bibitem [{\citenamefont {Mielke}(2018)}]{Mielke2018}%
  \BibitemOpen
  \bibfield  {author} {\bibinfo {author} {\bibfnamefont {A.}~\bibnamefont
  {Mielke}},\ }\href {\doibase 10.1007/s10955-018-2030-0} {\bibfield  {journal}
  {\bibinfo  {journal} {Journal of Statistical Physics}\ }\textbf {\bibinfo
  {volume} {171}},\ \bibinfo {pages} {679} (\bibinfo {year}
  {2018})}\BibitemShut {NoStop}%
\bibitem [{\citenamefont {Vidal}\ \emph {et~al.}(2000)\citenamefont {Vidal},
  \citenamefont {Dou\ifmmode~\mbox{\c{c}}\else \c{c}\fi{}ot}, \citenamefont
  {Mosseri},\ and\ \citenamefont {Butaud}}]{VidalDoucot2000}%
  \BibitemOpen
  \bibfield  {author} {\bibinfo {author} {\bibfnamefont {J.}~\bibnamefont
  {Vidal}}, \bibinfo {author} {\bibfnamefont {B.}~\bibnamefont
  {Dou\ifmmode~\mbox{\c{c}}\else \c{c}\fi{}ot}}, \bibinfo {author}
  {\bibfnamefont {R.}~\bibnamefont {Mosseri}}, \ and\ \bibinfo {author}
  {\bibfnamefont {P.}~\bibnamefont {Butaud}},\ }\href {\doibase
  10.1103/PhysRevLett.85.3906} {\bibfield  {journal} {\bibinfo  {journal}
  {Phys. Rev. Lett.}\ }\textbf {\bibinfo {volume} {85}},\ \bibinfo {pages}
  {3906} (\bibinfo {year} {2000})}\BibitemShut {NoStop}%
\bibitem [{\citenamefont {Dou\ifmmode~\mbox{\c{c}}\else \c{c}\fi{}ot}\ and\
  \citenamefont {Vidal}(2002)}]{Doucot2002}%
  \BibitemOpen
  \bibfield  {author} {\bibinfo {author} {\bibfnamefont {B.}~\bibnamefont
  {Dou\ifmmode~\mbox{\c{c}}\else \c{c}\fi{}ot}}\ and\ \bibinfo {author}
  {\bibfnamefont {J.}~\bibnamefont {Vidal}},\ }\href {\doibase
  10.1103/PhysRevLett.88.227005} {\bibfield  {journal} {\bibinfo  {journal}
  {Phys. Rev. Lett.}\ }\textbf {\bibinfo {volume} {88}},\ \bibinfo {pages}
  {227005} (\bibinfo {year} {2002})}\BibitemShut {NoStop}%
\bibitem [{\citenamefont {Rizzi}\ \emph {et~al.}(2006)\citenamefont {Rizzi},
  \citenamefont {Cataudella},\ and\ \citenamefont {Fazio}}]{Rizzi2006JJ}%
  \BibitemOpen
  \bibfield  {author} {\bibinfo {author} {\bibfnamefont {M.}~\bibnamefont
  {Rizzi}}, \bibinfo {author} {\bibfnamefont {V.}~\bibnamefont {Cataudella}}, \
  and\ \bibinfo {author} {\bibfnamefont {R.}~\bibnamefont {Fazio}},\ }\href
  {\doibase 10.1103/PhysRevB.73.100502} {\bibfield  {journal} {\bibinfo
  {journal} {Phys. Rev. B}\ }\textbf {\bibinfo {volume} {73}},\ \bibinfo
  {pages} {100502} (\bibinfo {year} {2006})}\BibitemShut {NoStop}%
\bibitem [{\citenamefont {Daley}\ \emph {et~al.}(2009)\citenamefont {Daley},
  \citenamefont {Taylor}, \citenamefont {Diehl}, \citenamefont {Baranov},\ and\
  \citenamefont {Zoller}}]{Daley2009}%
  \BibitemOpen
  \bibfield  {author} {\bibinfo {author} {\bibfnamefont {A.~J.}\ \bibnamefont
  {Daley}}, \bibinfo {author} {\bibfnamefont {J.~M.}\ \bibnamefont {Taylor}},
  \bibinfo {author} {\bibfnamefont {S.}~\bibnamefont {Diehl}}, \bibinfo
  {author} {\bibfnamefont {M.}~\bibnamefont {Baranov}}, \ and\ \bibinfo
  {author} {\bibfnamefont {P.}~\bibnamefont {Zoller}},\ }\href {\doibase
  10.1103/PhysRevLett.102.040402} {\bibfield  {journal} {\bibinfo  {journal}
  {Phys. Rev. Lett.}\ }\textbf {\bibinfo {volume} {102}},\ \bibinfo {pages}
  {040402} (\bibinfo {year} {2009})}\BibitemShut {NoStop}%
\bibitem [{\citenamefont {Diehl}\ \emph
  {et~al.}(2010{\natexlab{a}})\citenamefont {Diehl}, \citenamefont {Baranov},
  \citenamefont {Daley},\ and\ \citenamefont {Zoller}}]{Daley2010}%
  \BibitemOpen
  \bibfield  {author} {\bibinfo {author} {\bibfnamefont {S.}~\bibnamefont
  {Diehl}}, \bibinfo {author} {\bibfnamefont {M.}~\bibnamefont {Baranov}},
  \bibinfo {author} {\bibfnamefont {A.~J.}\ \bibnamefont {Daley}}, \ and\
  \bibinfo {author} {\bibfnamefont {P.}~\bibnamefont {Zoller}},\ }\href
  {\doibase 10.1103/PhysRevLett.104.165301} {\bibfield  {journal} {\bibinfo
  {journal} {Phys. Rev. Lett.}\ }\textbf {\bibinfo {volume} {104}},\ \bibinfo
  {pages} {165301} (\bibinfo {year} {2010}{\natexlab{a}})}\BibitemShut
  {NoStop}%
\bibitem [{\citenamefont {Diehl}\ \emph
  {et~al.}(2010{\natexlab{b}})\citenamefont {Diehl}, \citenamefont {Baranov},
  \citenamefont {Daley},\ and\ \citenamefont {Zoller}}]{DiehlDaley2010}%
  \BibitemOpen
  \bibfield  {author} {\bibinfo {author} {\bibfnamefont {S.}~\bibnamefont
  {Diehl}}, \bibinfo {author} {\bibfnamefont {M.}~\bibnamefont {Baranov}},
  \bibinfo {author} {\bibfnamefont {A.~J.}\ \bibnamefont {Daley}}, \ and\
  \bibinfo {author} {\bibfnamefont {P.}~\bibnamefont {Zoller}},\ }\href
  {\doibase 10.1103/PhysRevB.82.064509} {\bibfield  {journal} {\bibinfo
  {journal} {Phys. Rev. B}\ }\textbf {\bibinfo {volume} {82}},\ \bibinfo
  {pages} {064509} (\bibinfo {year} {2010}{\natexlab{b}})}\BibitemShut
  {NoStop}%
\bibitem [{\citenamefont {Mazza}\ \emph {et~al.}(2010)\citenamefont {Mazza},
  \citenamefont {Rizzi}, \citenamefont {Lewenstein},\ and\ \citenamefont
  {Cirac}}]{MazzaRizzi2010}%
  \BibitemOpen
  \bibfield  {author} {\bibinfo {author} {\bibfnamefont {L.}~\bibnamefont
  {Mazza}}, \bibinfo {author} {\bibfnamefont {M.}~\bibnamefont {Rizzi}},
  \bibinfo {author} {\bibfnamefont {M.}~\bibnamefont {Lewenstein}}, \ and\
  \bibinfo {author} {\bibfnamefont {J.~I.}\ \bibnamefont {Cirac}},\ }\href
  {\doibase 10.1103/PhysRevA.82.043629} {\bibfield  {journal} {\bibinfo
  {journal} {Phys. Rev. A}\ }\textbf {\bibinfo {volume} {82}},\ \bibinfo
  {pages} {043629} (\bibinfo {year} {2010})}\BibitemShut {NoStop}%
\bibitem [{\citenamefont {Daley}\ and\ \citenamefont
  {Simon}(2014)}]{Daley2014}%
  \BibitemOpen
  \bibfield  {author} {\bibinfo {author} {\bibfnamefont {A.~J.}\ \bibnamefont
  {Daley}}\ and\ \bibinfo {author} {\bibfnamefont {J.}~\bibnamefont {Simon}},\
  }\href {\doibase 10.1103/PhysRevA.89.053619} {\bibfield  {journal} {\bibinfo
  {journal} {Phys. Rev. A}\ }\textbf {\bibinfo {volume} {89}},\ \bibinfo
  {pages} {053619} (\bibinfo {year} {2014})}\BibitemShut {NoStop}%
\bibitem [{\citenamefont {Sowi\ifmmode~\acute{n}\else \'{n}\fi{}ski}\ \emph
  {et~al.}(2015)\citenamefont {Sowi\ifmmode~\acute{n}\else \'{n}\fi{}ski},
  \citenamefont {Chhajlany}, \citenamefont {Dutta}, \citenamefont
  {Tagliacozzo},\ and\ \citenamefont {Lewenstein}}]{Lewenstein2015}%
  \BibitemOpen
  \bibfield  {author} {\bibinfo {author} {\bibfnamefont {T.}~\bibnamefont
  {Sowi\ifmmode~\acute{n}\else \'{n}\fi{}ski}}, \bibinfo {author}
  {\bibfnamefont {R.~W.}\ \bibnamefont {Chhajlany}}, \bibinfo {author}
  {\bibfnamefont {O.}~\bibnamefont {Dutta}}, \bibinfo {author} {\bibfnamefont
  {L.}~\bibnamefont {Tagliacozzo}}, \ and\ \bibinfo {author} {\bibfnamefont
  {M.}~\bibnamefont {Lewenstein}},\ }\href {\doibase
  10.1103/PhysRevA.92.043615} {\bibfield  {journal} {\bibinfo  {journal} {Phys.
  Rev. A}\ }\textbf {\bibinfo {volume} {92}},\ \bibinfo {pages} {043615}
  (\bibinfo {year} {2015})}\BibitemShut {NoStop}%
\bibitem [{\citenamefont {Peotta}\ and\ \citenamefont
  {T{\"o}rm{\"a}}(2015)}]{PeottaTorma2015}%
  \BibitemOpen
  \bibfield  {author} {\bibinfo {author} {\bibfnamefont {S.}~\bibnamefont
  {Peotta}}\ and\ \bibinfo {author} {\bibfnamefont {P.}~\bibnamefont
  {T{\"o}rm{\"a}}},\ }\href {\doibase 10.1038/ncomms9944} {\bibfield  {journal}
  {\bibinfo  {journal} {Nature communications}\ }\textbf {\bibinfo {volume}
  {6}},\ \bibinfo {pages} {8944} (\bibinfo {year} {2015})}\BibitemShut
  {NoStop}%
\bibitem [{\citenamefont {Tovmasyan}\ \emph {et~al.}(2016)\citenamefont
  {Tovmasyan}, \citenamefont {Peotta}, \citenamefont {T\"orm\"a},\ and\
  \citenamefont {Huber}}]{Huber2016}%
  \BibitemOpen
  \bibfield  {author} {\bibinfo {author} {\bibfnamefont {M.}~\bibnamefont
  {Tovmasyan}}, \bibinfo {author} {\bibfnamefont {S.}~\bibnamefont {Peotta}},
  \bibinfo {author} {\bibfnamefont {P.}~\bibnamefont {T\"orm\"a}}, \ and\
  \bibinfo {author} {\bibfnamefont {S.~D.}\ \bibnamefont {Huber}},\ }\href
  {\doibase 10.1103/PhysRevB.94.245149} {\bibfield  {journal} {\bibinfo
  {journal} {Phys. Rev. B}\ }\textbf {\bibinfo {volume} {94}},\ \bibinfo
  {pages} {245149} (\bibinfo {year} {2016})}\BibitemShut {NoStop}%
\bibitem [{\citenamefont {Kobayashi}\ \emph {et~al.}(2016)\citenamefont
  {Kobayashi}, \citenamefont {Okumura}, \citenamefont {Yamada}, \citenamefont
  {Machida},\ and\ \citenamefont {Aoki}}]{AokiKobayashi2016}%
  \BibitemOpen
  \bibfield  {author} {\bibinfo {author} {\bibfnamefont {K.}~\bibnamefont
  {Kobayashi}}, \bibinfo {author} {\bibfnamefont {M.}~\bibnamefont {Okumura}},
  \bibinfo {author} {\bibfnamefont {S.}~\bibnamefont {Yamada}}, \bibinfo
  {author} {\bibfnamefont {M.}~\bibnamefont {Machida}}, \ and\ \bibinfo
  {author} {\bibfnamefont {H.}~\bibnamefont {Aoki}},\ }\href {\doibase
  10.1103/PhysRevB.94.214501} {\bibfield  {journal} {\bibinfo  {journal} {Phys.
  Rev. B}\ }\textbf {\bibinfo {volume} {94}},\ \bibinfo {pages} {214501}
  (\bibinfo {year} {2016})}\BibitemShut {NoStop}%
\bibitem [{\citenamefont {Tovmasyan}\ \emph {et~al.}(2018)\citenamefont
  {Tovmasyan}, \citenamefont {Peotta}, \citenamefont {Liang}, \citenamefont
  {T{\"o}rm{\"a}},\ and\ \citenamefont {Huber}}]{Tovmasyan2018}%
  \BibitemOpen
  \bibfield  {author} {\bibinfo {author} {\bibfnamefont {M.}~\bibnamefont
  {Tovmasyan}}, \bibinfo {author} {\bibfnamefont {S.}~\bibnamefont {Peotta}},
  \bibinfo {author} {\bibfnamefont {L.}~\bibnamefont {Liang}}, \bibinfo
  {author} {\bibfnamefont {P.}~\bibnamefont {T{\"o}rm{\"a}}}, \ and\ \bibinfo
  {author} {\bibfnamefont {S.~D.}\ \bibnamefont {Huber}},\ }\href@noop {}
  {\bibfield  {journal} {\bibinfo  {journal} {ArXiv e-prints}\ } (\bibinfo
  {year} {2018})},\ \Eprint {http://arxiv.org/abs/1805.04529}
  {arXiv:1805.04529} \BibitemShut {NoStop}%
\bibitem [{\citenamefont {Hyrk\"as}\ \emph {et~al.}(2013)\citenamefont
  {Hyrk\"as}, \citenamefont {Apaja},\ and\ \citenamefont
  {Manninen}}]{Hyrkas2013}%
  \BibitemOpen
  \bibfield  {author} {\bibinfo {author} {\bibfnamefont {M.}~\bibnamefont
  {Hyrk\"as}}, \bibinfo {author} {\bibfnamefont {V.}~\bibnamefont {Apaja}}, \
  and\ \bibinfo {author} {\bibfnamefont {M.}~\bibnamefont {Manninen}},\ }\href
  {\doibase 10.1103/PhysRevA.87.023614} {\bibfield  {journal} {\bibinfo
  {journal} {Phys. Rev. A}\ }\textbf {\bibinfo {volume} {87}},\ \bibinfo
  {pages} {023614} (\bibinfo {year} {2013})}\BibitemShut {NoStop}%
\bibitem [{\citenamefont {Bercioux}\ \emph {et~al.}(2017)\citenamefont
  {Bercioux}, \citenamefont {Dutta},\ and\ \citenamefont {Rico}}]{Dutta2017}%
  \BibitemOpen
  \bibfield  {author} {\bibinfo {author} {\bibfnamefont {D.}~\bibnamefont
  {Bercioux}}, \bibinfo {author} {\bibfnamefont {O.}~\bibnamefont {Dutta}}, \
  and\ \bibinfo {author} {\bibfnamefont {E.}~\bibnamefont {Rico}},\ }\href
  {\doibase 10.1002/andp.201600262} {\bibfield  {journal} {\bibinfo  {journal}
  {Annalen der Physik}\ }\textbf {\bibinfo {volume} {529}},\ \bibinfo {pages}
  {1600262} (\bibinfo {year} {2017})},\ \bibinfo {note} {1600262}\BibitemShut
  {NoStop}%
\bibitem [{\citenamefont {Kremer}\ \emph {et~al.}(2018)\citenamefont {Kremer},
  \citenamefont {Petrides}, \citenamefont {Meyer}, \citenamefont {Heinrich},
  \citenamefont {Zilberberg},\ and\ \citenamefont {Szameit}}]{Kremer2018}%
  \BibitemOpen
  \bibfield  {author} {\bibinfo {author} {\bibfnamefont {M.}~\bibnamefont
  {Kremer}}, \bibinfo {author} {\bibfnamefont {I.}~\bibnamefont {Petrides}},
  \bibinfo {author} {\bibfnamefont {E.}~\bibnamefont {Meyer}}, \bibinfo
  {author} {\bibfnamefont {M.}~\bibnamefont {Heinrich}}, \bibinfo {author}
  {\bibfnamefont {O.}~\bibnamefont {Zilberberg}}, \ and\ \bibinfo {author}
  {\bibfnamefont {A.}~\bibnamefont {Szameit}},\ }\href@noop {} {\bibfield
  {journal} {\bibinfo  {journal} {ArXiv e-prints}\ } (\bibinfo {year}
  {2018})},\ \Eprint {http://arxiv.org/abs/1805.05209} {arXiv:1805.05209}
  \BibitemShut {NoStop}%
\bibitem [{\citenamefont {Mukherjee}\ \emph {et~al.}(2018)\citenamefont
  {Mukherjee}, \citenamefont {Di~Liberto}, \citenamefont {\"Ohberg},
  \citenamefont {Thomson},\ and\ \citenamefont {Goldman}}]{Mukherjee2018}%
  \BibitemOpen
  \bibfield  {author} {\bibinfo {author} {\bibfnamefont {S.}~\bibnamefont
  {Mukherjee}}, \bibinfo {author} {\bibfnamefont {M.}~\bibnamefont
  {Di~Liberto}}, \bibinfo {author} {\bibfnamefont {P.}~\bibnamefont
  {\"Ohberg}}, \bibinfo {author} {\bibfnamefont {R.~R.}\ \bibnamefont
  {Thomson}}, \ and\ \bibinfo {author} {\bibfnamefont {N.}~\bibnamefont
  {Goldman}},\ }\href {\doibase 10.1103/PhysRevLett.121.075502} {\bibfield
  {journal} {\bibinfo  {journal} {Phys. Rev. Lett.}\ }\textbf {\bibinfo
  {volume} {121}},\ \bibinfo {pages} {075502} (\bibinfo {year}
  {2018})}\BibitemShut {NoStop}%
\bibitem [{\citenamefont {Gladchenko}\ \emph {et~al.}(2009)\citenamefont
  {Gladchenko}, \citenamefont {Olaya}, \citenamefont {Dupont-Ferrier},
  \citenamefont {Dou\ifmmode~\mbox{\c{c}}\else \c{c}\fi{}ot}, \citenamefont
  {Ioffe},\ and\ \citenamefont {Gershenson}}]{DoucotJJ2009}%
  \BibitemOpen
  \bibfield  {author} {\bibinfo {author} {\bibfnamefont {S.}~\bibnamefont
  {Gladchenko}}, \bibinfo {author} {\bibfnamefont {D.}~\bibnamefont {Olaya}},
  \bibinfo {author} {\bibfnamefont {E.}~\bibnamefont {Dupont-Ferrier}},
  \bibinfo {author} {\bibfnamefont {B.}~\bibnamefont
  {Dou\ifmmode~\mbox{\c{c}}\else \c{c}\fi{}ot}}, \bibinfo {author}
  {\bibfnamefont {L.~B.}\ \bibnamefont {Ioffe}}, \ and\ \bibinfo {author}
  {\bibfnamefont {M.~E.}\ \bibnamefont {Gershenson}},\ }\href {\doibase
  10.1038/nphys1151} {\bibfield  {journal} {\bibinfo  {journal} {Nat Phys}\
  }\textbf {\bibinfo {volume} {5}},\ \bibinfo {pages} {48} (\bibinfo {year}
  {2009})}\BibitemShut {NoStop}%
\bibitem [{\citenamefont {Ioffe}\ and\ \citenamefont
  {Feigel'man}(2002)}]{Ioffe2002}%
  \BibitemOpen
  \bibfield  {author} {\bibinfo {author} {\bibfnamefont {L.~B.}\ \bibnamefont
  {Ioffe}}\ and\ \bibinfo {author} {\bibfnamefont {M.~V.}\ \bibnamefont
  {Feigel'man}},\ }\href {\doibase 10.1103/PhysRevB.66.224503} {\bibfield
  {journal} {\bibinfo  {journal} {Phys. Rev. B}\ }\textbf {\bibinfo {volume}
  {66}},\ \bibinfo {pages} {224503} (\bibinfo {year} {2002})}\BibitemShut
  {NoStop}%
\bibitem [{\citenamefont {Douçot}\ and\ \citenamefont
  {Ioffe}(2012)}]{Doucot2012}%
  \BibitemOpen
  \bibfield  {author} {\bibinfo {author} {\bibfnamefont {B.}~\bibnamefont
  {Douçot}}\ and\ \bibinfo {author} {\bibfnamefont {L.~B.}\ \bibnamefont
  {Ioffe}},\ }\href {http://stacks.iop.org/0034-4885/75/i=7/a=072001}
  {\bibfield  {journal} {\bibinfo  {journal} {Reports on Progress in Physics}\
  }\textbf {\bibinfo {volume} {75}},\ \bibinfo {pages} {072001} (\bibinfo
  {year} {2012})}\BibitemShut {NoStop}%
\bibitem [{\citenamefont {White}(1992)}]{DMRG1}%
  \BibitemOpen
  \bibfield  {author} {\bibinfo {author} {\bibfnamefont {S.~R.}\ \bibnamefont
  {White}},\ }\href {\doibase 10.1103/PhysRevLett.69.2863} {\bibfield
  {journal} {\bibinfo  {journal} {Phys. Rev. Lett.}\ }\textbf {\bibinfo
  {volume} {69}},\ \bibinfo {pages} {2863} (\bibinfo {year}
  {1992})}\BibitemShut {NoStop}%
\bibitem [{\citenamefont {White}(1993)}]{DMRG2}%
  \BibitemOpen
  \bibfield  {author} {\bibinfo {author} {\bibfnamefont {S.~R.}\ \bibnamefont
  {White}},\ }\href {\doibase 10.1103/PhysRevB.48.10345} {\bibfield  {journal}
  {\bibinfo  {journal} {Phys. Rev. B}\ }\textbf {\bibinfo {volume} {48}},\
  \bibinfo {pages} {10345} (\bibinfo {year} {1993})}\BibitemShut {NoStop}%
\bibitem [{\citenamefont {De~Chiara}\ \emph {et~al.}(2008)\citenamefont
  {De~Chiara}, \citenamefont {Rizzi}, \citenamefont {Rossini},\ and\
  \citenamefont {Montangero}}]{DMRG3}%
  \BibitemOpen
  \bibfield  {author} {\bibinfo {author} {\bibfnamefont {G.}~\bibnamefont
  {De~Chiara}}, \bibinfo {author} {\bibfnamefont {M.}~\bibnamefont {Rizzi}},
  \bibinfo {author} {\bibfnamefont {D.}~\bibnamefont {Rossini}}, \ and\
  \bibinfo {author} {\bibfnamefont {S.}~\bibnamefont {Montangero}},\ }\href
  {\doibase doi:10.1166/jctn.2008.011} {\bibfield  {journal} {\bibinfo
  {journal} {Journal of Computational and Theoretical Nanoscience}\ }\textbf
  {\bibinfo {volume} {5}},\ \bibinfo {pages} {1277} (\bibinfo {year}
  {2008})}\BibitemShut {NoStop}%
\bibitem [{\citenamefont {Tovmasyan}\ \emph {et~al.}(2013)\citenamefont
  {Tovmasyan}, \citenamefont {van Nieuwenburg},\ and\ \citenamefont
  {Huber}}]{Huber2013}%
  \BibitemOpen
  \bibfield  {author} {\bibinfo {author} {\bibfnamefont {M.}~\bibnamefont
  {Tovmasyan}}, \bibinfo {author} {\bibfnamefont {E.~P.~L.}\ \bibnamefont {van
  Nieuwenburg}}, \ and\ \bibinfo {author} {\bibfnamefont {S.~D.}\ \bibnamefont
  {Huber}},\ }\href {\doibase 10.1103/PhysRevB.88.220510} {\bibfield  {journal}
  {\bibinfo  {journal} {Phys. Rev. B}\ }\textbf {\bibinfo {volume} {88}},\
  \bibinfo {pages} {220510} (\bibinfo {year} {2013})}\BibitemShut {NoStop}%
\bibitem [{\citenamefont {Takayoshi}\ \emph {et~al.}(2013)\citenamefont
  {Takayoshi}, \citenamefont {Katsura}, \citenamefont {Watanabe},\ and\
  \citenamefont {Aoki}}]{Takayoshi2013}%
  \BibitemOpen
  \bibfield  {author} {\bibinfo {author} {\bibfnamefont {S.}~\bibnamefont
  {Takayoshi}}, \bibinfo {author} {\bibfnamefont {H.}~\bibnamefont {Katsura}},
  \bibinfo {author} {\bibfnamefont {N.}~\bibnamefont {Watanabe}}, \ and\
  \bibinfo {author} {\bibfnamefont {H.}~\bibnamefont {Aoki}},\ }\href {\doibase
  10.1103/PhysRevA.88.063613} {\bibfield  {journal} {\bibinfo  {journal} {Phys.
  Rev. A}\ }\textbf {\bibinfo {volume} {88}},\ \bibinfo {pages} {063613}
  (\bibinfo {year} {2013})}\BibitemShut {NoStop}%
\bibitem [{\citenamefont {Phillips}\ \emph {et~al.}(2015)\citenamefont
  {Phillips}, \citenamefont {De~Chiara}, \citenamefont {\"Ohberg},\ and\
  \citenamefont {Valiente}}]{PhillipsPRB2015}%
  \BibitemOpen
  \bibfield  {author} {\bibinfo {author} {\bibfnamefont {L.~G.}\ \bibnamefont
  {Phillips}}, \bibinfo {author} {\bibfnamefont {G.}~\bibnamefont {De~Chiara}},
  \bibinfo {author} {\bibfnamefont {P.}~\bibnamefont {\"Ohberg}}, \ and\
  \bibinfo {author} {\bibfnamefont {M.}~\bibnamefont {Valiente}},\ }\href
  {\doibase 10.1103/PhysRevB.91.054103} {\bibfield  {journal} {\bibinfo
  {journal} {Phys. Rev. B}\ }\textbf {\bibinfo {volume} {91}},\ \bibinfo
  {pages} {054103} (\bibinfo {year} {2015})}\BibitemShut {NoStop}%
\bibitem [{\citenamefont {Anisimovas}\ \emph {et~al.}(2016)\citenamefont
  {Anisimovas}, \citenamefont {Ra\ifmmode \check{c}\else
  \v{c}\fi{}i\ifmmode~\bar{u}\else \={u}\fi{}nas}, \citenamefont {Str\"ater},
  \citenamefont {Eckardt}, \citenamefont {Spielman},\ and\ \citenamefont
  {Juzeli{\=u}nas}}]{Anisimovas2016}%
  \BibitemOpen
  \bibfield  {author} {\bibinfo {author} {\bibfnamefont {E.}~\bibnamefont
  {Anisimovas}}, \bibinfo {author} {\bibfnamefont {M.}~\bibnamefont {Ra\ifmmode
  \check{c}\else \v{c}\fi{}i\ifmmode~\bar{u}\else \={u}\fi{}nas}}, \bibinfo
  {author} {\bibfnamefont {C.}~\bibnamefont {Str\"ater}}, \bibinfo {author}
  {\bibfnamefont {A.}~\bibnamefont {Eckardt}}, \bibinfo {author} {\bibfnamefont
  {I.~B.}\ \bibnamefont {Spielman}}, \ and\ \bibinfo {author} {\bibfnamefont
  {G.}~\bibnamefont {Juzeli{\=u}nas}},\ }\href {\doibase
  10.1103/PhysRevA.94.063632} {\bibfield  {journal} {\bibinfo  {journal} {Phys.
  Rev. A}\ }\textbf {\bibinfo {volume} {94}},\ \bibinfo {pages} {063632}
  (\bibinfo {year} {2016})}\BibitemShut {NoStop}%
\bibitem [{\citenamefont {Calabrese}\ and\ \citenamefont
  {Cardy}(2009)}]{CalabreseCardyReview2009}%
  \BibitemOpen
  \bibfield  {author} {\bibinfo {author} {\bibfnamefont {P.}~\bibnamefont
  {Calabrese}}\ and\ \bibinfo {author} {\bibfnamefont {J.}~\bibnamefont
  {Cardy}},\ }\href {http://stacks.iop.org/1751-8121/42/i=50/a=504005}
  {\bibfield  {journal} {\bibinfo  {journal} {Journal of Physics A:
  Mathematical and Theoretical}\ }\textbf {\bibinfo {volume} {42}},\ \bibinfo
  {pages} {504005} (\bibinfo {year} {2009})}\BibitemShut {NoStop}%
\bibitem [{\citenamefont {Laflorencie}(2016)}]{LaflorencieReview}%
  \BibitemOpen
  \bibfield  {author} {\bibinfo {author} {\bibfnamefont {N.}~\bibnamefont
  {Laflorencie}},\ }\href {\doibase 10.1016/j.physrep.2016.06.008} {\bibfield
  {journal} {\bibinfo  {journal} {Physics Reports}\ }\textbf {\bibinfo {volume}
  {646}},\ \bibinfo {pages} {1 } (\bibinfo {year} {2016})}\BibitemShut
  {NoStop}%
\bibitem [{\citenamefont {De~Chiara}\ and\ \citenamefont
  {Sanpera}(2018)}]{DeChiaraSanpera2018}%
  \BibitemOpen
  \bibfield  {author} {\bibinfo {author} {\bibfnamefont {G.}~\bibnamefont
  {De~Chiara}}\ and\ \bibinfo {author} {\bibfnamefont {A.}~\bibnamefont
  {Sanpera}},\ }\href {\doibase 10.1088/1361-6633/aabf61} {\bibfield  {journal}
  {\bibinfo  {journal} {Reports on Progress in Physics}\ }\textbf {\bibinfo
  {volume} {81}},\ \bibinfo {pages} {074002} (\bibinfo {year}
  {2018})}\BibitemShut {NoStop}%
\bibitem [{\citenamefont {Pannetier}\ \emph {et~al.}(1984)\citenamefont
  {Pannetier}, \citenamefont {Chaussy}, \citenamefont {Rammal},\ and\
  \citenamefont {Villegier}}]{Pannetier1984}%
  \BibitemOpen
  \bibfield  {author} {\bibinfo {author} {\bibfnamefont {B.}~\bibnamefont
  {Pannetier}}, \bibinfo {author} {\bibfnamefont {J.}~\bibnamefont {Chaussy}},
  \bibinfo {author} {\bibfnamefont {R.}~\bibnamefont {Rammal}}, \ and\ \bibinfo
  {author} {\bibfnamefont {J.~C.}\ \bibnamefont {Villegier}},\ }\href {\doibase
  10.1103/PhysRevLett.53.1845} {\bibfield  {journal} {\bibinfo  {journal}
  {Phys. Rev. Lett.}\ }\textbf {\bibinfo {volume} {53}},\ \bibinfo {pages}
  {1845} (\bibinfo {year} {1984})}\BibitemShut {NoStop}%
\bibitem [{\citenamefont {Lee}\ and\ \citenamefont {Lee}(1998)}]{Sooyeul1998}%
  \BibitemOpen
  \bibfield  {author} {\bibinfo {author} {\bibfnamefont {S.}~\bibnamefont
  {Lee}}\ and\ \bibinfo {author} {\bibfnamefont {K.-C.}\ \bibnamefont {Lee}},\
  }\href {\doibase 10.1103/PhysRevB.57.8472} {\bibfield  {journal} {\bibinfo
  {journal} {Phys. Rev. B}\ }\textbf {\bibinfo {volume} {57}},\ \bibinfo
  {pages} {8472} (\bibinfo {year} {1998})}\BibitemShut {NoStop}%
\bibitem [{\citenamefont {Polini}\ \emph {et~al.}(2005)\citenamefont {Polini},
  \citenamefont {Fazio}, \citenamefont {MacDonald},\ and\ \citenamefont
  {Tosi}}]{Polini2005}%
  \BibitemOpen
  \bibfield  {author} {\bibinfo {author} {\bibfnamefont {M.}~\bibnamefont
  {Polini}}, \bibinfo {author} {\bibfnamefont {R.}~\bibnamefont {Fazio}},
  \bibinfo {author} {\bibfnamefont {A.~H.}\ \bibnamefont {MacDonald}}, \ and\
  \bibinfo {author} {\bibfnamefont {M.~P.}\ \bibnamefont {Tosi}},\ }\href
  {\doibase 10.1103/PhysRevLett.95.010401} {\bibfield  {journal} {\bibinfo
  {journal} {Phys. Rev. Lett.}\ }\textbf {\bibinfo {volume} {95}},\ \bibinfo
  {pages} {010401} (\bibinfo {year} {2005})}\BibitemShut {NoStop}%
\bibitem [{\citenamefont {Travkin}\ \emph {et~al.}(2017)\citenamefont
  {Travkin}, \citenamefont {Diebel},\ and\ \citenamefont {Denz}}]{Travkin2000}%
  \BibitemOpen
  \bibfield  {author} {\bibinfo {author} {\bibfnamefont {E.}~\bibnamefont
  {Travkin}}, \bibinfo {author} {\bibfnamefont {F.}~\bibnamefont {Diebel}}, \
  and\ \bibinfo {author} {\bibfnamefont {C.}~\bibnamefont {Denz}},\ }\href
  {\doibase 10.1063/1.4990998} {\bibfield  {journal} {\bibinfo  {journal}
  {Applied Physics Letters}\ }\textbf {\bibinfo {volume} {111}},\ \bibinfo
  {pages} {011104} (\bibinfo {year} {2017})}\BibitemShut {NoStop}%
\bibitem [{\citenamefont {Aoki}\ \emph {et~al.}(1996)\citenamefont {Aoki},
  \citenamefont {Ando},\ and\ \citenamefont {Matsumura}}]{Aoki1996}%
  \BibitemOpen
  \bibfield  {author} {\bibinfo {author} {\bibfnamefont {H.}~\bibnamefont
  {Aoki}}, \bibinfo {author} {\bibfnamefont {M.}~\bibnamefont {Ando}}, \ and\
  \bibinfo {author} {\bibfnamefont {H.}~\bibnamefont {Matsumura}},\ }\href
  {\doibase 10.1103/PhysRevB.54.R17296} {\bibfield  {journal} {\bibinfo
  {journal} {Phys. Rev. B}\ }\textbf {\bibinfo {volume} {54}},\ \bibinfo
  {pages} {R17296} (\bibinfo {year} {1996})}\BibitemShut {NoStop}%
\bibitem [{\citenamefont {Kazymyrenko}\ \emph {et~al.}(2005)\citenamefont
  {Kazymyrenko}, \citenamefont {Dusuel},\ and\ \citenamefont
  {Dou\ifmmode~\mbox{\c{c}}\else \c{c}\fi{}ot}}]{Kazymyrenko2005}%
  \BibitemOpen
  \bibfield  {author} {\bibinfo {author} {\bibfnamefont {K.}~\bibnamefont
  {Kazymyrenko}}, \bibinfo {author} {\bibfnamefont {S.}~\bibnamefont {Dusuel}},
  \ and\ \bibinfo {author} {\bibfnamefont {B.}~\bibnamefont
  {Dou\ifmmode~\mbox{\c{c}}\else \c{c}\fi{}ot}},\ }\href {\doibase
  10.1103/PhysRevB.72.235114} {\bibfield  {journal} {\bibinfo  {journal} {Phys.
  Rev. B}\ }\textbf {\bibinfo {volume} {72}},\ \bibinfo {pages} {235114}
  (\bibinfo {year} {2005})}\BibitemShut {NoStop}%
\bibitem [{\citenamefont {Lopes}\ and\ \citenamefont {Dias}(2011)}]{Lopes2011}%
  \BibitemOpen
  \bibfield  {author} {\bibinfo {author} {\bibfnamefont {A.~A.}\ \bibnamefont
  {Lopes}}\ and\ \bibinfo {author} {\bibfnamefont {R.~G.}\ \bibnamefont
  {Dias}},\ }\href {\doibase 10.1103/PhysRevB.84.085124} {\bibfield  {journal}
  {\bibinfo  {journal} {Phys. Rev. B}\ }\textbf {\bibinfo {volume} {84}},\
  \bibinfo {pages} {085124} (\bibinfo {year} {2011})}\BibitemShut {NoStop}%
\bibitem [{\citenamefont {Tarruell}\ \emph {et~al.}(2012)\citenamefont
  {Tarruell}, \citenamefont {Greif}, \citenamefont {Uehlinger}, \citenamefont
  {Jotzu},\ and\ \citenamefont {Esslinger}}]{Tarruell2011}%
  \BibitemOpen
  \bibfield  {author} {\bibinfo {author} {\bibfnamefont {L.}~\bibnamefont
  {Tarruell}}, \bibinfo {author} {\bibfnamefont {D.}~\bibnamefont {Greif}},
  \bibinfo {author} {\bibfnamefont {T.}~\bibnamefont {Uehlinger}}, \bibinfo
  {author} {\bibfnamefont {G.}~\bibnamefont {Jotzu}}, \ and\ \bibinfo {author}
  {\bibfnamefont {T.}~\bibnamefont {Esslinger}},\ }\href
  {http://dx.doi.org/10.1038/nature10871} {\bibfield  {journal} {\bibinfo
  {journal} {Nature}\ }\textbf {\bibinfo {volume} {483}},\ \bibinfo {pages}
  {302} (\bibinfo {year} {2012})}\BibitemShut {NoStop}%
\bibitem [{\citenamefont {Gauthier}\ \emph {et~al.}(2016)\citenamefont
  {Gauthier}, \citenamefont {Lenton}, \citenamefont {Parry}, \citenamefont
  {Baker}, \citenamefont {Davis}, \citenamefont {Rubinsztein-Dunlop},\ and\
  \citenamefont {Neely}}]{Gauthier2016DMD}%
  \BibitemOpen
  \bibfield  {author} {\bibinfo {author} {\bibfnamefont {G.}~\bibnamefont
  {Gauthier}}, \bibinfo {author} {\bibfnamefont {I.}~\bibnamefont {Lenton}},
  \bibinfo {author} {\bibfnamefont {N.~M.}\ \bibnamefont {Parry}}, \bibinfo
  {author} {\bibfnamefont {M.}~\bibnamefont {Baker}}, \bibinfo {author}
  {\bibfnamefont {M.~J.}\ \bibnamefont {Davis}}, \bibinfo {author}
  {\bibfnamefont {H.}~\bibnamefont {Rubinsztein-Dunlop}}, \ and\ \bibinfo
  {author} {\bibfnamefont {T.~W.}\ \bibnamefont {Neely}},\ }\href {\doibase
  10.1364/OPTICA.3.001136} {\bibfield  {journal} {\bibinfo  {journal} {Optica}\
  }\textbf {\bibinfo {volume} {3}},\ \bibinfo {pages} {1136} (\bibinfo {year}
  {2016})}\BibitemShut {NoStop}%
\bibitem [{\citenamefont {Boada}\ \emph {et~al.}(2012)\citenamefont {Boada},
  \citenamefont {Celi}, \citenamefont {Latorre},\ and\ \citenamefont
  {Lewenstein}}]{Boada2012}%
  \BibitemOpen
  \bibfield  {author} {\bibinfo {author} {\bibfnamefont {O.}~\bibnamefont
  {Boada}}, \bibinfo {author} {\bibfnamefont {A.}~\bibnamefont {Celi}},
  \bibinfo {author} {\bibfnamefont {J.~I.}\ \bibnamefont {Latorre}}, \ and\
  \bibinfo {author} {\bibfnamefont {M.}~\bibnamefont {Lewenstein}},\ }\href
  {\doibase 10.1103/PhysRevLett.108.133001} {\bibfield  {journal} {\bibinfo
  {journal} {Phys. Rev. Lett.}\ }\textbf {\bibinfo {volume} {108}},\ \bibinfo
  {pages} {133001} (\bibinfo {year} {2012})}\BibitemShut {NoStop}%
\bibitem [{\citenamefont {Celi}\ \emph {et~al.}(2014)\citenamefont {Celi},
  \citenamefont {Massignan}, \citenamefont {Ruseckas}, \citenamefont {Goldman},
  \citenamefont {Spielman}, \citenamefont {Juzeli{\=u}nas},\ and\ \citenamefont
  {Lewenstein}}]{Celi2014}%
  \BibitemOpen
  \bibfield  {author} {\bibinfo {author} {\bibfnamefont {A.}~\bibnamefont
  {Celi}}, \bibinfo {author} {\bibfnamefont {P.}~\bibnamefont {Massignan}},
  \bibinfo {author} {\bibfnamefont {J.}~\bibnamefont {Ruseckas}}, \bibinfo
  {author} {\bibfnamefont {N.}~\bibnamefont {Goldman}}, \bibinfo {author}
  {\bibfnamefont {I.~B.}\ \bibnamefont {Spielman}}, \bibinfo {author}
  {\bibfnamefont {G.}~\bibnamefont {Juzeli{\=u}nas}}, \ and\ \bibinfo {author}
  {\bibfnamefont {M.}~\bibnamefont {Lewenstein}},\ }\href {\doibase
  10.1103/PhysRevLett.112.043001} {\bibfield  {journal} {\bibinfo  {journal}
  {Phys. Rev. Lett.}\ }\textbf {\bibinfo {volume} {112}},\ \bibinfo {pages}
  {043001} (\bibinfo {year} {2014})}\BibitemShut {NoStop}%
\bibitem [{\citenamefont {Ma}\ \emph {et~al.}(2011)\citenamefont {Ma},
  \citenamefont {Tai}, \citenamefont {Preiss}, \citenamefont {Bakr},
  \citenamefont {Simon},\ and\ \citenamefont {Greiner}}]{Ma2011}%
  \BibitemOpen
  \bibfield  {author} {\bibinfo {author} {\bibfnamefont {R.}~\bibnamefont
  {Ma}}, \bibinfo {author} {\bibfnamefont {M.~E.}\ \bibnamefont {Tai}},
  \bibinfo {author} {\bibfnamefont {P.~M.}\ \bibnamefont {Preiss}}, \bibinfo
  {author} {\bibfnamefont {W.~S.}\ \bibnamefont {Bakr}}, \bibinfo {author}
  {\bibfnamefont {J.}~\bibnamefont {Simon}}, \ and\ \bibinfo {author}
  {\bibfnamefont {M.}~\bibnamefont {Greiner}},\ }\href {\doibase
  10.1103/PhysRevLett.107.095301} {\bibfield  {journal} {\bibinfo  {journal}
  {Phys. Rev. Lett.}\ }\textbf {\bibinfo {volume} {107}},\ \bibinfo {pages}
  {095301} (\bibinfo {year} {2011})}\BibitemShut {NoStop}%
\bibitem [{\citenamefont {Eckardt}\ \emph {et~al.}(2010)\citenamefont
  {Eckardt}, \citenamefont {Hauke}, \citenamefont {Soltan-Panahi},
  \citenamefont {Becker}, \citenamefont {Sengstock},\ and\ \citenamefont
  {Lewenstein}}]{Eckardt2010}%
  \BibitemOpen
  \bibfield  {author} {\bibinfo {author} {\bibfnamefont {A.}~\bibnamefont
  {Eckardt}}, \bibinfo {author} {\bibfnamefont {P.}~\bibnamefont {Hauke}},
  \bibinfo {author} {\bibfnamefont {P.}~\bibnamefont {Soltan-Panahi}}, \bibinfo
  {author} {\bibfnamefont {C.}~\bibnamefont {Becker}}, \bibinfo {author}
  {\bibfnamefont {K.}~\bibnamefont {Sengstock}}, \ and\ \bibinfo {author}
  {\bibfnamefont {M.}~\bibnamefont {Lewenstein}},\ }\href
  {http://stacks.iop.org/0295-5075/89/i=1/a=10010} {\bibfield  {journal}
  {\bibinfo  {journal} {EPL (Europhysics Letters)}\ }\textbf {\bibinfo {volume}
  {89}},\ \bibinfo {pages} {10010} (\bibinfo {year} {2010})}\BibitemShut
  {NoStop}%
\bibitem [{\citenamefont {K\"uhner}\ \emph {et~al.}(2000)\citenamefont
  {K\"uhner}, \citenamefont {White},\ and\ \citenamefont
  {Monien}}]{MonienWhiteBHmodel}%
  \BibitemOpen
  \bibfield  {author} {\bibinfo {author} {\bibfnamefont {T.~D.}\ \bibnamefont
  {K\"uhner}}, \bibinfo {author} {\bibfnamefont {S.~R.}\ \bibnamefont {White}},
  \ and\ \bibinfo {author} {\bibfnamefont {H.}~\bibnamefont {Monien}},\ }\href
  {\doibase 10.1103/PhysRevB.61.12474} {\bibfield  {journal} {\bibinfo
  {journal} {Phys. Rev. B}\ }\textbf {\bibinfo {volume} {61}},\ \bibinfo
  {pages} {12474} (\bibinfo {year} {2000})}\BibitemShut {NoStop}%
\bibitem [{\citenamefont {Granato}\ and\ \citenamefont
  {Kosterlitz}(1986)}]{Granato1986}%
  \BibitemOpen
  \bibfield  {author} {\bibinfo {author} {\bibfnamefont {E.}~\bibnamefont
  {Granato}}\ and\ \bibinfo {author} {\bibfnamefont {J.~M.}\ \bibnamefont
  {Kosterlitz}},\ }\href {\doibase 10.1103/PhysRevB.33.4767} {\bibfield
  {journal} {\bibinfo  {journal} {Phys. Rev. B}\ }\textbf {\bibinfo {volume}
  {33}},\ \bibinfo {pages} {4767} (\bibinfo {year} {1986})}\BibitemShut
  {NoStop}%
\bibitem [{\citenamefont {Granato}\ \emph {et~al.}(1991)\citenamefont
  {Granato}, \citenamefont {Kosterlitz}, \citenamefont {Lee},\ and\
  \citenamefont {Nightingale}}]{Granato1991}%
  \BibitemOpen
  \bibfield  {author} {\bibinfo {author} {\bibfnamefont {E.}~\bibnamefont
  {Granato}}, \bibinfo {author} {\bibfnamefont {J.~M.}\ \bibnamefont
  {Kosterlitz}}, \bibinfo {author} {\bibfnamefont {J.}~\bibnamefont {Lee}}, \
  and\ \bibinfo {author} {\bibfnamefont {M.~P.}\ \bibnamefont {Nightingale}},\
  }\href {\doibase 10.1103/PhysRevLett.66.1090} {\bibfield  {journal} {\bibinfo
   {journal} {Phys. Rev. Lett.}\ }\textbf {\bibinfo {volume} {66}},\ \bibinfo
  {pages} {1090} (\bibinfo {year} {1991})}\BibitemShut {NoStop}%
\bibitem [{\citenamefont {Teitel}\ and\ \citenamefont
  {Jayaprakash}(1983)}]{Teitel1983}%
  \BibitemOpen
  \bibfield  {author} {\bibinfo {author} {\bibfnamefont {S.}~\bibnamefont
  {Teitel}}\ and\ \bibinfo {author} {\bibfnamefont {C.}~\bibnamefont
  {Jayaprakash}},\ }\href {\doibase 10.1103/PhysRevB.27.598} {\bibfield
  {journal} {\bibinfo  {journal} {Phys. Rev. B}\ }\textbf {\bibinfo {volume}
  {27}},\ \bibinfo {pages} {598} (\bibinfo {year} {1983})}\BibitemShut
  {NoStop}%
\bibitem [{\citenamefont {Halsey}(1985)}]{Halsey1985}%
  \BibitemOpen
  \bibfield  {author} {\bibinfo {author} {\bibfnamefont {T.~C.}\ \bibnamefont
  {Halsey}},\ }\href {http://stacks.iop.org/0022-3719/18/i=12/a=008} {\bibfield
   {journal} {\bibinfo  {journal} {Journal of Physics C: Solid State Physics}\
  }\textbf {\bibinfo {volume} {18}},\ \bibinfo {pages} {2437} (\bibinfo {year}
  {1985})}\BibitemShut {NoStop}%
\bibitem [{\citenamefont {Korshunov}(1986)}]{Korshunov1986}%
  \BibitemOpen
  \bibfield  {author} {\bibinfo {author} {\bibfnamefont {S.~E.}\ \bibnamefont
  {Korshunov}},\ }\href {\doibase 10.1007/BF01010570} {\bibfield  {journal}
  {\bibinfo  {journal} {Journal of Statistical Physics}\ }\textbf {\bibinfo
  {volume} {43}},\ \bibinfo {pages} {17} (\bibinfo {year} {1986})}\BibitemShut
  {NoStop}%
\bibitem [{\citenamefont {Li}\ and\ \citenamefont {Cieplak}(1994)}]{Li1994}%
  \BibitemOpen
  \bibfield  {author} {\bibinfo {author} {\bibfnamefont {M.~S.}\ \bibnamefont
  {Li}}\ and\ \bibinfo {author} {\bibfnamefont {M.}~\bibnamefont {Cieplak}},\
  }\href {\doibase 10.1103/PhysRevB.50.955} {\bibfield  {journal} {\bibinfo
  {journal} {Phys. Rev. B}\ }\textbf {\bibinfo {volume} {50}},\ \bibinfo
  {pages} {955} (\bibinfo {year} {1994})}\BibitemShut {NoStop}%
\bibitem [{\citenamefont {Ejima}\ \emph {et~al.}(2012)\citenamefont {Ejima},
  \citenamefont {Fehske}, \citenamefont {Gebhard}, \citenamefont
  {zu~M{\"u}nster}, \citenamefont {Knap}, \citenamefont {Arrigoni},\ and\
  \citenamefont {von~der Linden}}]{Ejima2012}%
  \BibitemOpen
  \bibfield  {author} {\bibinfo {author} {\bibfnamefont {S.}~\bibnamefont
  {Ejima}}, \bibinfo {author} {\bibfnamefont {H.}~\bibnamefont {Fehske}},
  \bibinfo {author} {\bibfnamefont {F.}~\bibnamefont {Gebhard}}, \bibinfo
  {author} {\bibfnamefont {K.}~\bibnamefont {zu~M{\"u}nster}}, \bibinfo
  {author} {\bibfnamefont {M.}~\bibnamefont {Knap}}, \bibinfo {author}
  {\bibfnamefont {E.}~\bibnamefont {Arrigoni}}, \ and\ \bibinfo {author}
  {\bibfnamefont {W.}~\bibnamefont {von~der Linden}},\ }\href {\doibase
  10.1103/PhysRevA.85.053644} {\bibfield  {journal} {\bibinfo  {journal} {Phys.
  Rev. A}\ }\textbf {\bibinfo {volume} {85}},\ \bibinfo {pages} {053644}
  (\bibinfo {year} {2012})}\BibitemShut {NoStop}%
\bibitem [{\citenamefont {Giamarchi}(2004)}]{Giamarchi2004}%
  \BibitemOpen
  \bibfield  {author} {\bibinfo {author} {\bibfnamefont {T.}~\bibnamefont
  {Giamarchi}},\ }\href@noop {} {\emph {\bibinfo {title} {Quantum Physics in
  One Dimension}}},\ Vol.\ \bibinfo {volume} {121}\ (\bibinfo  {publisher}
  {Oxford university press},\ \bibinfo {address} {Oxford},\ \bibinfo {year}
  {2004})\BibitemShut {NoStop}%
\bibitem [{\citenamefont {Vidal}\ \emph {et~al.}(2003)\citenamefont {Vidal},
  \citenamefont {Latorre}, \citenamefont {Rico},\ and\ \citenamefont
  {Kitaev}}]{Vidal2003}%
  \BibitemOpen
  \bibfield  {author} {\bibinfo {author} {\bibfnamefont {G.}~\bibnamefont
  {Vidal}}, \bibinfo {author} {\bibfnamefont {J.~I.}\ \bibnamefont {Latorre}},
  \bibinfo {author} {\bibfnamefont {E.}~\bibnamefont {Rico}}, \ and\ \bibinfo
  {author} {\bibfnamefont {A.}~\bibnamefont {Kitaev}},\ }\href {\doibase
  10.1103/PhysRevLett.90.227902} {\bibfield  {journal} {\bibinfo  {journal}
  {Phys. Rev. Lett.}\ }\textbf {\bibinfo {volume} {90}},\ \bibinfo {pages}
  {227902} (\bibinfo {year} {2003})}\BibitemShut {NoStop}%
\bibitem [{\citenamefont {Calabrese}\ and\ \citenamefont
  {Cardy}(2004)}]{Calabrese2004}%
  \BibitemOpen
  \bibfield  {author} {\bibinfo {author} {\bibfnamefont {P.}~\bibnamefont
  {Calabrese}}\ and\ \bibinfo {author} {\bibfnamefont {J.}~\bibnamefont
  {Cardy}},\ }\href {\doibase 10.1088/1742-5468/2004/06/P06002} {\bibfield
  {journal} {\bibinfo  {journal} {Journal of Statistical Mechanics: Theory and
  Experiment}\ }\textbf {\bibinfo {volume} {2004}},\ \bibinfo {pages} {P06002}
  (\bibinfo {year} {2004})}\BibitemShut {NoStop}%
\bibitem [{\citenamefont {De~Chiara}\ \emph {et~al.}(2011)\citenamefont
  {De~Chiara}, \citenamefont {Lewenstein},\ and\ \citenamefont
  {Sanpera}}]{DeChiaraPRB2011}%
  \BibitemOpen
  \bibfield  {author} {\bibinfo {author} {\bibfnamefont {G.}~\bibnamefont
  {De~Chiara}}, \bibinfo {author} {\bibfnamefont {M.}~\bibnamefont
  {Lewenstein}}, \ and\ \bibinfo {author} {\bibfnamefont {A.}~\bibnamefont
  {Sanpera}},\ }\href {\doibase 10.1103/PhysRevB.84.054451} {\bibfield
  {journal} {\bibinfo  {journal} {Phys. Rev. B}\ }\textbf {\bibinfo {volume}
  {84}},\ \bibinfo {pages} {054451} (\bibinfo {year} {2011})}\BibitemShut
  {NoStop}%
\bibitem [{\citenamefont {Li}\ and\ \citenamefont {Haldane}(2008)}]{LiHaldane}%
  \BibitemOpen
  \bibfield  {author} {\bibinfo {author} {\bibfnamefont {H.}~\bibnamefont
  {Li}}\ and\ \bibinfo {author} {\bibfnamefont {F.~D.~M.}\ \bibnamefont
  {Haldane}},\ }\href {\doibase 10.1103/PhysRevLett.101.010504} {\bibfield
  {journal} {\bibinfo  {journal} {Phys. Rev. Lett.}\ }\textbf {\bibinfo
  {volume} {101}},\ \bibinfo {pages} {010504} (\bibinfo {year}
  {2008})}\BibitemShut {NoStop}%
\bibitem [{\citenamefont {Pollmann}\ \emph {et~al.}(2010)\citenamefont
  {Pollmann}, \citenamefont {Turner}, \citenamefont {Berg},\ and\ \citenamefont
  {Oshikawa}}]{Oshikawa}%
  \BibitemOpen
  \bibfield  {author} {\bibinfo {author} {\bibfnamefont {F.}~\bibnamefont
  {Pollmann}}, \bibinfo {author} {\bibfnamefont {A.~M.}\ \bibnamefont
  {Turner}}, \bibinfo {author} {\bibfnamefont {E.}~\bibnamefont {Berg}}, \ and\
  \bibinfo {author} {\bibfnamefont {M.}~\bibnamefont {Oshikawa}},\ }\href
  {\doibase 10.1103/PhysRevB.81.064439} {\bibfield  {journal} {\bibinfo
  {journal} {Phys. Rev. B}\ }\textbf {\bibinfo {volume} {81}},\ \bibinfo
  {pages} {064439} (\bibinfo {year} {2010})}\BibitemShut {NoStop}%
\bibitem [{\citenamefont {Deng}\ and\ \citenamefont {Santos}(2011)}]{ESDeng}%
  \BibitemOpen
  \bibfield  {author} {\bibinfo {author} {\bibfnamefont {X.}~\bibnamefont
  {Deng}}\ and\ \bibinfo {author} {\bibfnamefont {L.}~\bibnamefont {Santos}},\
  }\href {\doibase 10.1103/PhysRevB.84.085138} {\bibfield  {journal} {\bibinfo
  {journal} {Phys. Rev. B}\ }\textbf {\bibinfo {volume} {84}},\ \bibinfo
  {pages} {085138} (\bibinfo {year} {2011})}\BibitemShut {NoStop}%
\bibitem [{\citenamefont {De~Chiara}\ \emph {et~al.}(2012)\citenamefont
  {De~Chiara}, \citenamefont {Lepori}, \citenamefont {Lewenstein},\ and\
  \citenamefont {Sanpera}}]{DeChiaraPRL2012}%
  \BibitemOpen
  \bibfield  {author} {\bibinfo {author} {\bibfnamefont {G.}~\bibnamefont
  {De~Chiara}}, \bibinfo {author} {\bibfnamefont {L.}~\bibnamefont {Lepori}},
  \bibinfo {author} {\bibfnamefont {M.}~\bibnamefont {Lewenstein}}, \ and\
  \bibinfo {author} {\bibfnamefont {A.}~\bibnamefont {Sanpera}},\ }\href
  {\doibase 10.1103/PhysRevLett.109.237208} {\bibfield  {journal} {\bibinfo
  {journal} {Phys. Rev. Lett.}\ }\textbf {\bibinfo {volume} {109}},\ \bibinfo
  {pages} {237208} (\bibinfo {year} {2012})}\BibitemShut {NoStop}%
\bibitem [{\citenamefont {Lepori}\ \emph {et~al.}(2013)\citenamefont {Lepori},
  \citenamefont {De~Chiara},\ and\ \citenamefont {Sanpera}}]{ESDeChiara2013}%
  \BibitemOpen
  \bibfield  {author} {\bibinfo {author} {\bibfnamefont {L.}~\bibnamefont
  {Lepori}}, \bibinfo {author} {\bibfnamefont {G.}~\bibnamefont {De~Chiara}}, \
  and\ \bibinfo {author} {\bibfnamefont {A.}~\bibnamefont {Sanpera}},\ }\href
  {\doibase 10.1103/PhysRevB.87.235107} {\bibfield  {journal} {\bibinfo
  {journal} {Phys. Rev. B}\ }\textbf {\bibinfo {volume} {87}},\ \bibinfo
  {pages} {235107} (\bibinfo {year} {2013})}\BibitemShut {NoStop}%
\bibitem [{\citenamefont {L{\"a}uchli}(2013)}]{Lauchli2013Operator}%
  \BibitemOpen
  \bibfield  {author} {\bibinfo {author} {\bibfnamefont {A.~M.}\ \bibnamefont
  {L{\"a}uchli}},\ }\href@noop {} {\bibfield  {journal} {\bibinfo  {journal}
  {ArXiv e-prints}\ } (\bibinfo {year} {2013})},\ \Eprint
  {http://arxiv.org/abs/1303.0741} {arXiv:1303.0741} \BibitemShut {NoStop}%
\end{thebibliography}%
\vspace{0.5cm}
\appendix*
\renewcommand\thefigure{A.\arabic{figure}}    
\section{Increasing the robustness of the PLL phase} \manuallabel{sec:app}{Appendix}
\setcounter{figure}{0}  
In order to increase the robustness of the PLL phase the Hamiltonian can be formulated using a tunnelling modulation instead of the magnetic flux used previously (see Sec.~\ref{sectmod}). In order to do this the Hamiltonian is again given by:
\begin{eqnarray} 
\label{BHeqncos}
\hat{H}_{BH}&=& \hat{H}_0 + \hat{H}_U  \equiv 
\hat{H}_0 + \frac{U}{2}\sum_j \sum_{\alpha} \hat{n}_{j,\alpha}(\hat{n}_{j,\alpha}-1) \\
\hat{H}_0 & = & - J \sum_j \sum_\ell \sum_{\alpha,\beta} 
T^{(\ell)}_{\alpha ,\beta} \ \hat{b}^{\dagger}_{j+\ell,\alpha} \hat{b}^{\phantom{\dagger}}_{j,\beta} \ ,
\end{eqnarray}\\

\begin{figure}[h!]
\includegraphics[width=0.97\columnwidth,trim={0.05cm 1.1cm 0.05cm 2.5cm}]{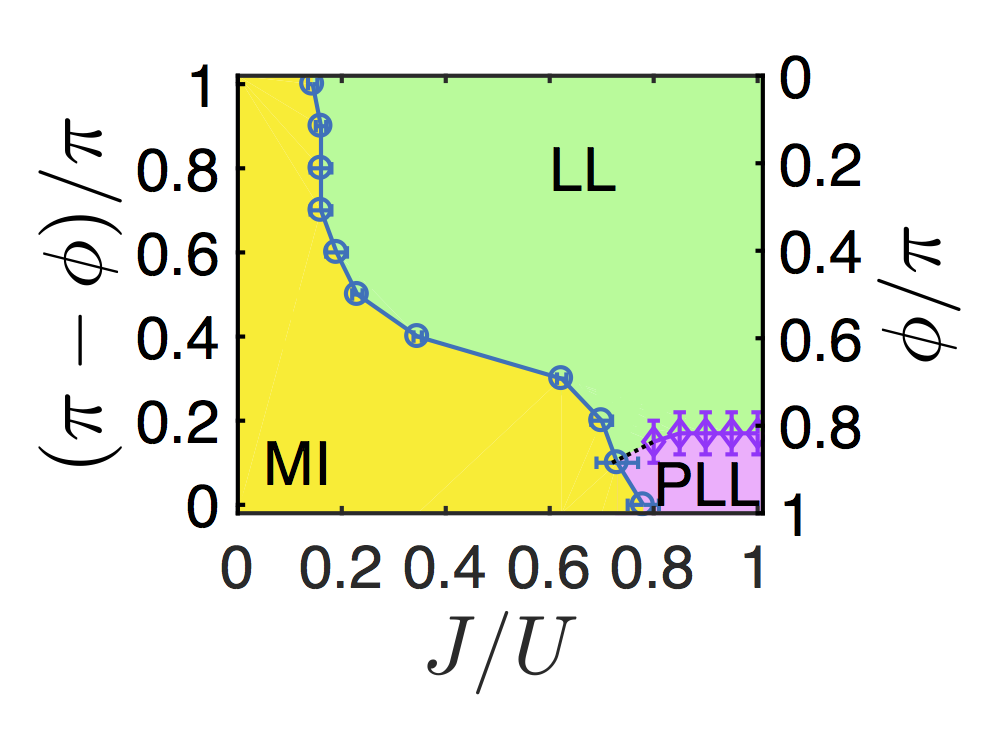}
\caption{The phase diagram for an infinite estimated length of the Hamiltonian Eq.~\eqref{BHeqncos} for amplitude modulation of $\cos(\phi)$ against parameters $J/U$. The MI, LL and PLL regions are as labelled.The MI-LL  and the MI-PLL transitions are again obtained from the compressibility of the energy gap (see Sec.~\ref{sectphase}). The LL and PLL phases are characterised by the decay of the correlation functions as illustrated in Sec.~\ref{sectPLL}).}
\label{appcosphase}
\end{figure}

where the labels are as previously defined (Sec.~\ref{sectmod}).
In order to modulate the amplitude, a $\cos(\phi)$ factor is included in the $C-B$ leg of each diamond (as shown by the dashed lines in Fig.~\ref{model}). This means the hopping matrices are now instead:
\begin{equation}\label{eq:tunnellcos}
{T}^{(0)} = \left(\begin{array}{ccc}
0&1&0\\
1&0&1\\
0&1&0\\
\end{array} \right),
\quad
{T}^{(+1)}=\left(\begin{array}{ccc}
0&1&0\\
0&0& \cos(\phi)\\
0&0&0\\
\end{array} \right),
\quad
{T}^{(-1)}= \left({T}^{(+1)}\right)^\dagger
\end{equation}

It should be noted that this modulation will be exactly the same Hamiltonian in either extreme case, i.e. fully unfrustrated ($\phi=0$) and fully frustrated ($\phi=\pi$). The differences and advantages to an experimental replication only occur when exploring imperfect frustration. This advantage is evident in the increased size and therefore robustness of the PLL region (see Fig. \ref{appcosphase}). As mentioned in the paper, this can be performed experimentally using digital micromirror devices (DMDs) or single atom microscopes. The reason for the increased region of PLL can be connected to the curvature of the single particle momentum bands. When a magnetic flux is applied a slight shift from full frustration i.e ($\phi=\pi-\epsilon$), results in a an almost immediate loss of the flatband property of the bands. Using the $\cos{\phi}$ adaptation, the bands retain their flat property (i.e almost full frustration) for small shifts away from $\phi=\pi$, so the PLL remains intact.

\end{document}